
\input epsf

\magnification\magstep1


\overfullrule=0pt  
\hbadness=10000      
\vbadness=10000

\font\eightit=cmti8  
\font\eightrm=cmr8 \font\eighti=cmmi8                 
\font\eightsy=cmsy8 
\font\sixrm=cmr6
\font\sevenrm=cmr7

\def\eightpoint{\normalbaselineskip=10pt 
\def\rm{\eightrm\fam0} \let\it\eightit
\textfont0=\eightrm \scriptfont0=\sixrm 
\textfont1=\eighti \scriptfont1=\seveni
\textfont2=\eightsy \scriptfont2=\sevensy 
\normalbaselines \eightrm
\parindent=1em}

\def\sevenpoint{\normalbaselineskip=8pt 
\def\rm{\sevenrm\fam0} \let\it\sevenit
\textfont0=\sevenrm \scriptfont0=\sixrm 
\textfont1=\seveni \scriptfont1=\seveni
\textfont2=\sevensy \scriptfont2=\sevensy 
\normalbaselines \sevenrm
\parindent=0.5em}

\def\eq#1{{\noexpand\rm(#1)}}          
\def\eqprime#1{{\noexpand\rm(#1')}}          
\newcount\eqcounter                    
\eqcounter=0                           
\def\numeq{\global\advance\eqcounter by 1\eq{\the\eqcounter}}           
\def\relativeq#1{{\advance\eqcounter by #1\eq{\the\eqcounter}}}
\def\relativeqprime#1{{\advance\eqcounter by #1\eqprime{\the\eqcounter}}}
\def\lasteqprime{\relativeqprime0}     

\def\namelasteq#1{\global\edef#1{{\eq{\the\eqcounter}}}}  

\def\A{{\rm A}}
\def\Aslash{{A\mkern-9mu/}}    
\def\a{\alpha}                        
\def\arrowsim#1{{
  \buildrel\leftrightarrow\over{#1} }} 
\def\b{\beta}                          
\def\cite#1{{\rm[#1]}}                 
\def\Dslash{{ {\rm D}\mkern-11mu/}}    
\def\eps{\varepsilon}                  
\def\ga{\gamma}                     
\def\ghat{\hat g}                      
\def\gbar{\bar g}                      
\def\gamhat{\hat\gamma}                
\def\gambar{\bar\gamma}                
\def\gam5{\gamma_5}
\def\Gapcero{{\tilde\Gamma}^{(0)}}     
\def\intd{\int\! {\rm d}}              
\def\L{{\rm L}}                        
\def\leftslashedarrow{{{\buildrel\leftarrow\over\pr}\mkern-12mu/}{\!}}
\def\pr{\partial}                      
\def\PL{{P_\L}}                        
\def\PR{{P_\R}}                        
\def\kslash{{k\mkern-8mu/}{\!}}        
\def\pslash{{p\mkern-8mu/}{\!}}        
\def\prslash{{\partial\mkern-9mu/}}    
\def\R{{\rm R}}                        
\def\rightslashedarrow{{{\buildrel\leftarrow\over\pr}\mkern-12mu/}{\!}}
\def\square{\vbox{\hrule\hbox{\vrule height5.2pt \hskip 5.2pt
  \vrule}\hrule}}                      
\def\SS{{\cal S}}     
\def\Tr{{\rm Tr}}                      
\def\ti{\tilde}
\def\un{{\rm I}\!{\rm I}}              

\newif\ifstartsec                   

\outer\def\section#1{\vskip 0pt plus .15\vsize \penalty -250
\vskip 0pt plus -.15\vsize \bigskip \startsectrue
\message{#1}\centerline{\bf#1}\nobreak\noindent}

\def\subsection#1{\ifstartsec\medskip\else\bigskip\fi \startsecfalse
\noindent{\it#1}\penalty100\medskip}

\def\refno#1. #2\par{\smallskip\item{\rm\lbrack#1\rbrack}#2\par}

\hyphenation{geo-me-try}
\hyphenation{Ma-the-ma-thi-ca}
\hyphenation{in-de-pen-dent}

\def\Ash{1}
\def\HV{2}
\def\BMabc{3}
\def\BonneauABC{4}
\def\Collins{5}
\def\Min{6}
\def\BonneauRemarks{7}
\def\naive{8}
\def\BonneauReview{9}
\def\facts{10}
\def\Kreimer{11}
\def\Delbourgo{12}
\def\Tonin{13}
\def\PerniciABC{14}
\def\Hepp{15}
\def\Colnor{16}
\def\bmym{17}
\def\nochiral{18}
\def\axial{19}
\def\Korner{20}
\def\Ferrari{21}
\def\Weiglein{22}
\def\Susy{23}
\def\Epstein{24}
\def\QAP{25}
\def\BRSah{26}
\def\BRSann{27}
\def\Piguet{28}
\def\ARSM{29}
\def\GrassiAH{30}
\def\GrassiPAR{31}
\def\abhk{32}
\def\GrossJackiw{33}
\def\Mathematica{34}
\def\Tracer{35}



\vskip 1cm

\centerline{\bf BRS symmetry restoration of
                chiral Abelian Higgs-Kibble theory}
\centerline{\bf in dimensional renormalization with a non-anticommuting 
$\gam5$}

\bigskip

\centerline{\rm D. S\'anchez-Ruiz*}
\medskip
\centerline{\eightit Departamento de F{\'\i}sica Te\'orica I,			Universidad Complutense, 28040 Madrid, Spain}
\vfootnote*{email: {\tt domingo@toboso.fis.ucm.es}}

\bigskip\bigskip

\begingroup\narrower\narrower
\eightpoint
The one-loop renormalization of the Abelian Higgs-Kibble model
in a general 't Hooft gauge and with chiral fermions
is fully worked out within 
dimensional renormalization scheme with a non-anticommuting
$\gam5$. The anomalous terms 
introduced in the Slavnov-Taylor identities by the minimal subtraction 
algorithm are calculated and the asymmetric counterterms needed to restore the 
BRS symmetry, if the anomaly cancellation conditions are met, are
computed. The computations draw heavily from
regularized action principles and Algebraic Renormalization theory.
\par
\endgroup 

\medskip

\section{1. Introduction}

Due to the
availability of high precision tests of the Standard Model
in particle accelerators,
it is mandatory to compute higher order quantum corrections
and therefore to investigate thoroughly consistent and
systematic renormalization schemes. 
Dimensional Regularization \cite{\Ash,\HV } is the standard regularization 
method applied to particle physics.
Both its axiomatic and properties were rigorously 
established long ago 
\cite{\BMabc,\BonneauABC,\Collins }.
Its success
as a practical regularization method stems from the fact that 
in vector-like non-supersymmetric  gauge theories preserves 
enough properties 
so that the minimal subtraction (MS) scheme \cite{\Min}
leads to a renormalized gauge invariant theory \cite{\HV}.
But the electroweak interactions of the Standard Model are chiral, and,
unfortunately,  in Dimensional Renormalization 
the algebraic properties of $\gamma_5$ cannot be maintained consistently
as we move away from four dimensions \cite{\BonneauRemarks }. 
Thus, in the so called ``Naive'' prescription of Dimensional
Renormalization (NDR), commonly used in
multiloop computations in the Standard Model, one assumes
$[\gam5,\gamma_{\mu}]=0$ and the
cyclicity of the trace \cite{\naive}. 
These assumptions have the consequence
\cite{\BMabc,\Collins,\BonneauRemarks,\BonneauReview}
that $\Tr[\gamma_{\mu_1}\ldots\gamma_{\mu_4}\gam5]$
is identically 0 in the dimensionally regularized theory ($d\ne4$),
which is incompatible with the property in four dimensions
$\Tr\,[\gamma^{\mu_1}\cdots\gamma^{\mu_4}\gam5]=
 i\, \hbox{Tr}\,\un \,\,\epsilon_{\mu_1\ldots\mu_4}$ 
in theories where a not null tensor $\epsilon$
is needed.  

Therefore it seems unavoidable to find ambiguities as we use
NDR to carry out computations in any
chiral theory with fermionic loops with an odd number
of $\gam5$'s. However, it is 
commonly claimed that with usual manipulations the resulting
ambiguities are proportional to the
coefficient of the (chiral gauge) anomaly
--- null in the Standard Model --- and the ``naive'' prescription is
then thought to be safe \cite{\facts}.
Calculations to low orders in perturbation theory
support this idea, but there is not a rigorous
proof of it  valid to all orders. 
Higher order computations within the SM
have already reached a point where
the possible inconsistencies of NDR
cannot be simply forgotten.
To avoid these algebraic
inconsistencies it has been suggested not to assume
the cyclicity of the trace \cite{\facts, \Kreimer}, but
in this case extra checks or proofs are needed.
The only dimensional dimensional regularization scheme
(sometimes named as BMHV)
which is known rigorously \cite{\BonneauRemarks} to be consistent
in presence of
$\gamma_5$ is the original one devised by 't Hooft and Veltman
\cite{\HV} and later systematized by Breitenlohner and
Maison \cite{\BMabc} following
the definition of $\gamma_5$ in \cite{\Delbourgo}. 
A similar scheme is considered in \cite{\Tonin} and
an extension has been
recently proposed in \cite{\PerniciABC}.

In BMHV scheme, besides the ``$d$-dimensional'' metrics
$g_{\mu\nu}$, a new one is introduced \cite{\BMabc},
$\ghat_{\mu\nu}$, which can be
considered as a ``$(d-4)$-dimensional covariant''. Defining
$\gbar^{\mu\nu} = g^{\mu\nu} - \ghat^{\mu\nu}$,
$\gbar^{\mu\nu}$ can be thought of as a projector over the ``4-dimensional
space'' and $\ghat^{\mu\nu}$ as a projector over the ``$(d-4)$-dimensional''
one.
Moreover
the $\epsilon$ tensor is  considered to be
a ``4 dimensional covariant'' object, because it is assumed to
satisfy
$
   \epsilon_{\mu_1\ldots\mu_4} \epsilon_{\nu_1\ldots\nu_4} =
        -\sum_{\pi\in\, S_4} \hbox{sign}\, \pi \prod_{i=1}^4 \,
      \gbar_{\mu_i\nu_{\pi(i)}},
$. 
With the definition $\gam5={i \over 4!} \epsilon_{\mu_1\ldots\mu_4}
        \gamma^{\mu_1}\cdots\gamma^{\mu_4}$
and the {\it assumption of cyclicity\/} of the symbol Tr,
it can be proved algebraically \cite{\BMabc}
$$\eqalignno{
&\Tr\,[\gamma^{\mu_1}\cdots\gamma^{\mu_4}\gam5]=
 \Tr\,[\gambar^{\mu_1}\cdots\gambar^{\mu_4}\gam5]=
 i\, \hbox{Tr}\,\un \,\,\epsilon_{\mu_1\ldots\mu_4};               \cr
&\{\gam5,\gamma^\mu\}=\{\gam5,\gamhat^\mu\}=2\gam5 \gamhat^\mu=\,
        2\gamhat^\mu \gam5, \;\;
\{\gam5,\gambar^\mu\}=[\gam5,\gamhat^\mu]=0,             
 &\numeq\cr
} \namelasteq\EqBMHV
$$
which is the same algebra of symbols than of the original
one of 't Hooft and Veltman
\cite{\HV,\Collins}. Notice that $\gam5$ no longer anticommutes
with $\gamma_\mu$ as it does in NDR.

The four dimensional projection of the minimal subtraction (MS)
of the singular part of the dimensionally regularized Feynman diagrams,
their  divergences having been also minimally subtracted in advance,
leads to a renormalized quantum field theory 
\cite{\BMabc,\BonneauABC} 
that satisfies Hepp axioms \cite{\Hepp} of renormalization theory,
field equations, the action principles and Zimmerman-Bonneau identities
\cite{\BMabc,\BonneauABC,\Colnor}.
The reader can find a review in \cite{\bmym}.

Not many computations have been done in the BMHV scheme. Most of
them involve theories without a chiral gauge symmetry \cite{\nochiral},
where there is no room for the introduction of
spurious anomalies but subtleties related with
evanescent operators \cite{\Collins} appear,
or quantities related with axial currents \cite{\Delbourgo,\axial}, 
where it gives correctly the essential axial current anomalies.
Several practical computations,
taking care of the fulfillment of
Slavnov Taylor Identities (STI), but
involving only some restricted set of diagrams, 
have been done in the case of the Standard Model
\cite{\Korner,\Ferrari,\Weiglein} or in Supersymmetric QED
\cite{\Susy}. In \cite{\Weiglein},
the authors report
a relevant finite difference between the results
of NDR and BMHV in two-loop diagrams of the Standard
Model containing triangle subdiagrams.

In BMHV scheme the regularization breaks the gauge symmetry,
and the MS procedure gives Green's functions which does not
fulfill the STIs, essential ingredients to guarantee unitarity 
and gauge independence. The axioms of local relativistic quantum
field theories allows for local ambiguities \cite{\Hepp,\Epstein}
to be removed by imposing renormalization conditions. The Quantum
Action Principle \cite{\QAP} tells us that any symmetry
breaking term generated in the renormalization process is local
at the lowest non-vanishing order,
and the question here is whether  
it is possible to remove the breaking terms of the STIs
by adding to the classical action appropriate local finite counterterms. 
(The removable breaking terms are called
{\it spurious anomalies}) \cite{\BonneauReview}.
Algebraic Renormalization theory
\cite{\BRSah,\BRSann,\Piguet}
establish elegantly how and when this process can be accomplished: when
there are no obstructions in the form of ({\it essential}) anomalies.
This impose anomaly cancellation conditions on the field
representations.
These anomalies belong to the cohomology of the Slavnov-Taylor 
operator that governs the chiral gauge symmetry at the quantum level.
On the other hand, non-physical or spurious anomalies correspond to trivial
objects in the cohomology.

Algebraic Renormalization has been recently applied to theoretical
studies of Standard Model \cite{\ARSM} and important progress
towards practical uses of non-invariant
regularization schemes \cite{\GrassiAH,\GrassiPAR}
have been achieved.
In \cite{\bmym} a systematic computation of
the finite one-loop counterterms needed to restore
the STI of non-Abelian
gauge theories without scalar fields in
BMHV dimensional renormalization was done.
The modified action which gives gauge invariant results in MS
of BMHV scheme was explicitly given.
Specially use of the rigorous identity (see \cite{\bmym}
for notations and a deduction based on the 
Action Principles of Dimensional Regularization \cite{\BMabc})
$$
\eqalignno{
\SS_{d}(\Gamma_{\rm DReg})\equiv&\intd^d x\,\,
(s_d\varphi){\delta\Gamma_{\rm DReg}\over\delta\varphi} +
{\delta\Gamma_{\rm DReg}\over\delta K_\Phi} 
{\delta\Gamma_{\rm DReg}\over\delta\Phi}=     \cr
=&\,{s_d S_0}\cdot\Gamma_{\rm DReg} + 
{s_d S_{\rm ct}^{(n)}}\cdot\Gamma_{\rm DReg} +
\intd^dx\,\bigl[{\delta S_{\rm ct}^{(n)}\over \delta K_\Phi(x)}
 \cdot\Gamma_{\rm DReg}\bigr]
      {\delta\Gamma_{\rm DReg}\over\delta\Phi(x)}    &\numeq\cr
} \namelasteq\EqRegSTBreakingct
$$
was made, which allowed at one loop level for a direct computation
of the breaking in terms of finite parts of diagrams 
with an insertion of an evanescent operator, thus
avoiding the evaluation of the l.h.s.\ of the STIs.

Multiloop extension of these techniques and the explicit form
of the two-loop modified action 
for the Standard Model which would give BRS invariant results
with a MS procedure would be valuable.
For reasons of simplicity and as a previous step, 
the present paper is devoted to a simpler model,
the Abelian Higgs-Kibble 
model with chiral fermions \cite{\abhk,\GrossJackiw,\BRSah,\GrassiAH}
in a general gauge of the 't Hooft class.
Compared with the models of \cite{\bmym}, this model
presents the new feature of 
spontaneous symmetry breaking, which makes more involved the
restoration process of the (hidden)
BRS symmetry. Moreover, it is
free of IR problems and it is very easy
to write in an exhaustive manner the
list of the monomials needed, to all orders,
in the restoration procedure of its BRS symmetry. This
makes the  model a perfect training ground
for future work on the Standard Model and  two-loop computations. Since 
Algebraic Renormalization is not well known to
practitioners, we will show in detail each step of the algorithm.

The layout of this paper is as follows. In Section 2 we give the
classical action, fields and symmetry of the model. Section 3
is devoted to a quick reminder of the general theory of 
algebraic renormalization, suited to this model and  which 
is independent of the order of the renormalization procedure.
In Section 4 the BMHV dimensional regularization of the model
is presented. In Section 5 we use the techniques of \cite{\bmym}
to compute the breaking at one loop. Of course, the correct
form of the anomaly are thus obtained and in Section 6, following
the lines of section 3, the finite counterterms needed to restore
the BRS symmetry in the anomaly free case are finally computed.
In Appendix A the explicit matrix form of the linearized
Slavnov-Taylor operator, needed to do the computation, is given.
Appendix B is devoted to a pedagogical explicit computation of
the BRS cohomology of order one.
In Appendix C the results for each diagram contributing to
the one loop breaking are presented in full detail.  

\section{2. Classical action}

The CP symmetric four-dimensional classical action is the following
$$
\eqalignno{
S_{\rm inv} = \intd^4x \;\Big\{
  &-{1\over 4 g^2}\, F_{\mu\nu} F^{\mu\nu} +
        (D_\mu \phi^\dagger)(D^\mu\phi) + \mu^2 \phi^\dagger\phi -
        \lambda (\phi^\dagger\phi)^2  \cr 
  &+\sum_{k\in I\cup J}\Bigl(
          \bar\psi_k[i\prslash + 
           \Aslash (e_{{\rm L}k} \PL +e_{{\rm R}k}\PR)]\psi_k\Bigr) \cr
  &-\sum_{i\in I}\Bigl( \sqrt2 f_i \phi\,\bar\psi_i\PR\psi_i \,+\, 
                        \sqrt2 f_i\phi^\dagger\,\bar\psi_i\PL\psi_i \Bigr) \cr
  &-\sum_{j\in J}\Bigl( \sqrt2 f_j \phi\,\bar\psi_j\PL\psi_j \,+\,
                        \sqrt2 f_j\phi^\dagger\,\bar\psi_j\PR\psi_j \Bigr)
         \Big\}\,.              &\numeq\cr
}
$$
Here $A_\mu$ is an abelian gauge field, $\phi$ a scalar complex field,
$\psi_i$ and $\psi_j$ are families of Dirac fermion fields,
$F_{\mu\nu}\equiv\pr_\mu A_\nu - \pr_\nu A_\mu$ and
$D_\mu \phi\equiv (\pr_\mu -i A_\mu)\phi\,$.
This action is invariant also under the
abelian gauge transformations
$\delta A_\mu=\pr_\mu \omega$, $\delta \phi=i\omega\phi$,
$\delta\psi_k=i\omega(e_{{\rm L}k}\PL+e_{{\rm R}k}\PR)\psi_k$
provided that the fermion charges satisfy*
\vfootnote*{\begingroup\eightpoint
If the field redefinitions $A_\mu \;->\; 2\, g\, A_\mu,$
 $\omega \;->\; 2\, g \,\omega$ are done
 the model coincides with the model of section IV.A
 of \cite{\GrossJackiw}, with
 $\psi$ ($\psi^\prime$) being
 our fermions of type $r=+1$ ($r=-1$) and their charge $f$
 our $(2 e_{{\rm R}}+1)g/2$
 ($(2 e_{{\rm R}}-1)g/2$). 
\endgroup}
$$
\eqalignno{
e_{{\rm L}i}\,&=\,e_{{\rm R}i}+1,\;\; {\rm if }\;i\in I \cr 
e_{{\rm L}j}\,&=\,e_{{\rm R}j}-1,\;\; {\rm if }\;j\in J &\numeq\cr 
}\namelasteq\EqConstraints
$$

\begingroup
\parindent=0pt \leftskip=1cm \rightskip=1cm \baselineskip=11pt
\topinsert     

\centerline{\vbox{\eightrm\offinterlineskip
  \def\tabrule{\noalign{\hrule}}
  \def\hpt{height3pt&\omit&&\omit&&\omit&&\omit&&\omit&&\omit&&\omit&&\omit&&
		\omit&&\omit&&\omit&}
 \halign{\vrule#&\strut\quad\hfil#\hfil\quad&\vrule width 1.2 pt#&
			&\strut\quad\hfil#\hfil\quad&\vrule#\cr
 \tabrule\hpt\cr
 &&& $\!\!s\!\!$ && $\!x_\mu\!$ && $\!\!\phi_{1(2)}\!\!\!\!$ && $\!A_\mu\!$
         && $\!\!\psi\!\!\!$ && $c$ 
         && $\bar c$ && $B$ && $\!\!K_{\phi_{1(2)}}\!\!\!\!$   
         && $K_\psi\!$  &\cr
 \hpt\cr
 \noalign{\hrule height 1.2pt}
 \hpt\cr
 & $\!$Gh. n.$\!\!\!$  && $\!$1$\!$ && 0  && 0 && 0  && 0  
         && 1  && -1 && 0  && -1  && -1 &\cr
 \hpt\cr \tabrule \hpt\cr
 & $\!$Dimen.$\!\!$ && 0  && -1  && 1 && 1 && $\!$3/2$\!$&& 0 
         && 2 && 2 && 3  && $\!$5/2$\!$&\cr
 \hpt\cr \tabrule \hpt\cr
 & $\!\!$Comm.$\!\!\!$  && $\!$-1$\!$ && $\!$	+1$\!$ && $\!$+1$\!$ && $\!$+1$\!$ && -1 && -1 
                && -1$\!$ && $\!$+1$\!$ 
                && -1$\!$  && +1 &\cr
 \hpt\cr \tabrule \hpt\cr
 & $\!$CP$\!\!$  && $\!$ $s\!$ && $x^\mu\!$ && $\!\!$(-)$\phi_{1(2)}\!\!\!\!$ 
                 && -$A^\mu\!$ 
                 && $\!\!D\bar\psi^{t\!}\!\!\!\!$ && -$c$ 
                && -$\bar c\!$ && $\!$-$B\!$ 
                &&$\!\!$(-)$K_{\phi_{1(2)}}\!\!\!\!$  
                &&$\!\!\!$-$K_{\bar\psi}^{t}D^{-1\!}\!\!\!\!$&\cr
 \hpt\cr \tabrule}}}
\vskip 12pt
{\bf Table 1:}
{\eightrm Ghost number, dimension, conmmutativity and
CP transformation of fields, coordinates and the BRS operator. In third
row, +1 (-1) means 
that the symbol commutes (anticommutes) and in last
one, $D\equiv\gamma^0 C$, where $C$ is the usual conjugation matrix.
Note that the chosen CP properties of the ghosts make the
BRS operator CP invariant}
\vskip 0.4cm
\endinsert     
\endgroup

\smallskip

We shall consider spontaneous symmetry breaking due to a nonvanishing vev 
of the real component of $\phi$. Let us now  write the
classical BRS invariant action
$$
S_{\rm cl} = S_{\rm inv} + S_{\rm gf} + S_{\rm ext},
   \eqno\numeq\namelasteq\EqClassicalact
$$
where
$$
\eqalignno{
S_{\rm inv} = \intd^4x \;\Big\{
	&-{1\over 4 g^2}\, F_{\mu\nu} F^{\mu\nu} +
        \left[
        (D_\mu \phi^{+})(D^\mu\phi) + \mu^2 \phi^{+}\phi -
        \lambda (\phi^{+}\phi)^2  
        \right]_{
         \phi={1\over\sqrt2}(\phi_1 + v + i\phi_2)
           } \cr 
	&+\bar\psi\,[i\prslash + \Aslash \bigl((\theta +r) \PL 
                                              +\theta\PR\bigr)]\,\psi
         - f\,[(v+\phi_1)\,\bar\psi\psi + 
          i \,r\, \phi_2\,\bar\psi\gam5\psi]    \Big\}              \cr
S_{\rm gf} = 
\intd^4x \; &s\;\Bigl[{1\over2}\,\xi \,\bar c\,B+ 
           \,\bar c\,\Sigma\,\Bigl]=\cr
\intd^4x \; &\,{\xi\over2}\,B^2+ 
           \,B(\pr_\mu A^\mu + \rho\,\phi_2)-
              \bar c \,[\pr^\mu \pr_\mu \,+\,\rho(\phi_1+v)]\,c,      \cr
S_{\rm ext} = \intd^4x \; & {K_{\phi_1}}\, s\phi_1 +
         {K_{\phi_2}}\, s\phi_2 +
         {K_\psi}\, s\psi + s\bar\psi\,{K_{\bar\psi}},        
      &\numeq\cr
}
$$
In this action, $v$ is a parameter with dimension of mass,
$\phi_1$ and $\phi_2$ are real scalar fields, $c$ is the ghost field,
$\bar c$ the antighost field,
$B$ is the Lautrup-Nakanishi field, $K_\Phi$ are the external fields
coupled to the corresponding
BRS variations
\vfootnote\dag{\begingroup\eightpoint
``Antifields'' or sources to the trivial BRS variations of $A_\mu$
and $c$ could also have easily been introduced. This would  make more clear the
study and interpretation of the cohomology of $\tilde b$ at order 0, but
they are neither relevant nor necessary for the practical purposes we pursue 
in this paper.\endgroup}
and  $\Sigma\,=\, \pr_\mu A^\mu + \rho\,\phi_2$ is the
most general  gauge-fixing functional which is linear,
preserves CP symmetry,  has the appropriate  ghost number and is consistent 
with power counting (see Table 1). $\xi$ and $\rho$
are the gauge parameters.
We have omitted the index labeling the  fermion families, a notation
that  we shall keep  unless otherwise stated.
We have set $e_{{\rm R}k}\equiv\theta_k$ and
$e_{{\rm L}k}\equiv\theta_k + r_k$.
Therefore, $\{r_k\}_{k\in I\cup J}$ are not free
parameters but convenient shorthands for
$+1$ if $k\in I$ and $-1$ if $k\in J$.

The BRS transformation is:
$$\eqalignno{
&s \,\phi_1= -\,\phi_2\;c,\qquad\qquad\qquad\qquad\; 
 s \,\phi_2=  \,\,(v\,+\,\phi_1)\,c,                  \cr
&s \psi= i \,c\, [(\theta +r)\PL + \theta\PR]\psi,\qquad 
 s\bar\psi=i\,\bar\psi[(\theta+r)\PR+\theta\PL]\,c,                  \cr
&s A_\mu = \pr_\mu c,\quad s\,c=0, \quad
  s \bar c=B, \quad s B=s K_{\phi_i} = s K_\psi = s K_{\bar\psi} = 0, 
&\numeq\cr
} \namelasteq\EqBRStrans
$$
which leaves
$S_{\rm cl}$ invariant due to the anticommutativity of $\gam5$ in four 
dimensions.

\topinsert

{\settabs 5\columns \def\graphwidth{1.0in} 
\eightpoint

\+\hfil$\vcenter{\epsfxsize=\graphwidth\epsffile{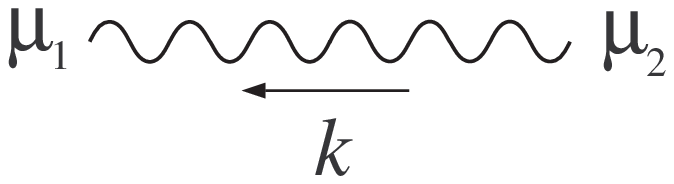}}$\hfil
 & $\vcenter{
     \hbox{$  G_{AA}^{(0)\;\mu_1\mu_2} (k)=
  \;  
       (-i) g^2\;      
       \left[ {1\over k^2-(vg)^2} 
                \big(g^{\mu_1\mu_2} -{k^{\mu_1}k^{\mu_2}\over k^2}\big)
         +\xi^\prime 
         {k^2 - \rho^2 / \xi \over 
             (k^2 - \rho v)^2}
           {k^{\mu_1}k^{\mu_2}\over k^2}\right]
           $}
     }$\cr
\+\hfil$\vcenter{\epsfxsize=\graphwidth\epsffile{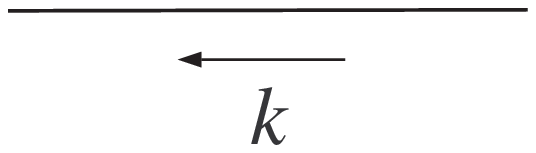}}$\hfil
 & $\vcenter{
     \hbox{$  G_{\phi_1\phi_1}^{(0)} (k)=
  \;  
       {i \over k^2-2\lambda v^2} 
           $}
     }$
 &\hfil$\vcenter{\epsfxsize=\graphwidth\epsffile{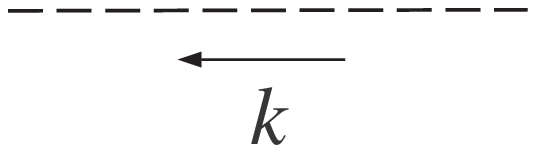}}$\hfil
 & $\vcenter{
     \hbox{$  G_{\phi_2\phi_2}^{(0)} (k)=
  \;  
       i \;      
       {k^2 - \xi v^2\over (k^2-\rho v)^2} 
           $}
     }$\cr
\+\hfil$\vcenter{\epsfxsize=\graphwidth\epsffile{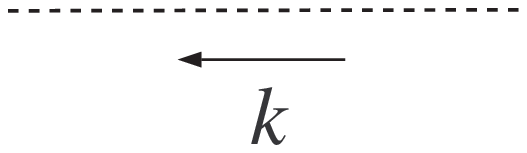}}$\hfil
 & $\vcenter{
     \hbox{$  G_{BB}^{(0)} (k)=\;0
             $}
     }$
 &\hfil$\vcenter{\epsfxsize=\graphwidth\epsffile{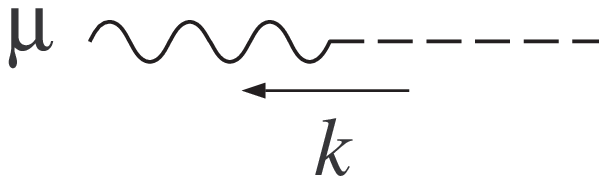}}$\hfil
 & $\vcenter{
     \hbox{$  G_{A\phi_2}^{(0)\;\mu} (k)=
  \;  
       g {\rho-\xi v\over
 k^2 - \rho v}\,k^\mu
           $}
     }$\cr
\+\hfil$\vcenter{\epsfxsize=\graphwidth\epsffile{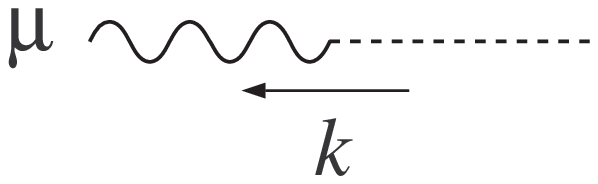}}$\hfil
 & $\vcenter{
     \hbox{$  G_{AB}^{(0)\;\mu} (k)=
  \;  
        {-k^\mu\over k^2 - \rho v}
           $}
     }$
 &\hfil$\vcenter{\epsfxsize=\graphwidth\epsffile{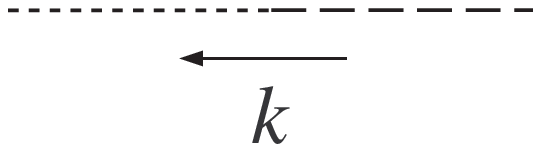}}$\hfil
 & $\vcenter{
     \hbox{$  G_{B{\phi_2}}^{(0)} (k)=
  \;  
      {(-i)}\,{v\over k^2 - \rho v}
           $}
     }$\cr
\+\hfil$\vcenter{\epsfxsize=\graphwidth\epsffile{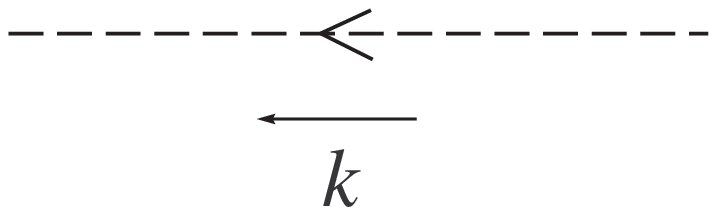}}$\hfil
  & $\vcenter{
        \hbox{$  G_{c \bar c}^{(0)}(k)=
    \;     \;\;{i\over k^2 - \rho v} 
              $}
   }$
 &\hfil$\vcenter{\epsfxsize=\graphwidth\epsffile{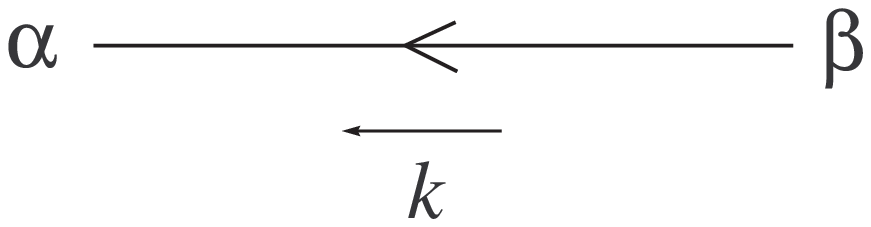}}$\hfil
 & $\vcenter{
        \hbox{$  G_{\psi \bar \psi}^{(0)}(k)=
    \;     \;{i\over \kslash - f v}{\big|}_{\a\b}
              $}
   }$\cr
}
\vskip 12pt
\narrower\noindent {\bf Figure 1:}
{\eightpoint Free field  propagators if $\mu^2=\lambda v^2$ . For
convenience, the
following abbreviation has been defined: $\xi^\prime=\xi / g^2$.
Momentum $k$ flows from
second to first field of Green functions. $\mu$, $\mu_1$ and $\mu_2$
are Lorentz indices. $\a$ and $\b$ are the indices of Dirac matrices and
will be ommitted.}

\vskip 0.1cm
\endinsert

If $\mu^2=\lambda\,v^2$ there are no linear terms
in the classical action \EqClassicalact. This is    
equivalent to the renormalization condition which sets
to zero the vev of the field
$\phi_1$. The quadratic terms in \EqClassicalact\ 
give  the propagators, which are shown in Figure 1 for 
$\mu^2=\lambda\,v^2$.
Note that the propagators
with B fields only contribute to the 1PI Feynman diagrams at the
tree level  and that no
special value will be chosen for the $\rho$ parameter,
i.e.\ there will be $A\phi_2$ mixing at tree level and
beyond in the loop expasion.
Notice also that although the theory is abelian, the ghost fields are
not free:  they interact with $\phi_1$ trough the gauge-fixing
part of the action. This makes the BRST formalism very convenient
to use.

\section{3. Algebraic Renormalization}

The classical action \EqClassicalact\ is a solution, over the
space of CP-invariant integrated polynomials of ghost
number $0$ and mass dimension $4$, to the Slavnov-Taylor identity (STI)
$$
\eqalignno{
\intd^4 x\,\Big\{
&(\pr_\mu c){\delta S_{\rm cl}\over\delta A_\mu} +
    B {\delta S_{\rm cl}\over\delta \bar c} +
    {\delta S_{\rm cl}\over\delta K_{\phi_1}}{\delta S_{\rm cl}\over\delta \phi_1} +
    {\delta S_{\rm cl}\over\delta K_{\phi_2}}{\delta S_{\rm cl}\over\delta \phi_2} + 
   {\delta S_{\rm cl}\over\delta K_\psi}{\delta S_{\rm cl} \over\delta\psi} +
   {\delta S_{\rm cl}\over\delta K_{\bar\psi}}{\delta S_{\rm cl}\over\delta{\bar\psi}} 
   \Big\}=0\,,\, &\numeq\cr
}\namelasteq\EqSTI            
$$
which rules the BRS invariance of
the theory, and to the gauge-fixing equation
$$
\eqalignno{&
{\delta S_{\rm cl}\over\delta B} \;=\;
  \xi\,B + \pr_\mu A^\mu + \rho\,\phi_2\,, &\numeq\cr
} \namelasteq\EqGaugeFixing           
$$
which is the equation of motion of the Lagrange multiplier field $B$.

Any functional $\cal F$ satisfying both equations, also
satisfies the ghost equation:
$$
{\delta{\cal F}\over\delta\bar c} + \square\, c + 
    \rho{\delta{\cal F}\over\delta K_{\phi_2}} =0 \,. \eqno\numeq            
$$ \namelasteq\EqGhostEquation
After renormalization of the perturbative expansion, 
the 1PI generating functional $\Gamma$ will be required to
be, in the sense of a formal series of functionals in $\hbar$,
a deformation of $S_{\rm cl}$ constrained by the same equations.
If the renormalization procedure respects both the gauge-fixing
and ghost equations, $\Gamma$ will have the
form:
$$
\eqalignno{
\Gamma[\phi_1,\phi_2,A_\mu,&\psi,\bar\psi,c,\bar c, B,K_{\phi_1},
       K_{\phi_2},K_\psi,K_{\bar\psi}]=\cr
  &\intd^4x \; \,{\xi\over2}\,B^2+ 
           \,B(\pr_\mu A^\mu + \rho\,\phi_2)-
              \bar c \,\square \,c\;  \cr
 &+\;\tilde\Gamma[\phi_1,\phi_2,A_\mu,\psi,\bar\psi,c,K_{\phi_1},
    \ti K_{\phi_2}\!\equiv K_{\phi_2}\!-\rho\bar c ,K_\psi,K_{\bar\psi}]
  \,,&\numeq\cr
}
$$
so that the left hand side of the STI will read
$$
\eqalignno{
\ti {\cal S}(\ti\Gamma)\equiv
\intd^4 x\,\,\Big\{
&\,(\pr_\mu c){\delta\ti\Gamma\over\delta A_\mu} +
    {\delta \ti\Gamma\over\delta K_{\phi_1}}{\delta\ti\Gamma\over\delta \phi_1} +
    {\delta \ti\Gamma\over\delta\ti K_{\phi_2}}{\delta\ti\Gamma\over\delta \phi_2} +
   {\delta \ti\Gamma\over\delta K_\psi}{\delta\ti\Gamma\over\delta\psi} +
   {\delta \ti\Gamma\over\delta K_{\bar\psi}}{\delta\ti\Gamma\over\delta{\bar\psi}} 
   \Big\}\, &\numeq\cr
} 
$$
From now, a tilde will indicate dependence on the same fields as  
$\tilde\Gamma$ does.

Let us introduce the linearized Slavnov-Taylor operator:
$$
\eqalignno{
\ti{\cal S}_{\ti{\cal F}}\equiv
\intd^4 x\,\,\Big\{
&\,(\pr_\mu c){\delta\over\delta A_\mu} +
   {\delta \ti {\cal F}\over\delta\phi_1}{\delta\over\delta K_{\phi_1}} +
    {\delta \ti {\cal F}\over\delta K_{\phi_1}}{\delta\over\delta \phi_1} +
 {\delta \ti {\cal F}\over\delta\phi_2}{\delta\over\delta\ti K_{\phi_2}} +
    {\delta \ti {\cal F}\over\delta\ti K_{\phi_2}}{\delta\over\delta \phi_2} +\cr
&+ {\delta \ti {\cal F}\over\delta\psi}{\delta\over\delta K_\psi} +
   {\delta \ti {\cal F}\over\delta K_\psi}{\delta\over\delta\psi} +
   {\delta \ti {\cal F} \over\delta{\bar\psi}}{\delta\over\delta K_{\bar\psi}} +
   {\delta \ti {\cal F}\over\delta K_{\bar\psi}}{\delta\over\delta{\bar\psi}} 
   \Big\}\,, &\numeq\cr
}
$$
which has the nilpotency properties
$$
\eqalignno{
&\ti{\cal S}_{\ti{\cal F}}\,
 \ti{\cal S}(\ti{\cal F})=0\,,\;\;\forall\,
 \ti{\cal F}  \cr
&\ti{\cal S}_{\ti{\cal F}}\,\ti{\cal S}_{\ti{\cal F}}=0\,,\;\;
{\rm if}\;\, \ti{\cal S}(\ti{\cal F})=0\,. &\numeq\cr
}               \namelasteq\EqNilpotency
$$
For $\ti{\cal F}$ equal to the classical action $\ti S_{\rm cl}$ we have
the important linear operator
$$
\eqalignno{
\ti b &\equiv \ti{\cal S}_{\ti S_{\rm cl}}  \, . &\numeq\cr
}              \namelasteq\EqOperatorb
$$

The local part of
maximal dimension four of 
$\ti\Gamma$ at a determined order
in the perturbative expansion and prior to imposition of
the STI is an arbitrary linear combination of the following
basis of the space $\tilde{\cal V}_0$ of the 
integrated Lorentz scalar CP-invariant polynomials in
the fields and their derivatives with maximal canonical dimension 4 and
ghost number 0 (note that we choose the same  first twenty 
monomials as in \cite{\BRSah}):
\bigskip
{\settabs 3 \columns
\openup1\jot
\+${\ti e}_1\;\equiv\;\int\!{\phi_1}$,
 &${\ti e}_2\;\equiv\;\int\!{\phi_1}^2$,
 &${\ti e}_3\;\equiv\;\int\!{\phi_2}^2$,
\cr
\+${\ti e}_4\;\equiv\;\int\!{\phi_1}^3$,
 &${\ti e}_5\;\equiv\;\int\!\phi_1{\phi_2}^2$,
 &${\ti e}_6\;\equiv\;\int\!{\phi_1}^4$,
\cr
\+${\ti e}_7\;\equiv\;\int\!{\phi_2}^4$,
 &${\ti e}_8\;\equiv\;\int\!{\phi_1}^2{\phi_2}^2$,
 &${\ti e}_9\;\equiv\;\int\!(\pr_\mu\phi_1)(\pr^\mu\phi_1)$,    
\cr
\+${\ti e}_{10}\;\equiv\;\int\!(\pr_\mu\phi_2)(\pr^\mu\phi_2)$,
 &${\ti e}_{11}\;\equiv\;\int\!{\phi_2}(\pr_\mu A^\mu)$,
 &${\ti e}_{12}\;\equiv\;\int\!A_\mu \phi_1 (\pr^\mu \phi_2)$,    
\cr
\+${\ti e}_{13}\;\equiv\;\int\!A_\mu \phi_2 (\pr^\mu \phi_1)$,
 &${\ti e}_{14}\;\equiv\;\int\!A_\mu A^\mu$,
 &${\ti e}_{15}\;\equiv\;\int\!A_\mu A^\mu \phi_1$,    
\cr
\+${\ti e}_{16}\;\equiv\;\int\!A_\mu A^\mu {\phi_1}^2$,
 &${\ti e}_{17}\;\equiv\;\int\!A_\mu A^\mu {\phi_2}^2$,
 &${\ti e}_{18}\;\equiv\;\int\!(\pr_\mu A^\mu)^2$,    
\cr
\+${\ti e}_{19}\;\equiv\;\int\!(\pr_\mu A_\nu -\pr_\nu A_\mu)^2$,
 &${\ti e}_{20}\;\equiv\;\int\!{(A_\mu A^\mu)}^2$,
 &    
\cr
\+${\ti e}_{21}\;\equiv\;\int\!K_{\phi_1} \phi_2 c $,
 &${\ti e}_{22}\;\equiv\;\int\!{\ti K}_{\phi_2} c$,
 &${\ti e}_{23}\;\equiv\;\int\!{\ti K}_{\phi_2} \phi_1 c$,
\cr
}
{\settabs 2 \columns
\openup1\jot
\+${\ti e}_{24}\;\equiv\;\int\! \bar\psi \psi$,
\cr
\+${\ti e}_{25}\;\equiv\;\int\! \bar\psi i \prslash \PL \psi$,
 &${\ti e}_{26}\;\equiv\;\int\! \bar\psi i \prslash \PR \psi$,
\cr
\+${\ti e}_{27}\;\equiv\;\int\! \bar\psi   \Aslash \PL \psi$,
 &${\ti e}_{28}\;\equiv\;\int\! \bar\psi   \Aslash \PR \psi$,
\cr
\+${\ti e}_{29}\;\equiv\;\int\! \phi_1\,\bar\psi \psi\,=\,
         \phi_1\bar\psi\PR\psi + \phi_1\bar\psi\PL\psi$,
 &${\ti e}_{30}\;\equiv\;\int\! \phi_2\,\bar\psi\gam5 \psi\,=\,
         \phi_2\bar\psi\PR\psi - \phi_2\bar\psi\PL\psi$,
\cr
\+${\ti e}_{31}\;\equiv\;\int\! (K_\psi \PL \psi\,-\,
         \bar\psi\PR K_{\bar\psi})\,c$,
 &${\ti e}_{32}\;\equiv\;\int\! (K_\psi \PR \psi\,-\,
         \bar\psi\PL K_{\bar\psi})\,c$\hfill\numeq &
\cr
}\namelasteq\EqEbasis

\bigskip

That means that we have the freedom to add
to the starting action any action-like term of the form 
$\ti X=\sum_{i=1}^{32}\ti x^i \ti e_i$, each $\ti x^i$ being
a formal series in $\hbar$ of order $O(\hbar)$.

In a non-invariant renormalization scheme,
the renormalized STI will have a breaking:
$$
\ti {\cal S}(\ti\Gamma) = {\ti\Delta}\, . \, {\ti\Gamma} \eqno\numeq
$$
the r.h.s.\ of last equation being the insertion of a 
CP-invariant integrated local
operator of maximal dimension 4 and ghost number 1.

Now, 
in accordance with the algebraic theory of renormalization
and supposing that the breaking vanish
at lower orders of the perturbative  expansion,
the insertion at order $n$ is simply a local integrated polynomial
${\ti\Delta}\, . \, {\ti\Gamma}=\hbar^n{\ti\Delta}^{(n)} +O(\hbar^{n+1})$
which can be decomposed as a linear
combination of the following basis of the space $\tilde{\cal V}_1$ of 
CP invariant action-like
polynomials of maximal dimension 4 and ghost number 1 (again,
we choose a basis following the strategy in ref. \cite{\BRSah}):
\bigskip
{\settabs 3 \columns
\openup1\jot
\+${\ti u}_1\;\equiv\;\int\!{\phi_2}\,c$,
 &${\ti u}_2\;\equiv\;\int\!{\phi_1}\phi_2\,c$,
 &${\ti u}_3\;\equiv\;\int\!{\phi_2}^3\,c$,
\cr
\+${\ti u}_4\;\equiv\;\int\!{\phi_1}^2\phi_2\,c$,
 &${\ti u}_5\;\equiv\;\int\!(\square\phi_2)\,c$,
 &${\ti u}_6\;\equiv\;\int\!{\phi_1}^3\phi_2\,c$,
\cr
\+${\ti u}_7\;\equiv\;\int\!{\phi_1}{\phi_2}^3\,c$,
 &${\ti u}_8\;\equiv\;\int\!\phi_2(\square\phi_1)\,c$,
 &${\ti u}_9\;\equiv\;\int\!\phi_1(\square\phi_2)\,c$,
\cr
\+${\ti u}_{10}\;\equiv\;\int\!(\pr_\mu\phi_1)(\pr^\mu\phi_2)\,c$,
 &${\ti u}_{11}\;\equiv\;\int\!(\pr_\mu A^\mu)\,c$,
 &${\ti u}_{12}\;\equiv\;\int\!A_\mu (\pr^\mu\phi_1)\,c$,
\cr
\+${\ti u}_{13}\;\equiv\;\int\!(\pr_\mu A^\mu)\phi_1\,c$,
 &${\ti u}_{14}\;\equiv\;\int\!(\pr_\mu A^\mu){\phi_1}^2\,c$,
 &${\ti u}_{15}\;\equiv\;\int\!A_\mu \phi_1(\pr^\mu\phi_1)\,c$,
\cr
\+${\ti u}_{16}\;\equiv\;\int\!(\pr_\mu A^\mu){\phi_2}^2)\,c$,
 &${\ti u}_{17}\;\equiv\;\int\!A_\mu\phi_2(\pr^\mu\phi_2)\,c$,
 &${\ti u}_{18}\;\equiv\;\int\!\square(\pr_\mu A^\mu)\,c$,
\cr
\+${\ti u}_{19}\;\equiv\;\int\!A_\mu A_\nu(\pr^\mu A^\nu)\,c$,
 &${\ti u}_{20}\;\equiv\;\int\!A_\mu A^\nu \phi_2\,c$,
 &${\ti u}_{21}\;\equiv\;\int\!A_\mu A^\mu \phi_1 \phi_2\,c$,
\cr
\+${\ti u}_{22}\;\equiv\;\int\!A_\mu A^\mu(\pr^\nu A^\nu)\,c$,
\cr
\+${\ti u}_{23}\;\equiv\;\int\! \bar\psi\gam5\psi\,c$,
 &${\ti u}_{24}\;\equiv\;\int\! \pr_\mu(\bar\psi\gamma^\mu\PL\psi)\,c$,
 &${\ti u}_{25}\;\equiv\;\int\! \pr_\mu(\bar\psi\gamma^\mu\PR\psi)\,c$,
\cr
\+${\ti u}_{26}\;\equiv\;\int\! \phi_2\bar\psi\psi\,c$,
 &${\ti u}_{27}\;\equiv\;\int\! \phi_1\bar\psi\gam5\psi\,c$,
 &
\cr
\+${\ti u}_{28}\;\equiv\;\int\! \eps_{\mu_1\mu_2\mu_3\mu_4}
           (\pr^{\mu_1}A^{\mu_2}) \,(\pr^{\mu_3}A^{\mu_4})\,c\,$,
 &&\hfill\numeq &\cr
}\namelasteq\EqUbasis

\bigskip
\noindent that is $\ti\Delta^{(n)}=\sum_{j=1}^{28}\ti{\Delta}^j{}^{(n)}
\ti u_j$. Notice that, incidentally, external fields do not appear in
the basis due to power counting and the property of the abelian
ghosts $c\,c = 0$.

The projection of the breaking over the direction of the
last element of the basis
constitutes the anomaly of the theory, for it can be shown
that the linear system $\ti b \ti X{}^{(n)} = \ti\Delta {}^{(n)}$
has an (under-determined) solution if and only if the coefficient
$\ti\Delta^{28}{}^{(n)}$  vanishes. This can be established by using
cohomological methods: the first equation in 
\EqNilpotency\ leads to $\ti b \ti\Delta=0$, which is the famous
consistency condition, and,  the second equation in
\EqNilpotency\ implies
$\ti b^2 = 0$; hence, $\ti\Delta\equiv{\rm anomaly }
+ \ti b\ti X$, the anomaly belonging to  the ghostnumber one non-trivial 
cohomology space of the $\ti b$ operator. See
Appendix B for a proof of this statement by 
explicit computation.

Now, let us define the matrix elements of the operator $\ti b$
restricted to its action from $\tilde{\cal V}_0$ to $\tilde{\cal V}_1$
in the following manner $\ti b \ti e_i \equiv \ti b_0{}^j{}_i u_j$
( explicit values of these matrix elements for $S_{\rm cl}$ 
in \EqClassicalact\  are given in Appendix A). Then, 
the linear system
$$
\sum_{j=1}^{27}\ti b_0{}^j{}_i\,\ti x^i{}^{(n)}=\ti\Delta^j{}^{(n)},\;\;
   i=1,\cdots,32 \eqno\numeq
$$\namelasteq\EqLinearSystem 
has always a solution up
to an $O(\hbar^n)$ arbitrary linear combination of $\ti b$-invariants. These 
$\ti b$-invariants are  any basis of the kernel $\ti {\cal K}_0$ of the 
restricted linear operator $\ti b_0\equiv\ti b : \ti{\cal V}_0 \rightarrow
\ti{\cal V}_1$. 
We can choose for example
the following basis of $\ti {\cal K}_0$ for all values of $\theta$:
$$
\eqalignno{
\ti{\cal I}_1=&
  -{1\over4}\,\ti e_{19}\,=\,
  -{1\over4}\int F_{\mu\nu}F^{\mu\nu} \,,\cr
\ti{\cal I}_2=&
  v \ti e_1 + {\ti e_2 + \ti e_3 \over 2} =
 \int \phi^{\dag} \phi \,,\cr
\ti{\cal I}_3=&
  v^3 \ti e_1 + {3v^2\over2} \ti e_2 + {v^2\over2} \ti e_3 + 
  v (\ti e_4 + \ti e_5 )+ {\ti e_6 + \ti e_7\over4} +
  {\ti e_8\over2} =
  \int (\phi^{\dag} \phi)^2 \,,\cr
\ti{\cal I}_4=&v\tilde e_{24} + \tilde e_{29} + 
   i\,r\,\tilde e_{30}= \int
    v \bar\psi\psi + \phi_1\bar\psi\psi + 
    i\,r\,\phi_2\bar\psi\gam5\psi\,,\cr
\ti{\cal I}_5=&
 (\mu^2-3\lambda v^2)\ti e_1 - 3\lambda v \ti e_2 - \lambda v \ti e_3
 - \lambda v \ti e_3 - \lambda (\ti e_4 + \ti e_5) +\cr 
 &\ti e_{11} +
  v \ti e_{14} + \ti e_{15} + \ti e_{22} - f \ti e_{24}=
  \ti b \int K_{\phi_1}\,,\cr
\ti{\cal I}_6=&
 (\mu^2-\lambda v^2)v \ti e_1 +(\mu^2-3\lambda v^2) \ti e_2 
 - 3\lambda v \ti e_4
 - \lambda v \ti e_5 - \lambda (\ti e_6 + \ti e_8) +\cr 
 &\ti e_{9} -
 \ti e_{12} + \ti e_{13} + v \ti e_{15} +\ti e_{16} +
  \ti e_{21} + \ti e_{23} - f \ti e_{29}=
  \ti b \int K_{\phi_1}\,\phi_1\,,\cr
\ti{\cal I}_7=&
 (\mu^2-\lambda v^2) \ti e_3  
 - 2\lambda v \ti e_5
 - \lambda (\ti e_7 + \ti e_8) +\cr 
 &\ti e_{10} + v \ti e_{11} -\ti e_{12} +\ti e_{13} +\ti e_{17} -
  \ti e_{21} -v  \ti e_{22} - \ti e_{23} -i\,r\,f \ti e_{30}=
  \ti b \int K_{\phi_2}\,\phi_2\,,\cr
\ti{\cal I}_8=&
 - 2\ti e_{25}-2 (r+\theta)\ti e_{27}+
  v f \ti e_{24} + f \ti e_{29} +i\,r\,f \ti e_{30}=
  \ti b \int ( K_{\psi}\,\PL\psi + \bar\psi\PR K_{\bar\psi})\,,\cr
\ti{\cal I}_9=&
 - 2\ti e_{26}-2\theta\ti e_{28}+
  v f \ti e_{24} + f \ti e_{29} +i\,r\,f \ti e_{30}=
  \ti b \int ( K_{\psi}\,\PR\psi + \bar\psi\PL K_{\bar\psi})\,,\cr
\ti{\cal I}_{10}=&\,
 \theta (\ti e_9 + \ti e_{10} + v \ti e_{11} - \ti e_{12} +
         \ti e_{13} + \ti e_{21} - v \ti e_{22} - \ti e_{23} )
 - r \ti e_{26} - i\,r\theta \ti e_{32}\,,\cr
\ti{\cal I}_{11}=&\, 
 (\theta +r ) (\ti e_9 + \ti e_{10} + v \ti e_{11} - \ti e_{12} +
         \ti e_{13} + \ti e_{21} - v \ti e_{22} - \ti e_{23} )
 + r \ti e_{25} +
 i\,r(\theta + r) \ti e_{31}\,,\;\;\;\;\;\;\;\;\;\;\;\;
&\numeq\cr
}\namelasteq\EqInvariants
$$

which is fixed by choosing  an appropriate set of normalization conditions.

Notice that the rank of $\ti b_0$ is 
dim of $\ti{\cal V}_0-$ dim of $\ti{\cal K}_0=21 < $
dim of $\ti{\cal V}_1$, so that, in general,
for arbitrary values of 
the
breaking, 
the system \EqLinearSystem\
would be an incompatible one. Then its compatibility
when subtituting in it the  values of the breaking obtained by
explicit computation turns to be a not trivial
check of the correctness of the computation itself. 
 
Finally, if there is no anomaly, the breaking at order $n$ 
only consists of
cohomologically trivial terms, and, thus by adding
$\hbar^n \tilde S^{(n)}_{\rm fct}=-\hbar^n \ti X^{(n)}$ to the previous
action the breaking is canceled at the order $n$.

\section{4. Dimensionally regularized action}

If we want that the Regularized Action principle of Breitenlohner
and Maison be applicable,
we must define the regularized kinetic terms just with the same
forms as the four dimensional ones. The regularized kinetic terms are 
thus uniquely defined. Not so the interaction terms. For instance,
the Dirac matrix part of the fermion-gauge-boson vertex has 
the following equivalent forms in 4 dimensions: $\gamma^\mu\PL$ 
$=\PR\gamma^\mu$ $=\PR\gamma^\mu\PL$. But these forms
are not equal in the $d$-dimensional space-time of
Dimensional Regularization
because of the non-anticommutativity of $\gam5$. 
Of course, the generalization of the interaction to the 
Dimensional Regularization space is not unique, and any choice is {\it equally 
correct}. And yet, some choices will be more convenient than others.

Specially in the case at hand it would be far more convenient
to use a dimensionally regularized action which has the   
discrete simmetries of the four dimensional classical action. Indeed, if
the dimensionally regularized action were not  CP invariant, 
we would have to enlarge the basis of the relevant spaces presented in
section 3 with CP-non-invariant monomials. This would make the computations 
very lengthy. Even with the restriction of CP-symmetry the regularized action is not unique: there is always
the freedom of adding explicit evanescent operators, 
i.e.\ proportional to $d-4$. Here we shall adopt 
the simplest choice available and generalize in the obvious way  
to $d$-dimensional space-time 
the BRS variations and vertices of the action \EqClassicalact , ``barring'' the
boson-fermion vertex. For the latter vertex,  we shall use the following regularized form
$$
          \bar\psi\;[\,i\,\prslash + 
           \A_\mu \,(e_{{\rm L}}\bar\ga^\mu \PL \,+\,
                   e_{{\rm R}}\bar\ga^\mu \PR)\,]\;\psi
\eqno\numeq
$$
i.e. the CP-invariant or ``hermitian'' regularized form,
which can be cast in the following non-gauge invariant  expression
$$   
\eqalignno{
&i\,\bar\psi{\bar\Dslash}\psi +
{i}\,\bar\psi {\hat\prslash}\psi.     &\numeq\cr
}\namelasteq\nongaugeinvariant
$$

Hence, the regularized action, $S_0$, we shall start with will not be
BRS invariant. The regularized breaking, $s_d S_0$,  coming from the
the last term of eq.~\nongaugeinvariant , will read  thus:
$$
\eqalignno{
s_dS_0 &= s_d \intd^dx  \;
	{i}\,\bar\psi\gamhat^\mu \pr\!_\mu\psi
=\;\;\intd^dx \; {1\over2}\, c \;\Big\{                                 
    (r+2\theta)\,\pr_\mu (\bar\psi\gamhat^\mu\psi)\, +\,
     r\,\big(\bar\psi\gamhat^\mu\gam5\arrowsim\pr\!_\mu\psi\big) 
   \Big\}         \cr
&\equiv \hat\Delta\equiv \intd^dx\;\hat\Delta(x).            &\numeq\cr
}  \namelasteq\EqBreaking   
$$ 
The Feynman rule of 
the insertion of this anomalous breaking is given in fig.~2.
\midinsert
\def\graphwidth{1.5in}        
{\eightpoint
$$
\eqalign{\epsfxsize=\graphwidth\epsffile{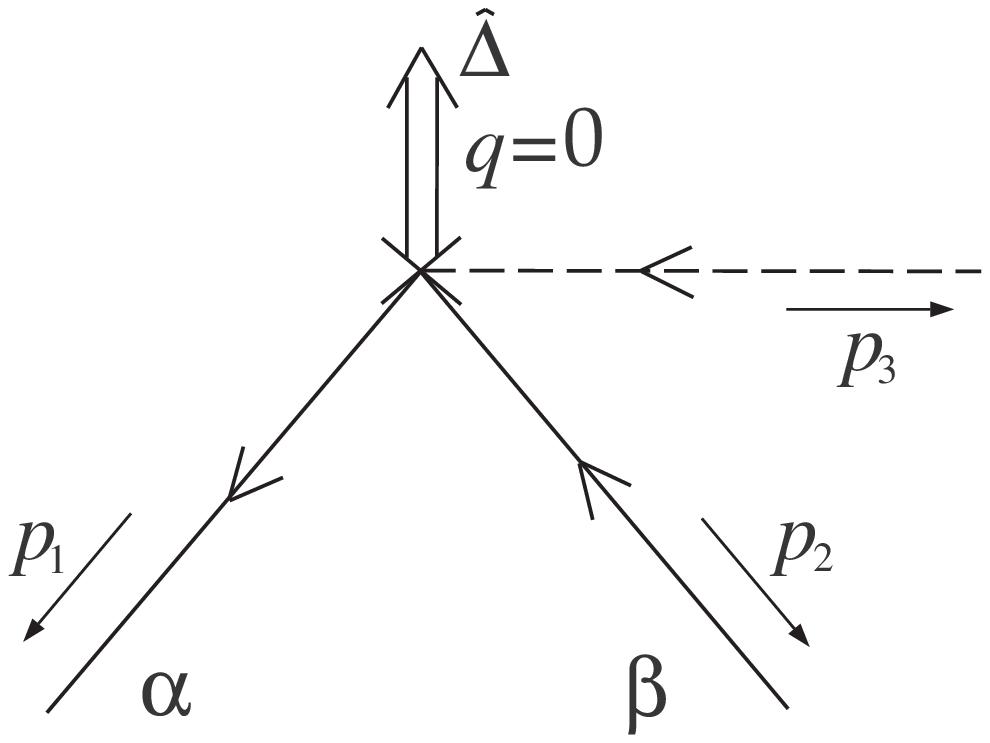}}
\quad\
\eqalign{
\equiv&\quad
   \Gapcero_{\psi\bar\psi c;\hat\Delta}
		{\,}_{\a\b}(p_1,p_2;q\equiv0)\quad=    \cr
=&\quad
{i\over2}\,[
 (r +2 \theta)\, (\hat\pslash_1+\hat\pslash_2)
\,-\,r\, (\hat\pslash_1-\hat\pslash_2)\gam5\,]{\big|}_{\a\b}
}
$$
}
\vskip -0.3cm
\narrower\noindent {\bf Figure 2.}
{\eightrm Feynman rule of the insertion of the integrated breaking
	eq.~\EqBreaking}
\vskip 0.4cm
\endinsert

The breaking is an (implicit) ``$(d-4)$-object'', {\it i.e.}\ an evanescent 
operator or an operator which vanishes in the four dimensional projection,
and, clearly, this would be also true for any other Dimensional
Regularization classical action we might have chosen. Other choices of 
regularized vertices can lead upon minimal subtraction to other values 
for renormalized Green functions; each value  corresponding to  a 
renormalization scheme. Hence, the different values we spoke of should
be related by finite counterterms.

In order not to deal with cumbersome propagators we set $\mu^2=\lambda v^2$ in
the regularized classical. Notice that for this choice of $\mu^2$
the starting action can be interpreted as the usual spontaneous
symmetry breaking action.

\section{5. One loop STI breaking in BMHV dimensional renormalization}

By using the action principles of Breitenlohner and Maison --which
basically state that the usual formal manipulations of path integrals
are allowed in the dimensionally regularized theory-- it
can be shown that the equation of motion holds in dimensional
regularization and renormalization \cite{\BMabc,\BonneauABC,\Collins}.
Therefore, the gauge-fixing \EqGaugeFixing\ and ghost
\EqGhostEquation\ equations holds for both the regularized and
MS renormalized 1PI generating functional $\Gamma$ if the
dimensionally regularized action of previous section is used,
or if it is modified by the addition of terms
independent of $B$ and 
depending on $\bar c$  and $K_{\phi_2}$ only
trough the combination $\tilde K_{\phi_2}=K_{\phi_2}-\rho\,\bar c$.
Hence, we will restrict the possible
finite counterterms of the regularized action to live in
the space
$\tilde{\cal V}_0$ whose basis was given in \EqEbasis. Of course,
as in the previous section, we have several possible ``$d$-dimensional''
generalizations of a given four-dimensional finite counternterm.
Two such generalizations will differ in a ``$d$-dimensional''
integrated evanescent operator of order $\hbar$, which modifies
the value of finite four-dimensional quantities only at order $\hbar^2$.
We choose the generalizations whose forms in the
``$d$-dimensional'' algebra
of covariants are exactly the same as in $\EqEbasis$.

In \cite{\bmym}, again by invoking the action principles, the identity 
\EqRegSTBreakingct\ was derived, and it was proved using it
that at the one-loop
level the breaking of the renormalized STI --- it's r.h.s --
simplifies to
$$
\eqalignno{
\tilde\Delta^{(1)}\,=&
 \Bigl[{\rm N}[\hat\Delta]\cdot\Gamma^{\rm R}\Bigr]^{(1)}+
b\,\tilde S_{\rm fct}^{(1)}\,,&\numeq\cr
}
$$\namelasteq\EqYMOneloopbreaking
where ${\rm N}[\hat\Delta]\cdot\Gamma_{\rm R}$ denotes
the insertion of a normal product as defined in
refs.~\cite{\BonneauABC,\Collins}: the minimally subtracted
generating functional of diagrams with an insertion of
the regularized operator $\hat\Delta$. Notice that
after algebraic manipulations of the
Feynman integrand of these diagrams
with an insertion of 
an evanescent operator,
an explicit $d-4$ factor
can appear in the numerators to be canceled with
the $d-4$ coming from the divergence of denominators,
 giving thus
a local renormalized value, as expected from the general
Algebraic Renormalization theory.

We could, of course, compute the breaking by evaluating the 
relevant zero and one-loop 1PI functions,
inserting them in the l.h.s of STIs and working out the functional
derivatives. But, it is clearly more efficient
to compute the breaking directly using
\EqYMOneloopbreaking.

With the aid of the
Bonneau identities of \cite{\BonneauABC}, 
the anomalous normal product \EqYMOneloopbreaking, i.e. a
normal product of an evanescent operator, can be decomposed
in terms of some basis of standard normal products, i.e.
normal products of non-evanescent operators. See
\cite{\BonneauABC,\bmym} for examples. But at lowest
order, this technique reads practically the same as
the direct computation of the one-loop finite part of
${\rm N}[\hat\Delta]\cdot\Gamma^{\rm R}$:

\begingroup\narrower
\noindent{\it i)} compute the finite part of all divergent
by power counting 1PI
diagrams with and insertion of $\hat\Delta$ and any quantum
or BRS external field as legs.

\noindent{\it ii)} set $\gbar^{\mu\nu}$ to $g^{\mu\nu}$ and
        $\ghat^{\mu\nu}$ to zero, i.e. set to zero every
    hatted object.

\noindent{\it iii)} find finite four-dimensional integrated operators
such as the Feynman rules of its tree-level insertions match the
results of {\it ii})

\endgroup

\medskip
We have carried out the procedure spelled out above in a completely automatic
manner by using own Mathematica $^{T\!M}$
\cite{\Mathematica} routines and the
Mathematica package ``Tracer'' \cite{\Tracer}, which manages
properly and carefully the BMHV gammas algebra.
The input of the programs
consists of the definition of Feynman rules and the expression of 
the diagrams in terms of symbolic Feynman rules. 
For the dimensionally regularized action of section 4, the one-loop 
contributions to the 1PI functions with a breaking insertion 
reads \dag\vfootnote\dag{
\begingroup\eightpoint
In order to avoid any typesetting mistake,
all following results of the paper
have been
automatically inserted from the \TeX\ output of the programs.
The code of the programs which do all computations and
generate the \TeX\ output can be found at arXiv:hep-th in the source
format of this preprint.
\endgroup}
(results after step {\it ii}$\,$ for
each relevant diagram are shown in Appendix C$\,$): 

\def\frac#1#2{{\displaystyle #1}\over{\displaystyle #2}}
\def\frac#1#2{{\displaystyle #1}\over{\displaystyle #2}}
$$ 
 \eqalignno{\cr\tilde\Gamma_{Ac;N[\hat\Delta]}^{{\rm R}(1)\;\mu_1 } (k_1)\,=\;&{\frac{{}\,\left( -6\,{f^2}\,{v^2} +  
        {k_1}^{2} \right) \, 
      {k_1}^{{\mu_1}}}{3}} 
 \,,\cr\tilde\Gamma_{AAc;N[\hat\Delta]}^{{\rm R}(1)\;\mu_1\mu_2 } (k_1,k_2)\,=\;&{\frac{-4\,i}{3}}\,{}\, 
   \left( 3\,{\theta} + {r} +  
     3\,{{{\theta}}^2}\,{r} \right) \, 
   \epsilon\left({k_1},{k_2},\{ {\mu_1}\} ,\{ \ 
 {\mu_2}\} \right) 
 \,,\cr\tilde\Gamma_{AAAc;N[\hat\Delta]}^{{\rm R}(1)\;\mu_1\mu_2\mu_3 } (k_1,k_2,k_3)\,=\;&{\frac{2\,{}\,\left( {k_1}^{{\mu_3}} +  
         {k_2}^{{\mu_3}} +  
         {k_3}^{{\mu_3}} \right) \, 
       g^{{\mu_1} {\mu_2}}}{3}} \cr & + {\frac{2\,{}\, 
       \left( {k_1}^{{\mu_2}} +  
         {k_2}^{{\mu_2}} +  
         {k_3}^{{\mu_2}} \right) \, 
       g^{{\mu_1} {\mu_3}}}{3}} \cr & + {\frac{2\,{}\,\left( {k_1}^{{\mu_1}} +  
         {k_2}^{{\mu_1}} +  
         {k_3}^{{\mu_1}} \right) \, 
       g^{{\mu_2} {\mu_3}}}{3}} 
 \,,\cr\tilde\Gamma_{{\phi_2} c;N[\hat\Delta]}^{{\rm R}(1)\; } (k_1)\,=\;&{\frac{4\,i}{3}}\,{}\,{f^2}\,v\, 
   \left( 6\,{f^2}\,{v^2} - {k_1}^{2} \right)  
 \,,\cr\tilde\Gamma_{{\phi_1}{\phi_2} c;N[\hat\Delta]}^{{\rm R}(1)\; } (k_1,k_2)\,=\;&{\frac{4\,i}{3}}\,{}\,{f^2}\, 
   \left( 18\,{f^2}\,{v^2} - 3\,{k_1}^{2} -  
     3\,{k_1} \cdot {k_2} - {k_2}^{2} \ 
 \right)  
 \,,\cr\tilde\Gamma_{{\phi_2}{\phi_2} {\phi_2} c;N[\hat\Delta]}^{{\rm R}(1)\; } (k_1,k_2,k_3)\,=\;&16\,i\,{}\,{f^4}\,v 
 \,,\cr\tilde\Gamma_{{\phi_1}{\phi_1}{\phi_2} c;N[\hat\Delta]}^{{\rm R}(1)\; } (k_1,k_2,k_3)\,=\;&48\,i\,{}\,{f^4}\,v 
 \,,\cr\tilde\Gamma_{{\phi_1}{\phi_1}{\phi_1}{\phi_2} c;N[\hat\Delta]}^{{\rm R}(1)\; } (k_1,k_2,k_3,k_4)\,=\;&48\,i\,{}\,{f^4} 
 \,,\cr\tilde\Gamma_{{\phi_1}{\phi_2}{\phi_2}{\phi_2} c;N[\hat\Delta]}^{{\rm R}(1)\; } (k_1,k_2,k_3,k_4)\,=\;&16\,i\,{}\,{f^4} 
 \,,\cr\tilde\Gamma_{A{\phi_1} c;N[\hat\Delta]}^{{\rm R}(1)\;\mu_1 } (k_1,k_2)\,=\;&-4\,{}\,{f^2}\,v\,{k_1}^{{\mu_1}} 
 \,,\cr\tilde\Gamma_{A{\phi_1}{\phi_1} c;N[\hat\Delta]}^{{\rm R}(1)\;\mu_1 } (k_1,k_2,k_3)\,=\;&-4\,{}\,{f^2}\,{k_1}^{{\mu_1}} 
 \,,\cr\tilde\Gamma_{A{\phi_2}{\phi_2} c;N[\hat\Delta]}^{{\rm R}(1)\;\mu_1 } (k_1,k_2,k_3)\,=\;&-4\,{}\,{f^2}\,\left( {k_1}^{{\mu_1}} +  
     2\,\left( {k_2}^{{\mu_1}} +  
        {k_3}^{{\mu_1}} \right)  \right)  
 \,,\cr\tilde\Gamma_{AA{\phi_2} c;N[\hat\Delta]}^{{\rm R}(1)\;\mu_1\mu_2 } (k_1,k_2,k_3)\,=\;&-8\,i\,{}\,{f^2}\,v\,g^{{\mu_1} {\mu_2}} 
 \,,\cr\tilde\Gamma_{AA{\phi_1}{\phi_2} c;N[\hat\Delta]}^{{\rm R}(1)\;\mu_1\mu_2 } (k_1,k_2,k_3,k_4)\,=\;&-8\,i\,{}\,{f^2}\,g^{{\mu_1} {\mu_2}} 
 \,,\cr\tilde\Gamma_{\psi\bar\psi c;N[\hat\Delta]}^{{\rm R}(1)\; } (k_1,k_2)\,=\;&{\frac{-\left( {}\,f\, 
         \left( 3\,{\rho}\,{r} +  
           4\,{g^2}\,{\theta}\, 
            \left( 1 + {\theta}\,{r} \right) \,v\, 
            \left( 5 + {\xi^\prime} \right)  \right) \,\gamma_5  \ 
 \right) }{6}} \cr & - {\frac{{}\,{g^2}\, 
       \left( 2\,{\theta} + {r} \right) \, 
       \left( 5 + {\xi^\prime} \right) \, \not\! {k_1} }{12}} \cr & - {\frac{{}\,{g^2}\, 
       \left( 2\,{\theta} + {r} \right) \, 
       \left( 5 + {\xi^\prime} \right) \, \not\! {k_2} }{12}} \cr & + {\frac{{}\,\left( -12\,{f^2}\,{r} +  
         {g^2}\, 
          \left( 2\,{\theta} + {r} +  
            2\,{{{\theta}}^2}\,{r} \right) \, 
          \left( 5 + {\xi^\prime} \right)  \right) \, 
        \not\! {k_1} \gamma_5 }{12}} \cr & + {\frac{{}\,\left( -12\,{f^2}\,{r} +  
         {g^2}\, 
          \left( 2\,{\theta} + {r} +  
            2\,{{{\theta}}^2}\,{r} \right) \, 
          \left( 5 + {\xi^\prime} \right)  \right) \, 
        \not\! {k_2} \gamma_5 }{12}} 
 \,,\cr\tilde\Gamma_{\psi\bar\psi {\phi_1} c;N[\hat\Delta]}^{{\rm R}(1)\; } (k_1,k_2,k_3)\,=\;&{\frac{-2\,{}\,f\,{g^2}\,{\theta}\, 
      \left( 1 + {\theta}\,{r} \right) \, 
      \left( 5 + {\xi^\prime} \right) \,\gamma_5 }{3}} 
 \,,\cr\tilde\Gamma_{\psi\bar\psi{\phi_2} c;N[\hat\Delta]}^{{\rm R}(1)\; } (k_1,k_2,k_3)\,=\;&{\frac{2\,i}{3}}\,{}\,f\,{g^2}\, 
   {\theta}\,\left( {\theta} + {r} \right) \, 
   \left( 5 + {\xi^\prime} \right) \un 
 \,,&\numeq\cr} 
 $$

\noindent where $\epsilon\left({k_1},{k_2},\{ {\mu_1}\} ,\{{\mu_2}\} \right)\equiv
  \epsilon_{\alpha\beta\mu_1\mu_2}\,k_1^\alpha k_2^\beta$ and
$\un$ is the unit of the spinor space.
Note that no significant simplification is achieved by using the standard 
choice of $R_\xi$-gauge $\rho\equiv\xi\, v $

Then, the coefficients, $\ti\Delta_j^{(1)}$, of the breaking in
the basis \EqUbasis\ of four dimensional integrated operators can be
automatically obtained with the aid of the formulas (step {\it iii}$\,$):
\bigskip
{\settabs 2 \columns
\openup1\jot
\+${\ti \Delta}_{1}^{(1)}\;=
  \tilde\Gamma_{{\phi_2} c;N[\hat\Delta]}^{{\rm R}(1)\; } 
     (k_1\!\equiv\!0)\,$,
 &${\ti \Delta}_{2}^{(1)}\;=
  \tilde\Gamma_{{\phi_1}{\phi_2} c;N[\hat\Delta]}^{{\rm R}(1)\; } 
     (k_1\!\equiv\!0,k_2\!\equiv0)\,$,
\cr
\+${\ti \Delta}_{3}^{(1)}\;=
  {1\over 3!}
  \tilde\Gamma_{{\phi_2}{}^3 c;N[\hat\Delta]}^{{\rm R}(1)\; }\,$,
 &${\ti \Delta}_{4}^{(1)}\;=
  {1\over 2!}
  \tilde\Gamma_{{\phi_1}{}^2{\phi_2} c;N[\hat\Delta]}^{{\rm R}(1)\; }\,$,
\cr
\+${\ti \Delta}_{5}^{(1)}\;=
  -{\rm Coeff.}\, {\rm of}\;{k_1}^2\, {\rm in}\;
  \tilde\Gamma_{{\phi_2} c;N[\hat\Delta]}^{{\rm R}(1)\; }
   (k_1)\,$,
 &${\ti \Delta}_{6}^{(1)}\;=
  {1\over 3!}
  \tilde\Gamma_{{\phi_1}{}^3 c;N[\hat\Delta]}^{{\rm R}(1)\; }\,$,
\cr
\+${\ti \Delta}_{7}^{(1)}\;=
  {1\over 3!}
  \tilde\Gamma_{{\phi_1}{\phi_2}{}^3 c;N[\hat\Delta]}^{{\rm R}(1)\; }\,$,
 &${\ti \Delta}_{8}^{(1)}\;=
  -{\rm Coeff.}\, {\rm of}\;{k_1}^2\,{\rm in}\;
  \tilde\Gamma_{{\phi_1}{\phi_2} c;N[\hat\Delta]}^{{\rm R}(1)\; }
   (k_1,k_2)\,$,
\cr
\+${\ti \Delta}_{9}^{(1)}\;=
  -{\rm Coeff.}\, {\rm of}\;{k_2}^2\,{\rm in}\;
  \tilde\Gamma_{{\phi_1}{\phi_2} c;N[\hat\Delta]}^{{\rm R}(1)\; }
   (k_1,k_2)\,$,
 &${\ti \Delta}_{10}^{(1)}\;=
  -{\rm Coeff.}\, {\rm of}\;k_1 \cdot k_2\,{\rm in}\;
  \tilde\Gamma_{{\phi_1}{\phi_2} c;N[\hat\Delta]}^{{\rm R}(1)\; }
   (k_1,k_2)\,$,
\cr
\+${\ti \Delta}_{11}^{(1)}\;=
  -i\,{\rm Coeff.}\, {\rm of}\;{k_1}^{\mu_1}\,{\rm in}\;
  \tilde\Gamma_{{A}c;N[\hat\Delta]}^{{\rm R}(1)\;\mu_1 }
   (k_1)\,$,
 &${\ti \Delta}_{12}^{(1)}\;=
  -i\,{\rm Coeff.}\, {\rm of}\;{k_2}^{\mu_1}\,{\rm in}\;
  \tilde\Gamma_{{A\phi_1}c;N[\hat\Delta]}^{{\rm R}(1)\;\mu_1 }
   (k_1,k_2)\,$,
\cr
\+${\ti \Delta}_{13}^{(1)}\;=
  -i\,{\rm Coeff.}\, {\rm of}\;{k_1}^{\mu_1}\,{\rm in}\;
  \tilde\Gamma_{{A\phi_1}c;N[\hat\Delta]}^{{\rm R}(1)\;\mu_1 }
   (k_1,k_2)\,$,
\cr
\+${\ti \Delta}_{14}^{(1)}\;=
  {-i\over2}\,{\rm Coeff.}\, {\rm of}\;{k_1}^{\mu_1}\,{\rm in}\;
  \tilde\Gamma_{{A{\phi_1}^2}c;N[\hat\Delta]}^{{\rm R}(1)\;\mu_1 }
   (k_1,k_2,k_3)\,$,
\cr
\+${\ti \Delta}_{15}^{(1)}\;=
  -i\,{\rm Coeff.}\, {\rm of}\;\big\{{k_2}^{\mu_1},
                                     {k_3}^{\mu_1}\big\}\,{\rm in}\;
 \tilde\Gamma_{{A{\phi_1}^2}c;N[\hat\Delta]}^{{\rm R}(1)\;\mu_1 }
   (k_1,k_2,k_3)\,$,
\cr
\+${\ti \Delta}_{16}^{(1)}\;=
  {-i\over2}\,{\rm Coeff.}\, {\rm of}\;{k_1}^{\mu_1}\,{\rm in}\;
  \tilde\Gamma_{{A{\phi_2}^2}c;N[\hat\Delta]}^{{\rm R}(1)\;\mu_1 }
   (k_1,k_2,k_3)\,$,
\cr
\+${\ti \Delta}_{17}^{(1)}\;=
  -i\,{\rm Coeff.}\, {\rm of}\;\big\{{k_2}^{\mu_1},
                                     {k_3}^{\mu_1}\big\}\,{\rm in}\;
  \tilde\Gamma_{{A{\phi_2}^2}c;N[\hat\Delta]}^{{\rm R}(1)\;\mu_1 }
   (k_1,k_2,k_3)\,$,
\cr
\+${\ti \Delta}_{18}^{(1)}\;=
  i\,{\rm Coeff.}\, {\rm of}\;{k_1}^{\mu_1}{k_1}^2\,{\rm in}\;
  \tilde\Gamma_{{A}c;N[\hat\Delta]}^{{\rm R}(1)\;\mu_1 }
   (k_1,k_2)\,$,
\cr
\+${\ti \Delta}_{19}^{(1)}\;=
  -i\,{\rm Coeff.}\, {\rm of}\;\big\{g^{\mu_1\mu_2}{k_2}^{\mu_3},
                                     g^{\mu_1\mu_2}{k_1}^{\mu_3},
                                     g^{\mu_1\mu_3}{k_1}^{\mu_2},
                                     g^{\mu_1\mu_2}{k_3}^{\mu_2},
                                     g^{\mu_2\mu_3}{k_2}^{\mu_1},
                                     g^{\mu_2\mu_3}{k_3}^{\mu_1}
                               \big\}\,$
  \cr\+$\qquad\qquad\qquad{\rm in}\;
 \tilde\Gamma_{{AAA}c;N[\hat\Delta]}^{{\rm R}(1)\;\mu_1\mu_2\mu_3 }
   (k_1,k_2,k_3)\,$,
\cr
\+${\ti \Delta}_{20}^{(1)}\;=
  {1\over2}\,{\rm Coeff.}\, {\rm of}\;g^{\mu_1\mu_2}\,{\rm in}\;
  \tilde\Gamma_{{AA{\phi_2}}c;N[\hat\Delta]}^{{\rm R}(1)\;
              \mu_1\mu_2 }\,$,
 &${\ti \Delta}_{21}^{(1)}\;=
  {1\over2}\,{\rm Coeff.}\, {\rm of}\;g^{\mu_1\mu_2}\,{\rm in}\;
  \tilde\Gamma_{{AA{\phi_1}{\phi_2}}c;N[\hat\Delta]}^{{\rm R}(1)\;
              \mu_1\mu_2 }\,$,
\cr
\+${\ti \Delta}_{22}^{(1)}\;=
  -i\,{\rm Coeff.}\, {\rm of}\;\big\{g^{\mu_1\mu_2}{k_3}^{\mu_3},
                                     g^{\mu_1\mu_3}{k_2}^{\mu_2},
                                     g^{\mu_2\mu_3}{k_1}^{\mu_1}
                               \big\}\,
 {\rm in}\;
 \tilde\Gamma_{{AAA}c;N[\hat\Delta]}^{{\rm R}(1)\;\mu_1\mu_2\mu_3 }
   (k_1,k_2,k_3)\,$,
\cr
\+${\ti \Delta}_{23}^{(1)}\;=
  {\rm Coeff.}\, {\rm of}\;\gam5\; {\rm in}\;\;
  \tilde\Gamma_{\psi\bar\psi c;N[\hat\Delta]}^{{\rm R}(1)}
     (k_1\!\equiv\!0,k_2\!\equiv0)\,$,
\cr
\+${\ti \Delta}_{24}^{(1)}\;=
  -i\,{\rm Coeff.}\, {\rm of}\;\big\{\kslash_1\PL,
                                \kslash_2\PL\big\}\; {\rm in}\;\;
  \tilde\Gamma_{\psi\bar\psi c;N[\hat\Delta]}^{{\rm R}(1)}
     (k_1,k_2)=$
\cr
\+$
\qquad
  -i\big(
    {\rm Coeff.}\, {\rm of}\;\kslash_1 -\;
    {\rm Coeff.}\, {\rm of}\;\kslash_1\gam5 \;\big) {\rm in}\;\;
  \tilde\Gamma_{\psi\bar\psi c;N[\hat\Delta]}^{{\rm R}(1)}
     (k_1,k_2)\,$,
\cr 
\+${\ti \Delta}_{25}^{(1)}\;=
  -i\,{\rm Coeff.}\, {\rm of}\;\big\{\kslash_1\PR,
                                \kslash_2\PR\big\}\; {\rm in}\;\;
  \tilde\Gamma_{\psi\bar\psi c;N[\hat\Delta]}^{{\rm R}(1)}
     (k_1,k_2)=$
\cr
\+$
\qquad
  -i\big(
    {\rm Coeff.}\, {\rm of}\;\kslash_1 +\;
    {\rm Coeff.}\, {\rm of}\;\kslash_1\gam5 \;\big) {\rm in}\;\;
  \tilde\Gamma_{\psi\bar\psi c;N[\hat\Delta]}^{{\rm R}(1)}
     (k_1,k_2)\,$,
\cr
\+${\ti \Delta}_{26}^{(1)}\;=
  \tilde\Gamma_{\psi\bar\psi\phi_2 c;N[\hat\Delta]}^{{\rm R}(1)}$,
 &${\ti \Delta}_{27}^{(1)}\;=
  {\rm Coeff.}\, {\rm of}\;\gam5\;{\rm in}\,
  \tilde\Gamma_{\psi\bar\psi\phi_1 c;N[\hat\Delta]}^{{\rm R}(1)}$,
\cr
\+${\ti \Delta}_{28}^{(1)}\;=
  {1\over2}{\rm Coeff.}\, {\rm of}\;
   \varepsilon^{\mu_1\mu_2\alpha\beta}\,{k_1}_\alpha{k_2}_\beta\;
   {\rm in}\,
  \tilde\Gamma_{AA c;N[\hat\Delta]}^{{\rm R}(1)\;\mu_1\mu_2}
    (k_1,k_2)$
  &\hfill\numeq &,
\cr
} \namelasteq\EqFormulaCoefficients

\noindent where, for example,
``${\rm Coeff.}\, {\rm of}\;\big\{{k_2}^{\mu_1},
                          {k_3}^{\mu_1}\big\}\,{\rm in}\;X$''
stands for
``coefficient of ${k_2}^{\mu_1}$ in $X$ or
  coefficient of ${k_3}^{\mu_1}$ in $X$''
(that is, they must be equal).

The results, consistent when several formulae for a coefficient are
possible, thus obtained read
\smallskip

\def\frac#1#2{{\displaystyle #1}\over{\displaystyle #2}}
\def\frac#1#2{{\displaystyle #1}\over{\displaystyle #2}}
  
 {\settabs 3 \columns 
 \openup1\jot 
 \+$(4\pi)^2{\tilde \Delta}_1^{(1)}\;= \;
 -8\,{}\,{f^4}\,{v^3} 
   $, 
  &$(4\pi)^2{\tilde \Delta}_2^{(1)}\;= \;
 -24\,{}\,{f^4}\,{v^2} 
   $, 
  &$(4\pi)^2{\tilde \Delta}_3^{(1)}\;= \;
 {\frac{-8\,{}\,{f^4}\,v}{3}} 
   $, 
 \cr 
 \+$(4\pi)^2{\tilde \Delta}_4^{(1)}\;= \;
 -24\,{}\,{f^4}\,v 
   $, 
  &$(4\pi)^2{\tilde \Delta}_5^{(1)}\;= \;
 {\frac{-4\,{}\,{f^2}\,v}{3}} 
   $, 
  &$(4\pi)^2{\tilde \Delta}_6^{(1)}\;= \;
 -8\,{}\,{f^4} 
   $, 
 \cr 
 \+$(4\pi)^2{\tilde \Delta}_7^{(1)}\;= \;
 {\frac{-8\,{}\,{f^4}}{3}} 
   $, 
  &$(4\pi)^2{\tilde \Delta}_8^{(1)}\;= \;
 -4\,{}\,{f^2} 
   $, 
  &$(4\pi)^2{\tilde \Delta}_9^{(1)}\;= \;
 {\frac{-4\,{}\,{f^2}}{3}} 
   $, 
 \cr 
 \+$(4\pi)^2{\tilde \Delta}_{10} ^{(1)}\;= \;
 -4\,{}\,{f^2} 
   $, 
  &$(4\pi)^2{\tilde \Delta}_{11}^{(1)}\;= \;
 -2\,{}\,{f^2}\,{v^2} 
   $, 
  &$(4\pi)^2{\tilde \Delta}_{12}^{(1)}\;= \;
 0 
   $, 
 \cr 
 \+$(4\pi)^2{\tilde \Delta}_{13}^{(1)}\;= \;
 -4\,{}\,{f^2}\,v 
   $, 
  &$(4\pi)^2{\tilde \Delta}_{14}^{(1)}\;= \;
 -2\,{}\,{f^2} 
   $, 
  &$(4\pi)^2{\tilde \Delta}_{15}^{(1)}\;= \;
 0 
   $, 
 \cr 
 \+$(4\pi)^2{\tilde \Delta}_{16}^{(1)}\;= \;
 -2\,{}\,{f^2} 
   $, 
  &$(4\pi)^2{\tilde \Delta}_{17}^{(1)}\;= \;
 -8\,{}\,{f^2} 
   $, 
  &$(4\pi)^2{\tilde \Delta}_{18}^{(1)}\;= \;
 {\frac{-{1}}{3}} 
   $, 
 \cr 
 \+$(4\pi)^2{\tilde \Delta}_{19}^{(1)}\;= \;
 {\frac{2\,{}}{3}} 
   $, 
  &$(4\pi)^2{\tilde \Delta}_{20}^{(1)}\;= \;
 4\,{}\,{f^2}\,v 
   $, 
  &$(4\pi)^2{\tilde \Delta}_{21}^{(1)}\;= \;
 4\,{}\,{f^2} 
   $, 
 \cr 
 \+$(4\pi)^2{\tilde \Delta}_{22}^{(1)}\;= \;
 {\frac{{1}}{3}} 
   $, 
 \cr 
 \+$(4\pi)^2{\tilde \Delta}_{23}^{(1)}\;= \;
 {\frac{-i}{6}}\,{}\,f\, 
   \left( 3\,{\rho}\,{r} +  
     4\,{g^2}\,{\theta}\, 
      \left( 1 + {\theta}\,{r} \right) \,v\, 
      \left( 5 + {\xi^\prime} \right)  \right)  
   $, 
 \cr 
 \+$(4\pi)^2{\tilde \Delta}_{24}^{(1)}\;= \;
 {\frac{-\left( {}\, 
        \left( -6\,{f^2}\,{r} +  
          {g^2}\, 
           \left( 2\,{\theta} + {r} +  
             {{{\theta}}^2}\,{r} \right) \, 
           \left( 5 + {\xi^\prime} \right)  \right)  \right) }{6}} 
   $, 
 \cr 
 \+$(4\pi)^2{\tilde \Delta}_{25}^{(1)}\;= \;
 {\frac{{}\,{r}\, 
      \left( -6\,{f^2} + {g^2}\,{{{\theta}}^2}\, 
         \left( 5 + {\xi^\prime} \right)  \right) }{6}} 
   $, 
 \cr 
 \settabs 2 \columns 
 \+$(4\pi)^2{\tilde \Delta}_{26}^{(1)}\;= \;
 {\frac{-2\,{}\,f\,{g^2}\,{\theta}\, 
      \left( {\theta} + {r} \right) \, 
      \left( 5 + {\xi^\prime} \right) }{3}} 
   $, 
  &$(4\pi)^2{\tilde \Delta}_{27}^{(1)}\;= \;
 {\frac{-2\,i}{3}}\,{}\,f\,{g^2}\, 
   {\theta}\,\left( 1 + {\theta}\,{r} \right) \, 
   \left( 5 + {\xi^\prime} \right)  
   $, 
  & 
 \cr 
 \+$(4\pi)^2{\tilde \Delta}_{28}^{(1)}\;= \;
 {\frac{2\,{}\,\left( 3\,{\theta} +  
        {r} + 3\,{{{\theta}}^2}\,{r} \ 
 \right) }{3}} 
  $. 
  & \hfill\numeq &\cr 
  \+\cr 
 }

\smallskip
This breaking is simplified a bit with the choice of
gauge $\xi^\prime\equiv -5 $.

Note that if only a fermion is present of
type, for example, $r=+1$, then the anomaly coefficient is not zero
for any value of $\theta$ and that by adding fermions of the
same type, the coefficient anomaly can not be canceled. Fermions
of both types are needed. For example, there is
cancellation of the anomaly in the case of two fermions with
$\theta_1=\theta_2=0$ and $r_1=+1$, $r_2=-1$ or in
the case of two fermions with
$\theta_1=1$, $\theta_2=-1$ and $r_1=+1$, $r_2=-1$.
Remembering the definitions $e_{{\rm R}k}\equiv\theta_k$ and
$e_{{\rm L}k}\equiv\theta_k + r_k$ with $r_k=\pm 1$, the
obtained coefficient of
the anomaly can be written in the more familiar form:
$$
\ti\Delta_{28}{}^{(1)}=
{1 \over (4\pi)^2} {2\over3} 
  \sum_{k\in I\cup J} \left( e_{{\rm L}\,k}^3 - e_{{\rm R}\,k}^3 \right)\,
 \eqno\numeq
$$
but the constraints \EqConstraints\ should never be forgotten.

\section{6. Restoration of BRS symmetry: finite counterterms}

We know from the algebraic theory of renormalization presented
in third section that the linear system \EqLinearSystem\ has
to be compatible, but its solution is not unique. Facts which can not
be trivially deduced from \EqFormulaCoefficients.
Using the values of the coefficients ${\tilde \Delta}_i^{(1)}$,
found in the previous section, this turns to be the case
and one of the solutions is:
\smallskip

\def\frac#1#2{{\displaystyle #1}\over{\displaystyle #2}}
\def\frac#1#2{{\displaystyle #1}\over{\displaystyle #2}}
  
 {\settabs 3 \columns 
 \openup1\jot 
 \+$(4\pi)^2{\tilde x}_{0,1}^{(1)}\;= \;
 8\,{}\,{f^4}\,{v^3} 
   $, 
  &$(4\pi)^2{\tilde x}_{0,2}^{(1)}\;= \;
 12\,{}\,{f^4}\,{v^2} 
   $, 
  &$(4\pi)^2{\tilde x}_{0,3}^{(1)}\;= \;
 0 
   $, 
 \cr 
 \+$(4\pi)^2{\tilde x}_{0,4}^{(1)}\;= \;
 8\,{}\,{f^4}\,v 
   $, 
  &$(4\pi)^2{\tilde x}_{0,5}^{(1)}\;= \;
 0 
   $, 
  &$(4\pi)^2{\tilde x}_{0,6}^{(1)}\;= \;
 2\,{}\,{f^4} 
   $, 
 \cr 
 \+$(4\pi)^2{\tilde x}_{0,7}^{(1)}\;= \;
 {\frac{-2\,{}\,{f^4}}{3}} 
   $, 
  &$(4\pi)^2{\tilde x}_{0,8}^{(1)}\;= \;
 0 
   $, 
  &$(4\pi)^2{\tilde x}_{0,9}^{(1)}\;= \;
 0 
   $, 
 \cr 
 \+$(4\pi)^2{\tilde x}_{0,10} ^{(1)}\;= \;
 {\frac{2\,{}\,{f^2}}{3}} 
   $, 
  &$(4\pi)^2{\tilde x}_{0,11}^{(1)}\;= \;
 0 
   $, 
  &$(4\pi)^2{\tilde x}_{0,12}^{(1)}\;= \;
 0 
   $, 
 \cr 
 \+$(4\pi)^2{\tilde x}_{0,13}^{(1)}\;= \;
 4\,{}\,{f^2} 
   $, 
  &$(4\pi)^2{\tilde x}_{0,14}^{(1)}\;= \;
 {}\,{f^2}\,{v^2} 
   $, 
  &$(4\pi)^2{\tilde x}_{0,15}^{(1)}\;= \;
 2\,{}\,{f^2}\,v 
   $, 
 \cr 
 \+$(4\pi)^2{\tilde x}_{0,16}^{(1)}\;= \;
 {}\,{f^2} 
   $, 
  &$(4\pi)^2{\tilde x}_{0,17}^{(1)}\;= \;
 3\,{}\,{f^2} 
   $, 
  &$(4\pi)^2{\tilde x}_{0,18}^{(1)}\;= \;
 {\frac{-{1}}{6}} 
   $, 
 \cr 
 \+$(4\pi)^2{\tilde x}_{0,19}^{(1)}\;= \;
 0 
   $, 
  &$(4\pi)^2{\tilde x}_{0,20}^{(1)}\;= \;
 {\frac{-{1}}{12}} 
   $, 
  &$(4\pi)^2{\tilde x}_{0,21}^{(1)}\;= \;
 0 
   $, 
 \cr 
 \+$(4\pi)^2{\tilde x}_{0,22}^{(1)}\;= \;
 0 
   $, 
  &$(4\pi)^2{\tilde x}_{0,23}^{(1)}\;= \;
 0 
   $, 
  & 
 \cr 
 \+$(4\pi)^2{\tilde x}_{0,24}^{(1)}\;= \;
 {\frac{{}\,f\,\left( 3\,{\rho}\,{r} +  
        4\,{g^2}\,{\theta}\, 
         \left( 1 + {\theta}\,{r} \right) \,v\, 
         \left( 5 + {\xi^\prime} \right)  \right) }{6\,{r}}} 
   $, 
  & 
  & 
 \cr 
 \+$(4\pi)^2{\tilde x}_{0,25}^{(1)}\;= \;
 0 
   $, 
  &$(4\pi)^2{\tilde x}_{0,26}^{(1)}\;= \;
 0 
   $, 
  & 
 \cr 
 \+$(4\pi)^2{\tilde x}_{0,27}^{(1)}\;= \;
 {\frac{{}\,\left( -6\,{f^2}\,{r} +  
        {g^2}\,\left( 2\,{\theta} +  
           {r} + {{{\theta}}^2}\,{r} \ 
 \right) \,\left( 5 + {\xi^\prime} \right)  \right) }{6}} 
   $, 
  & 
  & 
 \cr 
 \+$(4\pi)^2{\tilde x}_{0,28}^{(1)}\;= \;
 {\frac{-\left( {}\,{r}\, 
        \left( -6\,{f^2} + {g^2}\,{{{\theta}}^2}\, 
           \left( 5 + {\xi^\prime} \right)  \right)  \right) }{6}} 
   $, 
  & 
  & 
 \cr 
 \+$(4\pi)^2{\tilde x}_{0,29}^{(1)}\;= \;
 {\frac{2\,{}\,f\,{g^2}\,{\theta}\, 
      \left( {\theta} + {r} \right) \, 
      \left( 5 + {\xi^\prime} \right) }{3}} 
   $, 
 \cr 
 \+$(4\pi)^2{\tilde x}_{0,30}^{(1)}\;= \;
 0 
   $, 
  &$(4\pi)^2{\tilde x}_{0,31}^{(1)}\;= \;
 0 
   $, 
  &$(4\pi)^2{\tilde x}_{0,32}^{(1)}\;= \;
 0 
   $.  \hfill\numeq & 
 \cr 
 \+\cr 
 }

Therefore, the general solution for the finite counterterms
up to one-loop order reads
$$
   \hbar\,\tilde S^{(1)}_{\rm fct}\;=\;-\,
     \hbar\,\sum_{i=1}^{32} \tilde x^{(1)}_{0,i}
 \tilde e_i \; + \
  \hbar\,\sum_{l=1}^{11} c^{(1)}_{l}\,{\cal I}_{l}\,,  \eqno\numeq
$$
with the basis $\tilde e_i$ being given by \EqEbasis\ and the
symmetric terms ${\cal I}_{l}$ by \EqInvariants. 
Therefore, the parametric family of regularized
actions $S_1\equiv S_0 + 
 \hbar\,\tilde S^{(1)}_{\rm fct}$, with
$\ti K_{\phi_2}=K_{\phi_2} - \rho\,\bar c$,
gives, by minimal subtraction in the BMHV scheme, all possible
CP-symmetric renormalized
theories, compatible with the tree-level action \EqClassicalact\ and
power counting renormalizability,
and satisfying up to one-loop level both the
STI \EqSTI\ and the gauge-fixing equation
\EqGaugeFixing.

Notice that, if $\theta\ne0$ and $\theta +r \ne0$
($\theta=0$ or $\theta +r=0$),
 there is a
seven (eight) dimensional family of
solutions, or equivalently,
of normalization conditions, which does not
imply finite counterterms depending on BRS external fields. The restriction
to this family would certainly simplify the two-loop analysis
of \EqRegSTBreakingct.
In a more general regulator independent context,
the simplificatory power given by the freedom
in the choice of normalization conditions 
have been stressed in \cite{\GrassiAH,\GrassiPAR}.

Finally, note that although the starting classical action of order $\hbar^0$
was chosen to satisfy $\mu^2=\lambda v^2$ so that the monomial
$\int\!{\phi_1}$ does not appear in the action,
we have the freedom to impose any $O(\hbar^n)$ mean value on the field
thanks to the trivial finite counterterms $\ti{\cal I}_2$,
$\ti{\cal I}_3$ and $\ti{\cal I}_6$. Setting
that value to zero would just define one of the normalization conditions
mentioned at the end of section 3.

\section{7. Conclusions}

{
Algebraic Renormalization theory has been mainly used for
demonstrative purposes (but see \cite{\GrassiAH,\GrassiPAR}).
In this paper, we have shown a pedagogical example which makes
manifest that the theoretical tools provided by the Algebraic
Renormalization theory
are of utmost importance in order to 
blindly carry out computations in a non-invariant
renormalization procedure such as the BMHV scheme is for chiral
gauge theories. Such non-invariant renormalization procedures
seems to be unavoidable in a near future for doing
trustable high-precision test of relevant quantum field
theories like the Standard Model, and mastery of these
techniques will be needed.

Although at first sight the method looks cumbersome for
the practitioners, we want to stress that once the
general expression for a non-invariant modified action
have been found at some order of the perturbative expansion
the automatic evaluation of renormalized 
diagrams satisfying the symmetries of the theory is not much more
difficult that the conventional procedures, 
because we need to do only a minimal subtraction of
all Feynman integrals obtained from the Feynman rules of
the given modified action. Certainly,
the gamma algebra is a bit
more tedious and there are more Feynman rules in the modified
action that in the conventional one, but nowadays all this is
perfectly admissible for the current computer codes.

The simplicity of the abelian Higgs-Kibble model allows for
explicit and order independent expressions for the possible
counterterms. 
Not obscuring thus the main steps of the algebraic method and making
it very suitable for a future study at two-loop order. This study
could be easily extended to the physically relevant Standard
Model. The arbitrariness of the choice of the regularizations
of vertices and finite counterterms could be a key to simplify
the computations at higher loop orders of the r.h.s.\ of
\EqRegSTBreakingct\ using for example similar
techniques to the ones in \cite{\Tonin}.
}

\section{Acknowledgments}

We thank C.P. Mart{\'\i}n for many useful discussions
regarding algebraic and dimensional renormalization and for
a careful reading of the manuscript. 
Correspondence with P.A. Grassi and partial financial support
by  Consejer{\'\i}a de Educaci\'on y
Cultura of Castilla-La Mancha 
during first stages of this research are also acknowledged.
 
\bigskip

\section{Appendix A. Matrix elements  of the linearized ST operator 
 $\ti b_0$ }

{
{
Using the definition \EqOperatorb,
the $\tilde b$--variations of the
fields are
$$
\eqalignno{
\tilde b\,A_\mu =& s A_\mu = \pr_\mu c\,, \cr
\tilde b\,\phi_1 =& s \phi_1 = -\phi_2 c\,, \cr
\tilde b\,\phi_2 =& s \phi_2 = (v + \phi_1) c\,, \cr
\tilde b\,\psi =& s \psi = 
  i c \left[(\theta+r)\PL + \theta\PR\right]\,\psi\,,\cr
\tilde b\,\bar\psi =& s \bar\psi = 
  i \bar\psi  \left[(\theta+r)\PR + \theta\PL\right]\,c\,,\cr
\tilde b\, c = & 0\,, \cr
\tilde b\, K_{\phi_1} =& {\delta\tilde\Gamma^0\over\delta\phi_1}=
  {\delta S_0\over\delta\phi_1}=\hbox{\rm e.o.m. of }\phi_1= \cr
  =& 
  -\square\,\phi_1 - (\pr_\mu A^\mu)\phi_2 - 2A^\mu (\pr_\mu \phi_2)
  + A_\mu A^\mu (v +\phi_1) + \mu^2 (v + \phi_1)\cr
&-
  \lambda [(v+\phi_1)^2 + \phi_2{}^2] (v+\phi_1) + \ti K_{\phi_2} c -
  f \bar\psi\psi\,,      \cr
\tilde b\,\tilde K_{\phi_2} =& {\delta\tilde\Gamma^0\over\delta\phi_2}=
    {\delta S_0\over\delta\phi_2}=\hbox{\rm e.o.m. of }\phi_2 -\rho B= \cr
   =& 
  -\square\,\phi_2 + (\pr_\mu A^\mu)(v+\phi_1) + 2A^\mu (\pr_\mu \phi_1)
  + A_\mu A^\mu \phi_2 + \mu^2\phi_2\cr
&-
   \lambda [(v+\phi_1)^2 + \phi_2{}^2] \phi_2 - K_{\phi_1} c -
   i\,r f \bar\psi\gam5\psi\,,      \cr
\tilde b\,K_\psi=&{\delta\tilde\Gamma^0\over\delta\psi} =
   \hbox{\rm e.o.m. of }\bar\psi= \cr
=& \bar\psi\left[i\leftslashedarrow 
    - \Aslash ((\theta +r)\PL + \theta\PR)\right] 
     +f\left[ (v +\phi_1)\bar\psi + i\,r\,\phi_2\bar\psi\gam5\right] \cr
&- i\, c\,K_\psi\left[(\theta+r)\PL + \theta\PR\right]\,, \cr
\tilde b\,K_{\bar\psi}=&{\delta\tilde\Gamma^0\over\delta\bar\psi} =
   \hbox{\rm e.o.m. of }\psi= \cr
=& \left[i\rightslashedarrow 
    + \Aslash ((\theta +r)\PL + \theta\PR)\right]\,\psi 
     -f\left[ (v +\phi_1)\psi + i\,r\,\phi_2\gam5\psi\right] \cr
&+ i\, c\,\left[(\theta+r)\PR + \theta\PL\right]\,K_{\bar\psi}\,. 
      &(A.1)\cr } 
$$

Therefore, applying these variations to the basis \EqEbasis\ of
$\tilde{\cal V}_0$ and expanding the results in 
the basis \EqUbasis\ of $\tilde{\cal V}_1$,
the matrix elements of this restriction of $\ti b$,
defined as $\ti b \ti e_i \equiv \ti b_0{}^j{}_i u_j$,
are easily found.}

\def\frac#1#2{{#1}\over{#2}}
The first 23 rows and 20 columns of matrix 
$\left\{\ti b_0{}^j{}_i\right\}_{1\le i\le 32 \atop 1\le j\le28}$ are:
\medskip
\begingroup
\sevenpoint
$
\left(\matrix{ -1 & 0 & 2\,v & 0 & 0 & 0 & 0 & 0 & 0 & 0 & 0 & 0 & 0 & 
  0 & 0 & 0 & 0 & 0 & 0 & 0 \cr 0 & -2 & 2 & 0 & 2\,v & 0 & 0 & 0
   & 0 & 0 & 0 & 0 & 0 & 0 & 0 & 0 & 0 & 0 & 0 & 0 \cr 0 & 0 & 0
   & 0 & -1 & 0 & 4\,v & 0 & 0 & 0 & 0 & 0 & 0 & 0 & 0 & 0 & 0
   & 0 & 0 & 0 \cr 0 & 0 & 0 & -3 & 2 & 0 & 0 & 2\,v & 0 & 0 & 0
   & 0 & 0 & 0 & 0 & 0 & 0 & 0 & 0 & 0 \cr 0 & 0 & 0 & 0 & 0 & 0
   & 0 & 0 & 0 & -2\,v & 1 & 0 & 0 & 0 & 0 & 0 & 0 & 0 & 0 & 0
   \cr 0 & 0 & 0 & 0 & 0 & -4 & 0 & 2 & 0 & 0 & 0 & 0 & 0 & 0 & 
  0 & 0 & 0 & 0 & 0 & 0 \cr 0 & 0 & 0 & 0 & 0 & 0 & 4 & -2 & 0
   & 0 & 0 & 0 & 0 & 0 & 0 & 0 & 0 & 0 & 0 & 0 \cr 0 & 0 & 0 & 0
   & 0 & 0 & 0 & 0 & 2 & 0 & 0 & 0 & -1 & 0 & 0 & 0 & 0 & 0 & 0
   & 0 \cr 0 & 0 & 0 & 0 & 0 & 0 & 0 & 0 & 0 & -2 & 0 & -1 & 0
   & 0 & 0 & 0 & 0 & 0 & 0 & 0 \cr 0 & 0 & 0 & 0 & 0 & 0 & 0 & 0
   & 0 & 0 & 0 & -1 & -1 & 0 & 0 & 0 & 0 & 0 & 0 & 0 \cr 0 & 0
   & 0 & 0 & 0 & 0 & 0 & 0 & 0 & 0 & v & 0 & 0 & -2 & 0 & 0 & 0
   & 0 & 0 & 0 \cr 0 & 0 & 0 & 0 & 0 & 0 & 0 & 0 & 0 & 0 & 0 & 
  -v & v & 0 & -2 & 0 & 0 & 0 & 0 & 0 \cr 0 & 0 & 0 & 0 & 0 & 0
   & 0 & 0 & 0 & 0 & 1 & -v & 0 & 0 & -2 & 0 & 0 & 0 & 0 & 0
   \cr 0 & 0 & 0 & 0 & 0 & 0 & 0 & 0 & 0 & 0 & 0 & -1 & 0 & 0 & 
  0 & -2 & 0 & 0 & 0 & 0 \cr 0 & 0 & 0 & 0 & 0 & 0 & 0 & 0 & 0
   & 0 & 0 & -1 & 1 & 0 & 0 & -4 & 0 & 0 & 0 & 0 \cr 0 & 0 & 0
   & 0 & 0 & 0 & 0 & 0 & 0 & 0 & 0 & 0 & 1 & 0 & 0 & 0 & -2 & 0
   & 0 & 0 \cr 0 & 0 & 0 & 0 & 0 & 0 & 0 & 0 & 0 & 0 & 0 & -1 & 
  1 & 0 & 0 & 0 & -4 & 0 & 0 & 0 \cr 0 & 0 & 0 & 0 & 0 & 0 & 0
   & 0 & 0 & 0 & 0 & 0 & 0 & 0 & 0 & 0 & 0 & 2 & 0 & 0 \cr 0 & 0
   & 0 & 0 & 0 & 0 & 0 & 0 & 0 & 0 & 0 & 0 & 0 & 0 & 0 & 0 & 0
   & 0 & 0 & -8 \cr 0 & 0 & 0 & 0 & 0 & 0 & 0 & 0 & 0 & 0 & 0 & 
  0 & 0 & 0 & -1 & 0 & 2\,v & 0 & 0 & 0 \cr 0 & 0 & 0 & 0 & 0 & 0
   & 0 & 0 & 0 & 0 & 0 & 0 & 0 & 0 & 0 & -2 & 2 & 0 & 0 & 0 \cr 
  0 & 0 & 0 & 0 & 0 & 0 & 0 & 0 & 0 & 0 & 0 & 0 & 0 & 0 & 0 & 0
   & 0 & 0 & 0 & -4 \cr 0 & 0 & 0 & 0 & 0 & 0 & 0 & 0 & 0 & 0 & 
  0 & 0 & 0 & 0 & 0 & 0 & 0 & 0 & 0 & 0 \cr }\right) 
$
\endgroup
\bigskip
\noindent the columns 21 to 32 are
\medskip
\begingroup
\sevenpoint
$
\left(\matrix{ v\,\left( {\mu^2} - {\lambda}\,{v^2} \right)  & 
  {\mu^2} - {\lambda}\,{v^2} & 0 & 0 & 0 & 0 & 0 & 0
   & 0 & 0 & 0 & 0 \cr {\mu^2} - 3\,{\lambda}\,{v^2} & 
  -2\,{\lambda}\,v & {\mu^2} - {\lambda}\,{v^2}
   & 0 & 0 & 0 & 0 & 0 & 0 & 0 & 0 & 0 \cr 
  -\left( {\lambda}\,v \right)  & -{\lambda} & 0 & 0 & 0
   & 0 & 0 & 0 & 0 & 0 & 0 & 0 \cr -3\,{\lambda}\,v & 
  -{\lambda} & -2\,{\lambda}\,v & 0 & 0 & 0 & 0 & 0
   & 0 & 0 & 0 & 0 \cr 0 & -1 & 0 & 0 & 0 & 0 & 0 & 0 & 0 & 0 & 
  0 & 0 \cr -{\lambda} & 0 & -{\lambda} & 0 & 0 & 0
   & 0 & 0 & 0 & 0 & 0 & 0 \cr -{\lambda} & 0 & 
  -{\lambda} & 0 & 0 & 0 & 0 & 0 & 0 & 0 & 0 & 0 \cr -1 & 0
   & 0 & 0 & 0 & 0 & 0 & 0 & 0 & 0 & 0 & 0 \cr 0 & 0 & -1 & 0 & 
  0 & 0 & 0 & 0 & 0 & 0 & 0 & 0 \cr 0 & 0 & 0 & 0 & 0 & 0 & 0 & 
  0 & 0 & 0 & 0 & 0 \cr 0 & v & 0 & 0 & 0 & 0 & 0 & 0 & 0 & 0 & 
  0 & 0 \cr 0 & 2 & 0 & 0 & 0 & 0 & 0 & 0 & 0 & 0 & 0 & 0 \cr 0
   & 1 & v & 0 & 0 & 0 & 0 & 0 & 0 & 0 & 0 & 0 \cr 0 & 0 & 1 & 0
   & 0 & 0 & 0 & 0 & 0 & 0 & 0 & 0 \cr 0 & 0 & 2 & 0 & 0 & 0 & 0
   & 0 & 0 & 0 & 0 & 0 \cr -1 & 0 & 0 & 0 & 0 & 0 & 0 & 0 & 0 & 
  0 & 0 & 0 \cr -2 & 0 & 0 & 0 & 0 & 0 & 0 & 0 & 0 & 0 & 0 & 0
   \cr 0 & 0 & 0 & 0 & 0 & 0 & 0 & 0 & 0 & 0 & 0 & 0 \cr 0 & 0 & 
  0 & 0 & 0 & 0 & 0 & 0 & 0 & 0 & 0 & 0 \cr v & 1 & 0 & 0 & 0 & 
  0 & 0 & 0 & 0 & 0 & 0 & 0 \cr 1 & 0 & 1 & 0 & 0 & 0 & 0 & 0 & 
  0 & 0 & 0 & 0 \cr 0 & 0 & 0 & 0 & 0 & 0 & 0 & 0 & 0 & 0 & 0 & 
  0 \cr 0 & -i\,f\,{r} & 0 & -i\,{r} & 0 & 0
   & 0 & 0 & 0 & v & -\left( f\,v \right)  & f\,v \cr 0 & 0 & 0 & 0
   & {\theta} + {r} & 0 & -1 & 0 & 0 & 0 & i & 0
   \cr 0 & 0 & 0 & 0 & 0 & {\theta} & 0 & -1 & 0 & 0 & 0
   & i \cr -f & 0 & 0 & 0 & 0 & 0 & 0 & 0 & -1 & 
  -i\,{r} & -i\,f\,{r} & i\,f\,{r}
   \cr 0 & 0 & -i\,f\,{r} & 0 & 0 & 0 & 0 & 0 & 
  -i\,{r} & 1 & -f & f \cr 0 & 0 & 0 & 0 & 0 & 0 & 0 & 
  0 & 0 & 0 & 0 & 0 \cr }\right) 
$
\endgroup
\bigskip
\noindent and the rest of its elements are 0.

}

\section{Appendix B. Explicit solution of the order one cohomology of $\ti b$ }

Let $\ti{\cal V}_i$ be the space of integrated Lorentz scalar
CP-invariant polynomials in the fields 
 $\phi_1$, $\phi_2$, $A_\mu$, $\psi$, $\bar\psi$, $\psi$, $c$,
$K_{\phi_1}$, $\ti{K}_{\phi_2}$, $K_\psi$ and $K_{\bar\psi}$
of maximal canonical dimension 4 and ghost number $i$.

We define
$\ti{\cal W}_{i+1} \equiv \ti b \ti{\cal V}_i$ and
$\ti{\cal K}_i \equiv \{\,
 \ti{\cal I} \in \ti{\cal V}_i \;/\; \ti b\, \ti{\cal I} =0\,
 \}$. Due to nilpotency of $\ti b$, 
$\ti{\cal W}_i  \subset \ti{\cal K}_i \subset  \ti{\cal V}_i$.

Solving explictly the cohomology of order one of $\ti b$ means
to find the elements of $\ti{\cal V}_1$ which are closed, i.e.
in $\ti{\cal K}_1$ but not exact, i.e. not
in $\ti{\cal W}_1$. Those not trivial elements of the cohomology
are termed {\it the anomaly }.

In order to do so, we introduce a basis for $\ti{\cal V}_2$:
{\settabs 3 \columns
\openup1\jot
\+${\ti v}_1\;\equiv\;\int\!{\phi_1}\,(\square\, c)\,c$,
 &${\ti v}_2\;\equiv\;\int\!{\phi_1}^2\,(\square\, c)\,c$,
 &${\ti v}_3\;\equiv\;\int\!{\phi_2}^2\,(\square\, c)\,c$,
\cr
\+${\ti v}_4\;\equiv\;\int\!A^\mu\phi_2\,(\pr_\mu c)\,c$,
 &${\ti v}_5\;\equiv\;\int\!A^\mu\phi_1\phi_2\,(\pr_\mu c)\,c$,
 &${\ti v}_6\;\equiv\;\int\!(\pr_\nu A^\nu) A^\mu\,(\pr_\mu c)\,c$,
\cr
\+${\ti v}_7\;\equiv\;\int\!(\pr^\nu A^\mu) A_\nu\,(\pr_\mu c)\,c$,
 &${\ti v}_8\;\equiv\;\int\!(\pr^\mu A^\nu) A_\nu\,(\pr_\mu c)\,c$,
 &\cr
}\namelasteq\EqVbasis
\noindent and the matrix of the restricted linear operator 
$\ti b_1\equiv \ti b : \ti{\cal V}_1\rightarrow \ti{\cal V}_2 $
as $\ti b \ti u_j \equiv \ti b_1{}^k{}_j\, \ti v_k$.
\def\frac#1#2{{#1}\over{#2}}
The columns 5 to 22 of matrix 
$\left\{\ti b_1{}^k{}_j\right\}_{1\le j\le 28 \atop 1\le k\le8}$ are

\medskip
\begingroup
\eightpoint
$
\left(\matrix{ -1 & 0 & 0 & 0 & v & -v & 0 & -1 & 1 & 0 & 0 & 0 & 0 & 0
   & 0 & 0 & 0 & 0 \cr 0 & 0 & 0 & 0 & 0 & -{\frac{1}{2}} & 0 & 0
   & 0 & 1 & -{\frac{1}{2}} & 0 & 0 & 0 & 0 & 0 & 0 & 0 \cr 0 & 0
   & 0 & 0 & 0 & {\frac{1}{2}} & 0 & 0 & 0 & 0 & 0 & 1 & 
  -{\frac{1}{2}} & 0 & 0 & 0 & 0 & 0 \cr 0 & 0 & 0 & 0 & 0 & 0 & 0
   & -1 & 0 & 0 & 0 & 0 & v & 0 & 0 & 2 & 0 & 0 \cr 0 & 0 & 0 & 
  0 & 0 & 0 & 0 & 0 & 0 & 0 & -1 & 0 & 1 & 0 & 0 & 0 & 2 & 0
   \cr 0 & 0 & 0 & 0 & 0 & 0 & 0 & 0 & 0 & 0 & 0 & 0 & 0 & 0 & 
  -1 & 0 & 0 & 2 \cr 0 & 0 & 0 & 0 & 0 & 0 & 0 & 0 & 0 & 0 & 0
   & 0 & 0 & 0 & 1 & 0 & 0 & -2 \cr 0 & 0 & 0 & 0 & 0 & 0 & 0 & 
  0 & 0 & 0 & 0 & 0 & 0 & 0 & 0 & 0 & 0 & 0 \cr }\right) 
$
\endgroup
\medskip
\noindent and the rest of its elements are 0.

A basis of the kernel $\ti{\cal K}_1$ of $\ti b_1$ is therefore:
{\settabs 3 \columns
\openup1\jot
\+$\ti{\cal J}_1\;=\;{\ti u}_{28}$,
 &$\ti{\cal J}_2\;=\;{\ti u}_1$,
 &$\ti{\cal J}_3\;=\;{\ti u}_2$,
\cr
\+$\ti{\cal J}_4\;=\;{\ti u}_3$,
 &$\ti{\cal J}_5\;=\;{\ti u}_4$,
 &$\ti{\cal J}_6\;=\;{\ti u}_6$,
\cr
\+$\ti{\cal J}_7\;=\;{\ti u}_7$,
 &$\ti{\cal J}_8\;=\;{\ti u}_{8}$,
 &$\ti{\cal J}_9\;=\;{\ti u}_{11}$,
\cr
\+$\ti{\cal J}_{10}\;=\;{\ti u}_{18}$,
 &$\ti{\cal J}_{11}\;=\;{\ti u}_{23}$,
 &$\ti{\cal J}_{12}\;=\;{\ti u}_{24}$,
\cr
\+$\ti{\cal J}_{13}\;=\;{\ti u}_{25}$,
 &$\ti{\cal J}_{14}\;=\;{\ti u}_{26}$,
 &$\ti{\cal J}_{15}\;=\;{\ti u}_{27}$,
\cr
\+$\ti{\cal J}_{16}\;=\;v\,{\ti u}_{5}+{\ti u}_{9}$,
 &$\ti{\cal J}_{17}\;=\;{\ti u}_{5}+{\ti u}_{13}$,
 &$\ti{\cal J}_{18}\;=\;2\,{\ti u}_{19}+{\ti u}_{22}$,
\cr
\+$\ti{\cal J}_{19}\;=\;-2{\ti u}_{5}+2{\ti u}_{12}+{\ti u}_{20}$,
 &$\ti{\cal J}_{20}\;=\;{\ti u}_{14}+2{\ti u}_{15}+{\ti u}_{21}$,
 &$\ti{\cal J}_{21}\;=\;2v{\ti u}_{5}\!-\!2{\ti u}_{10}\!-\!{\ti u}_{14}
                          \!+\!{\ti u}_{16}$,
\cr
\+$\ti{\cal J}_{22}\;=\;-2v{\ti u}_{5}+{\ti u}_{10}+v{\ti u}_{12}+ 
                          {\ti u}_{14}+ {\ti u}_{15}+ {\ti u}_{17}\,.$
 &
 &\cr
}\namelasteq\EqInvariantss
\medskip

Note that dim of $\ti{\cal K}_1-$ dim of $ \ti{\cal W}_1=1$, so
the anomaly is expanded by only one element of $\ti{\cal V}_1$.
Applying on each element of the basis of the kernel $\ti{\cal K}_1$
a linear independence test against the set of
linear independent columns of
the matrix ${\ti b_0}$, which expands the image of the operator
$\ti b_0$, it is inmediatly found that 
$\ti{\cal J}_1\;=\;{\ti u}_{28}$ is the anomaly, as affirmed
in section 3.

\section{Appendix C. Breaking 1 loop Feynman diagrams }

\def\eq#1{{{\noexpand\rm({\rm C.}#1)}}}          
\def\eqprime#1{{{\noexpand\rm({\rm C.}#1')}}}          
\eqcounter=0
\subsection{Notation}
$$
\eqalignno{
 (2\pi)^4\,\delta(k_1+\cdots&+k_m+k_{m+1})\,
  \tilde\Gamma_{X_1X_2\dots X_m c;N[\hat\Delta]}^{{\rm R}(1)\;\mu_1\dots\mu_p} 
  (k_1,\dots,k_m)=\cr
\int &dx_1\dots dx_{m+1}\;
 e^{i(k_1x_1+\cdots+k_{m+1}x_{m+1})}\,\times\cr
 &{\delta N[\hat\Delta]\cdot\Gamma^{{\rm R}(1)}
  [\phi_1,\phi_2,A,\psi,\bar\psi, c, \bar c,
   K_{\phi_1}, K_{\phi_2}, K_\psi, K_{\bar\psi}] \over
  \delta {X_1}_{\mu_1}(x_1)\dots \delta{X_p}_{\mu_p}(x_p)
  \delta{X_p}_{\mu_{p+1}}(x_{p+1})
  \dots\delta X_m(x_m)\, \delta c (x_{m+1})}\bigg|_{X\equiv0}
 \cr
}
$$
will stand for the minimally substracted one loop 1PI functions with one
insertion of the integrated breaking $N[\hat\Delta]$ and the
fields $ X_1X_2\dots X_m\, c $ as external legs. $X_i$ represents
any field of $\phi_1,\phi_2,A,\psi,\bar\psi$ (all 
1PI diagrams with at least a ghost or an anti-ghot or
and external BRS field are convergent by power counting,
and therefore null when taking into account the insertion
of the evanescent breaking operator). 
 $\mu_1\dots\mu_p$ are
the Lorentz indices of the corresponding bosons in $ X_1X_2\dots X_m$.
$k_1,\dots,k_m$ are the independent outgoing momenta of the fields
$ X_1X_2\dots X_m$.

\subsection{C.1 Bosonic diagrams}

\midinsert
{\settabs 4\columns \def\graphwidth{1.8in}   
 \def\graphwidthbis{1.5in}
\eightpoint
\+
         \hfill\hskip 0.26in
         \vbox{\epsfxsize=\graphwidthbis \epsffile{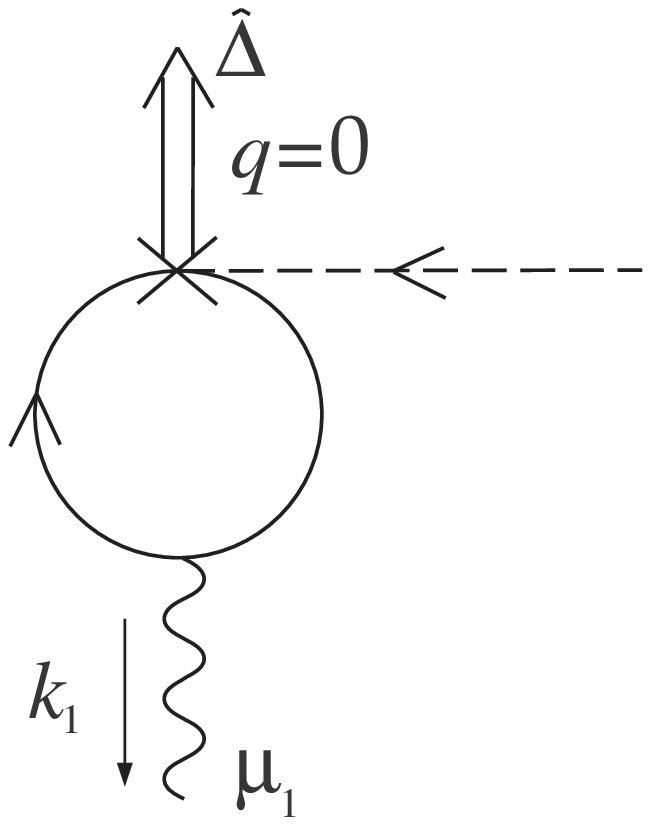}\vskip 0.02in}
         \hfill
	&\hfill\hskip 0.16in
         \vbox{\epsfxsize=\graphwidthbis \epsffile{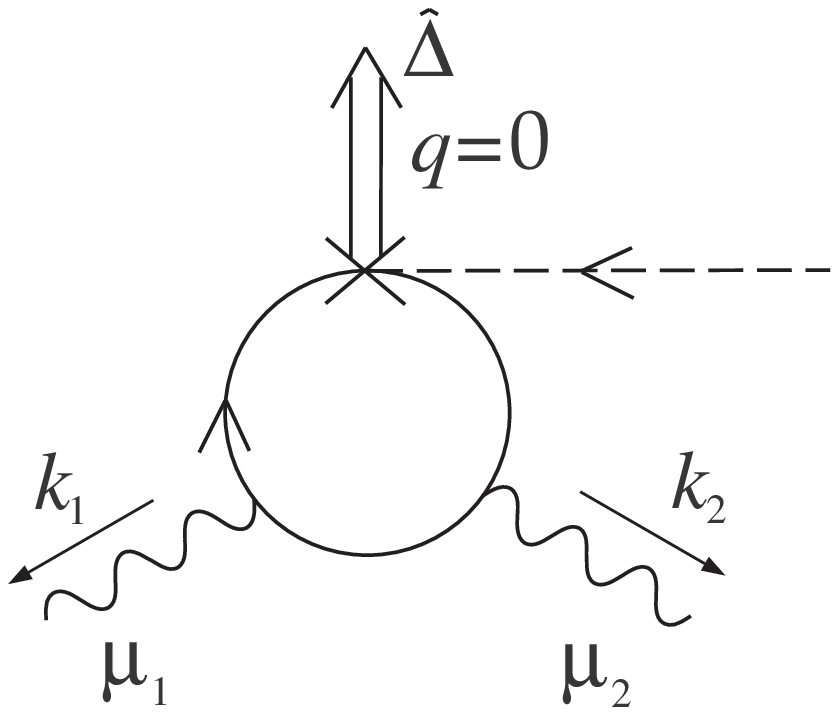}\vskip 0.21in}
         \hfill
	&\hfill
         \vbox{\epsfxsize=\graphwidth \epsffile{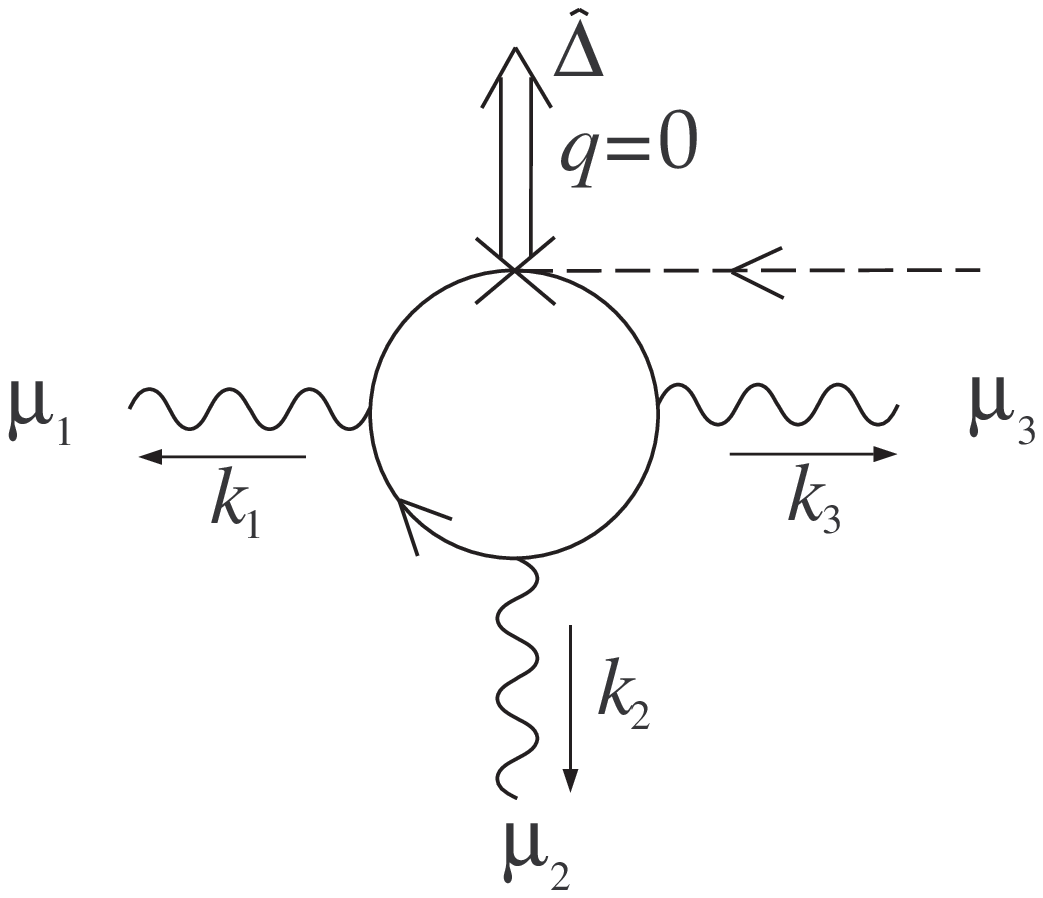}}
         \hfill 
	&\hfill
         \vbox{\epsfxsize=\graphwidth \epsffile{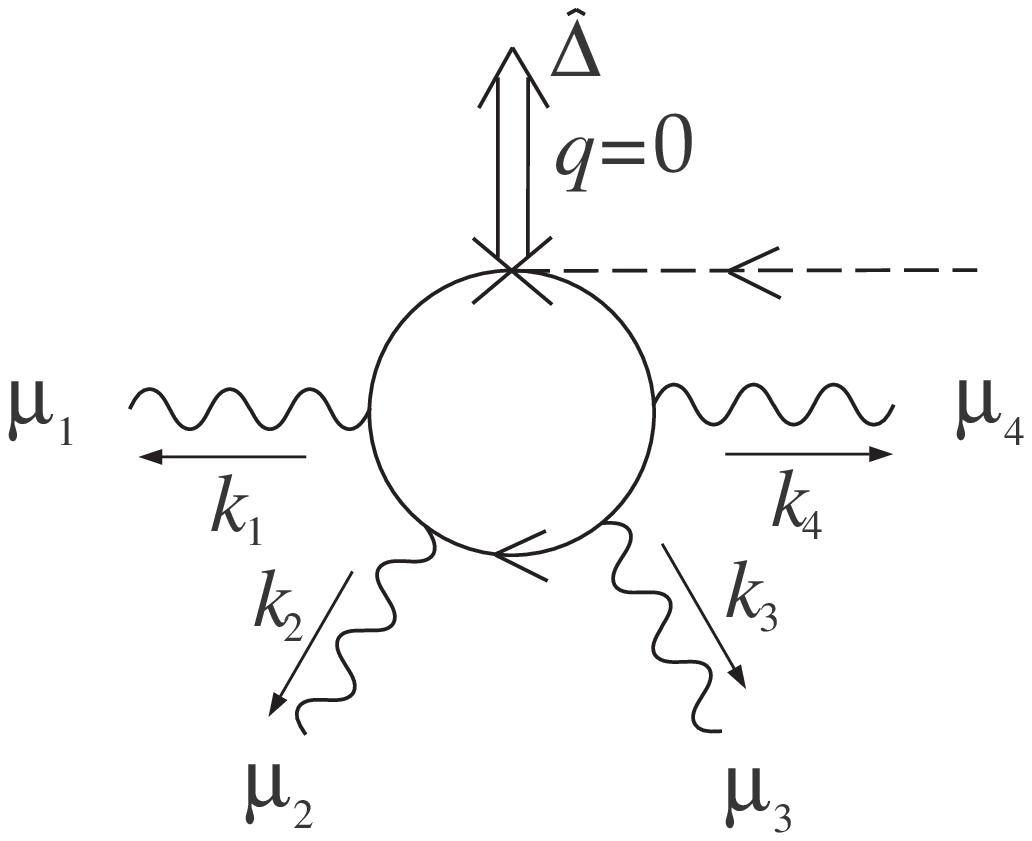}\vskip 0.03in}
         \hskip 0.14in \hfill &\cr
\+\hfill$(i)$\hfill
	&\hfill $(ii)_{12}$\hfill
	&\hfill $(iii)_{123}$\hfill
	& \hfill $(iv)_{1234}$\hfill &\cr
}

\vskip 12pt
\narrower\noindent {\bf Figure C.1:}
{\eightrm  Feynman diagrams with bosons needed to compute the 1PI breaking functions (C.1)}
%
%
\endinsert

{
$$
\eqalignno{\tilde\Gamma_{Ac;N[\hat\Delta]}^{{\rm R}(1)\;\mu_1 } (k_1)\,
 =\;&(i)\,,\cr
\tilde\Gamma_{AAc;N[\hat\Delta]}^{{\rm R}(1)\;\mu_1\mu_2 } (k_1,k_2)\,
 =\;&(ii)_{12} + (ii)_{21}\,,\cr
\tilde\Gamma_{AAAc;N[\hat\Delta]}^{{\rm R}(1)\;\mu_1\mu_2\mu_3 } 
  (k_1,k_2,k_3)\,=\;&
  (iii)_{123} + \hbox{\rm permutations of }123\,,\cr
\tilde\Gamma_{AAAAc;N[\hat\Delta]}^{{\rm R}(1)\;\mu_1\mu_2\mu_3\mu_4 } 
  (k_1,k_2,k_3,k_4)\,=\;&
  (iv)_{1234} + \hbox{\rm permutations of }1234\,.&\numeq\cr
}
$$
}

The renormalized result for each diagram of figure C.1 is:

\def\frac#1#2{{\displaystyle #1}\over{\displaystyle #2}}
\def\frac#1#2{{\displaystyle #1}\over{\displaystyle #2}}
$$ 
 \eqalignno{ 
 (4\pi)^2 \;(i)=&\;
 {\frac{-i}{3}}\,{}\, 
   \left( 6\,{f^2}\,{v^2} - {k_1}^{2} \right) \, 
   {k_1}^{{\mu_1}} 
  \,,\cr 
 (4\pi)^2 \;(ii)_{12}=&\;
 {\frac{2\,{}\,\left( 3\,{\theta} +  
         {r} + 3\,{{{\theta}}^2}\,{r} \ 
 \right) \,\epsilon\left({k_1},{k_2},\{ {\mu_1}\} \ 
 ,\{ {\mu_2}\} \right)}{3}} \cr & + {\frac{2\,i}{3}}\,{}\, 
    \left( 2\,{\theta} + {r} \right) \, 
    \left( {k_1}^{{\mu_1}}\, 
       {k_1}^{{\mu_2}} -  
      {k_2}^{{\mu_1}}\,{k_2}^{{\mu_2}} \ 
 \right)  \cr & - {\frac{i}{3}}\,{}\, 
    \left( 2\,{\theta} + {r} \right) \, 
    \left( {k_1}^{2} - {k_2}^{2} \right) \, 
    g^{{\mu_1} {\mu_2}} 
  \,,\cr 
 (4\pi)^2 \;(iii)_{123}=&\;
 {\frac{{}\,\left( 1 + 9\,{{{\theta}}^2} +  
         5\,{\theta}\,{r} +  
         6\,{{{\theta}}^3}\,{r} \right) \, 
       \epsilon\left({k_1},\{ {\mu_1}\} ,\{ \ 
 {\mu_2}\} ,\{ {\mu_3}\} \right)}{3}} \cr & + {\frac{\left( {} +  
         2\,{}\,{\theta}\,{r} \right) \, 
       \epsilon\left({k_2},\{ {\mu_1}\} ,\{ \ 
 {\mu_2}\} ,\{ {\mu_3}\} \right)}{3}} \cr & + {\frac{{}\,\left( 1 + 9\,{{{\theta}}^2} +  
         5\,{\theta}\,{r} +  
         6\,{{{\theta}}^3}\,{r} \right) \, 
       \epsilon\left({k_3},\{ {\mu_1}\} ,\{ \ 
 {\mu_2}\} ,\{ {\mu_3}\} \right)}{3}} \cr & + {\frac{i}{3}}\,{}\, 
    \left( 1 + {{{\theta}}^2} +  
      {\theta}\,{r} \right) \, 
    {k_1}^{{\mu_3}}\,g^{{\mu_1} {\mu_2}} \cr & + {\frac{i}{3}}\,{}\, 
    \left( 1 + 2\,{{{\theta}}^2} +  
      2\,{\theta}\,{r} \right) \, 
    {k_2}^{{\mu_3}}\,g^{{\mu_1} {\mu_2}} \cr & + {\frac{i}{3}}\,{}\, 
    \left( 1 + 3\,{{{\theta}}^2} +  
      3\,{\theta}\,{r} \right) \, 
    {k_3}^{{\mu_3}}\,g^{{\mu_1} {\mu_2}} \cr & - {\frac{i}{3}}\,{}\, 
    \left( 1 + 3\,{{{\theta}}^2} +  
      3\,{\theta}\,{r} \right) \, 
    {k_1}^{{\mu_2}}\,g^{{\mu_1} {\mu_3}} \cr & - {\frac{i}{3}}\,{}\, 
    \left( 1 + 6\,{{{\theta}}^2} +  
      6\,{\theta}\,{r} \right) \, 
    {k_2}^{{\mu_2}}\,g^{{\mu_1} {\mu_3}} \cr & - {\frac{i}{3}}\,{}\, 
    \left( 1 + 3\,{{{\theta}}^2} +  
      3\,{\theta}\,{r} \right) \, 
    {k_3}^{{\mu_2}}\,g^{{\mu_1} {\mu_3}} \cr & + {\frac{i}{3}}\,{}\, 
    \left( 1 + 3\,{{{\theta}}^2} +  
      3\,{\theta}\,{r} \right) \, 
    {k_1}^{{\mu_1}}\,g^{{\mu_2} {\mu_3}} \cr & + {\frac{i}{3}}\,{}\, 
    \left( 1 + 2\,{{{\theta}}^2} +  
      2\,{\theta}\,{r} \right) \, 
    {k_2}^{{\mu_1}}\,g^{{\mu_2} {\mu_3}} \cr & + {\frac{i}{3}}\,{}\, 
    \left( 1 + {{{\theta}}^2} +  
      {\theta}\,{r} \right) \, 
    {k_3}^{{\mu_1}}\,g^{{\mu_2} {\mu_3}} 
  \,,\cr 
 (4\pi)^2 \;(iv)_{1234}=&\;
 {\frac{-2\,{}\,{\theta}\, 
      \left( 1 + 6\,{{{\theta}}^2} +  
        4\,{\theta}\,{r} +  
        3\,{{{\theta}}^3}\,{r} \right) \, 
      \epsilon\left(\{ {\mu_1}\} ,\{ {\mu_2}\} ,\{ \ 
 {\mu_3}\} ,\{ {\mu_4}\} \right)}{3}} 
   \,. \;&\lasteqprime\cr 
 } 
 $$

Note that although the four boson diagram  is divergent by power counting,
the total result for the assocciated 1PI function must be zero due
to CP invariance of the regularized action and the dimensional
renormalization procedure. As a check of the automated
programs and of the preservation of discrete CP symmetry by
the renormalization scheme, we have computed
explicitly the value of the diagram $(iv)$ of figure C.1, which is not zero, 
and, as can easily
seen in last equation, the total result after suming all permutations
turns to be the expected result zero. That is, the discrete CP symmetry
is reflected in perturbative dimensinal renormalization as a
cancellation between permuted diagrams.

\subsection{C.2 Diagrams linear and quadractic in scalar fields}

\midinsert
{\settabs 3\columns \def\graphwidth{1.8in}   
 \def\graphwidthbis{1.5in}
\eightpoint
\+
         \hfill\hskip 0.46in
         \vbox{\epsfxsize=\graphwidthbis \epsffile{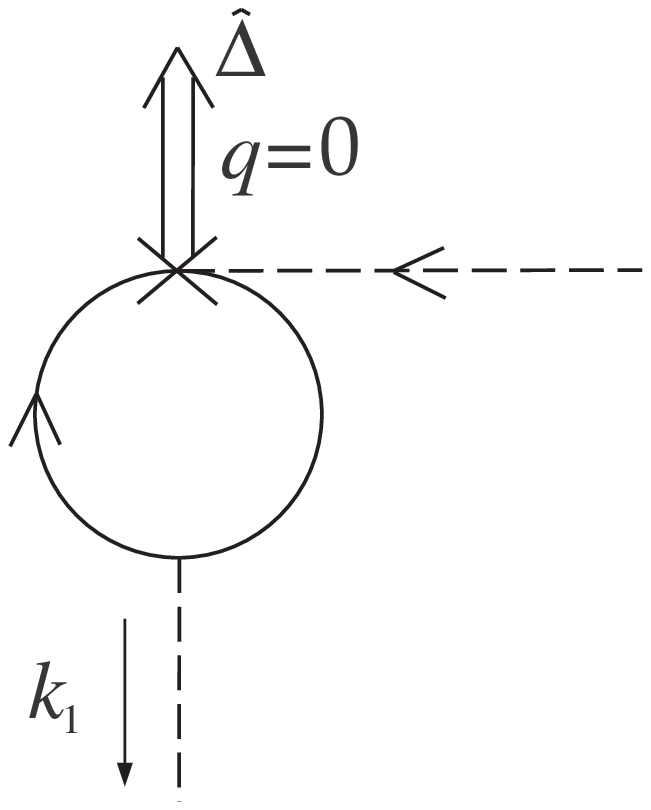}}
         \hfill
	&\hfill\hskip 0.32in
         \vbox{\epsfxsize=\graphwidth \epsffile{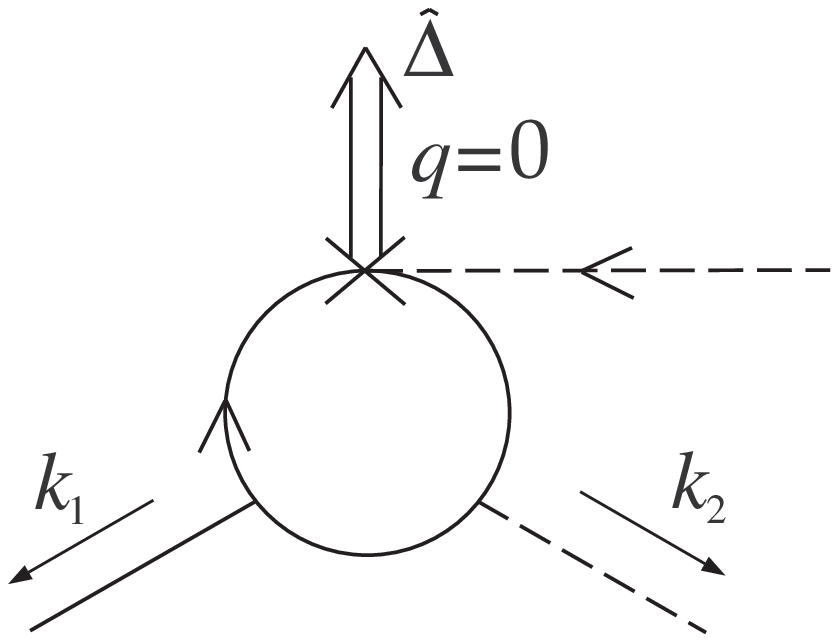}\vskip 0.13in}
         \hfill
	&\hfill\hskip 0.32in
         \vbox{\epsfxsize=\graphwidth \epsffile{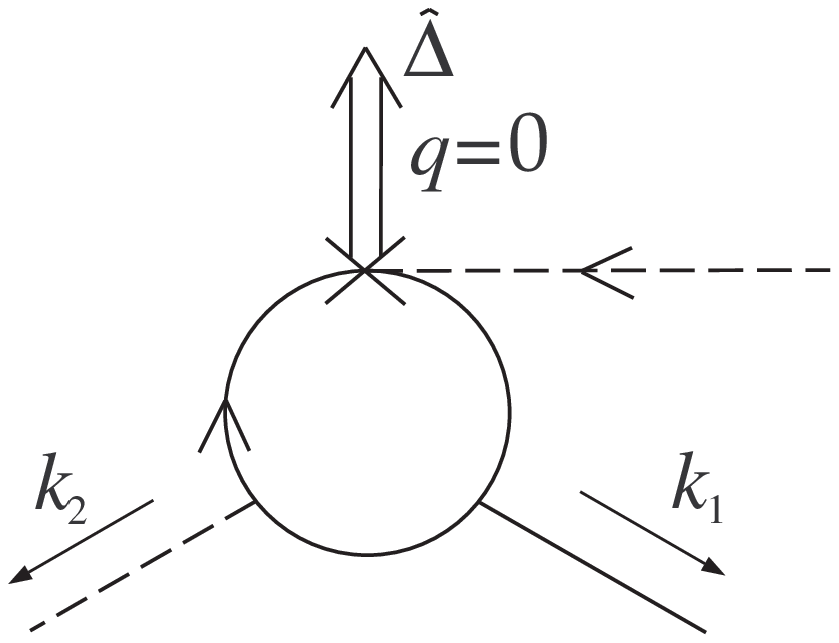}\vskip 0.13in}
         \hfill &\cr
\+       \hfill$(i)$\hfill
	&\hfill $(ii)$\hfill
	&\hfill $(iii)$\hfill &\cr
}

\vskip 12pt
\narrower\noindent {\bf Figure C.2:}
{\eightrm  Feynman diagrams linear or quadratic in scalar fields
needed to compute the 1PI breaking functions (C.2)}
%
%
\endinsert

{
$$
\eqalignno{
\tilde\Gamma_{\phi_2 c;N[\hat\Delta]}^{{\rm R}(1)} (k_1)\,
 =\;&(i)\,,\cr
\tilde\Gamma_{\phi_1\phi_2c;N[\hat\Delta]}^{{\rm R}(1)} (k_1,k_2)\,
 =\;&(ii) + (iii)\,.
  &\numeq\cr
}
$$
}

The renormalized result for each diagram of figure C.2 is:

\def\frac#1#2{{\displaystyle #1}\over{\displaystyle #2}}
\def\frac#1#2{{\displaystyle #1}\over{\displaystyle #2}}
$$ 
 \eqalignno{ 
 (4\pi)^2 \; (i)=&\;
 {\frac{4\,{}\,{f^2}\,v\, 
      \left( -6\,{f^2}\,{v^2} + {k_1}^{2} \right) }{3}} 
  \,,\cr 
 (4\pi)^2 \; (ii)=&\;
 {\frac{2\,{}\,{f^2}\, 
      \left( -18\,{f^2}\,{v^2} + 3\,{k_1}^{2} +  
        3\,{k_1} \cdot {k_2} + {k_2}^{2} \ 
 \right) }{3}} 
  \,,\cr 
 (4\pi)^2 \; (iii)=&\;
 {\frac{2\,{}\,{f^2}\, 
      \left( -18\,{f^2}\,{v^2} + 3\,{k_1}^{2} +  
        3\,{k_1} \cdot {k_2} + {k_2}^{2} \ 
 \right) }{3}} 
   \,. \; &\lasteqprime\cr 
 } 
 $$

\subsection{C.3 Diagrams with three scalar fields}

\midinsert
{\settabs 4\columns \def\graphwidth{1.8in}   
 \def\graphwidthbis{0.9in}
\eightpoint
\+
         \hfill\hskip 0.01in
         \vbox{\epsfxsize=\graphwidth \epsffile{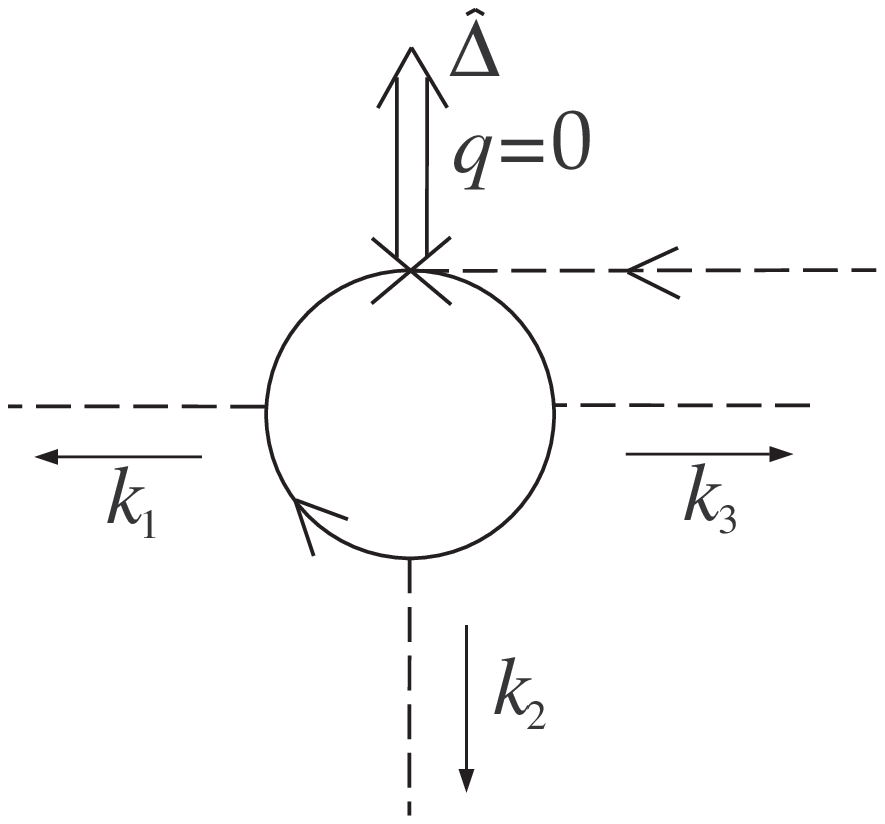}}
         \hfill
	&\hfill
         \vbox{\epsfxsize=\graphwidth \epsffile{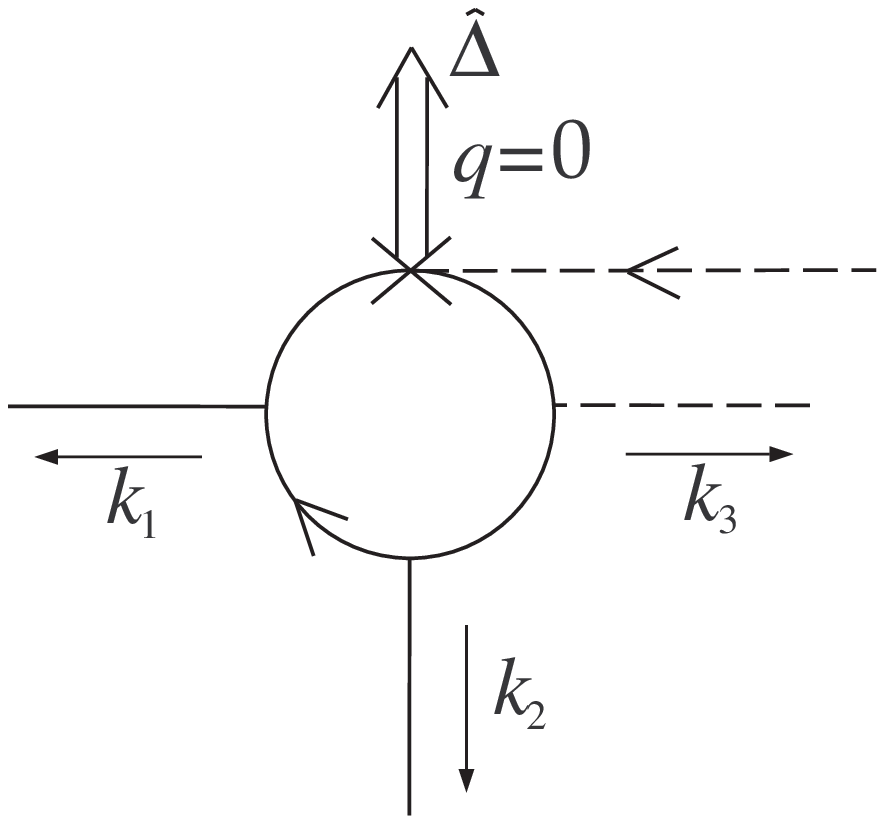}}
         \hfill
	&\hfill\hskip 0.13in
               \epsfxsize=\graphwidth \epsffile{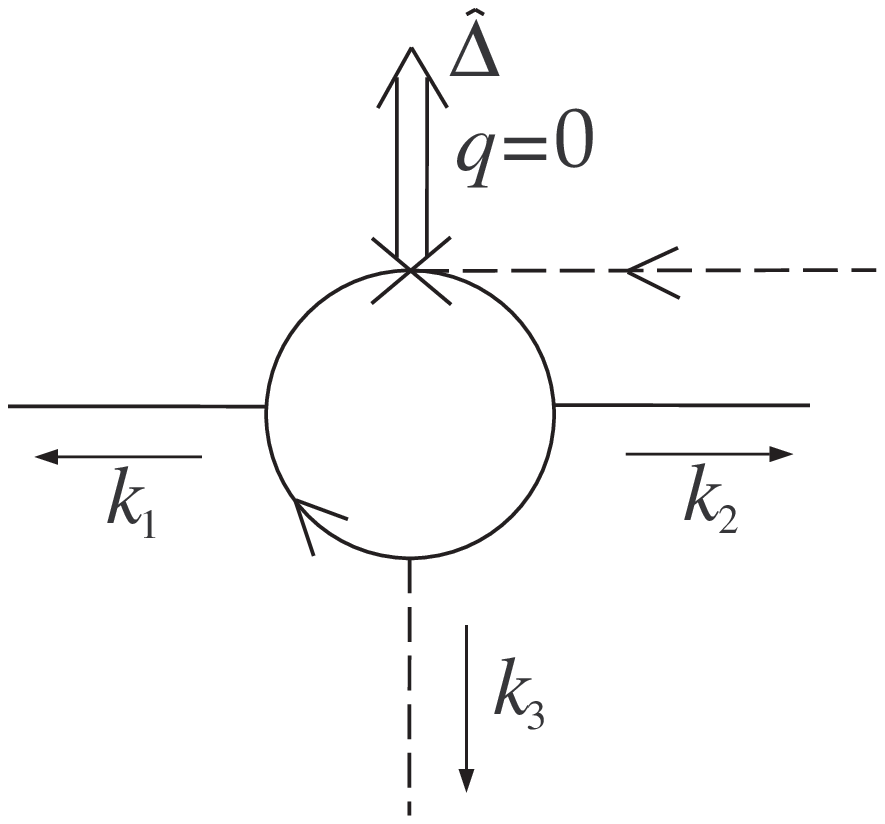}
         \hfill 
	&\hfill
         \vbox{\epsfxsize=\graphwidth \epsffile{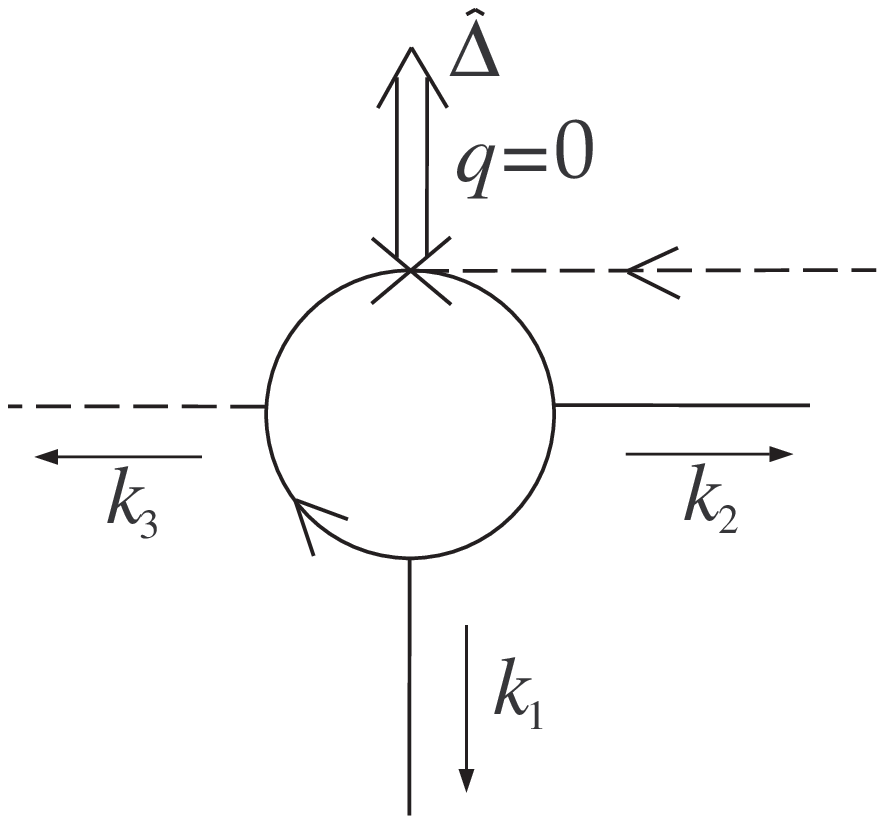}}
         \hskip 0.14in \hfill &\cr
\+\hfill$(i)_{123}$\hfill
	&\hfill $(ii)_{123}$\hfill
	&\hfill $(iii)_{123}$\hfill
	& \hfill $(iv)_{123}$\hfill &\cr
}

\vskip 12pt
\narrower\noindent {\bf Figure C.3:}
{\eightrm  Feynman diagrams with three scalar fields
needed to compute the 1PI breaking functions (C.3)}
%
%
\endinsert

{
$$
\eqalignno{\tilde\Gamma_{{\phi_2}{\phi_2}{\phi_2} c;N[\hat\Delta]}^{{\rm R}(1) }
  (k_1,k_2,k_3)\,
 =\;&(i)_{123}  + \hbox{\rm permut. of 123}\,,\cr
\tilde\Gamma_{{\phi_1}{\phi_1} \phi_2 c;N[\hat\Delta]}^{{\rm R}(1)}
  (k_1,k_2,k_3)\, =\;&
  (ii)_{123} + (iii)_{123} + (iv)_{123} + \hbox{\rm permut. of 12}\,,
  &\numeq\cr
}
$$
where, due to the locality and dimensionality of breaking terms, all the 
permutations should be obviously equal.
}

The renormalized result for each diagram of figure C.3 is:

\def\frac#1#2{{\displaystyle #1}\over{\displaystyle #2}}
\def\frac#1#2{{\displaystyle #1}\over{\displaystyle #2}}
$$ 
 \eqalignno{ 
 (4\pi)^2 \; (i)_{123}=&\;
 {\frac{-8\,{}\,{f^4}\,v}{3}} 
  \,,\cr 
 (4\pi)^2 \; (ii)_{123}=&\;
 -8\,{}\,{f^4}\,v 
  \,,\cr 
 (4\pi)^2 \; (iii)_{123}=&\;
 -8\,{}\,{f^4}\,v 
  \,,\cr 
 (4\pi)^2 \; (iv)_{123}=&\;
 -8\,{}\,{f^4}\,v 
   \,. \; &\lasteqprime\cr 
 } 
 $$

\subsection{C.4 Diagrams with four scalar fields}

\midinsert
{\settabs 4\columns \def\graphwidth{1.8in}   
 \def\graphwidthbis{0.9in}
\eightpoint
\+
         \hfill\hskip 0.09in
         \vbox{\epsfxsize=\graphwidth \epsffile{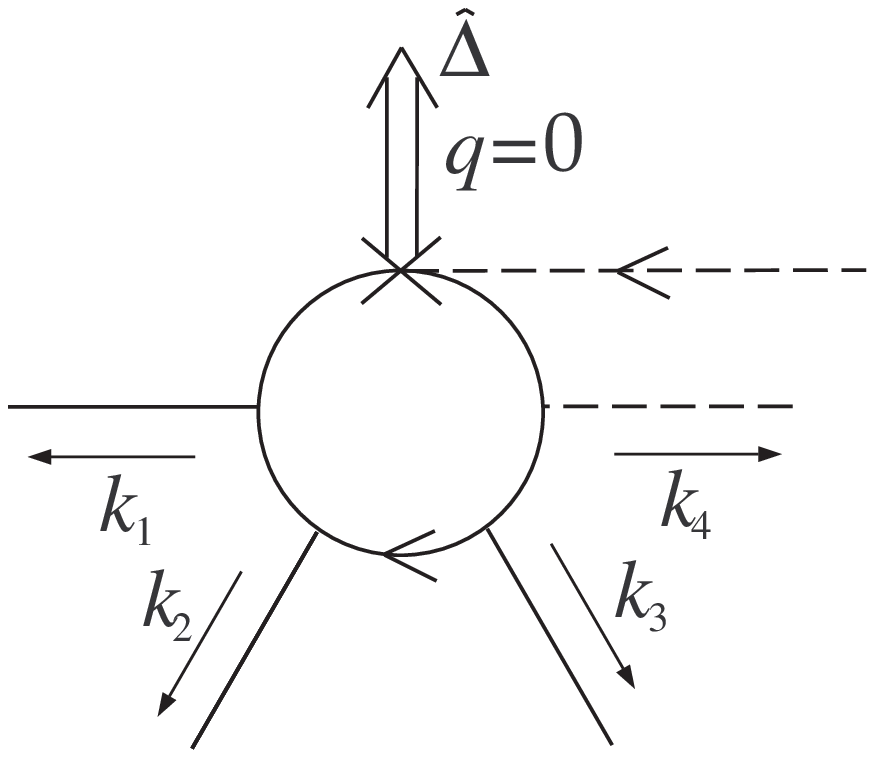}}
         \hfill
	&\hfill\hskip 0.05in
         \vbox{\epsfxsize=\graphwidth \epsffile{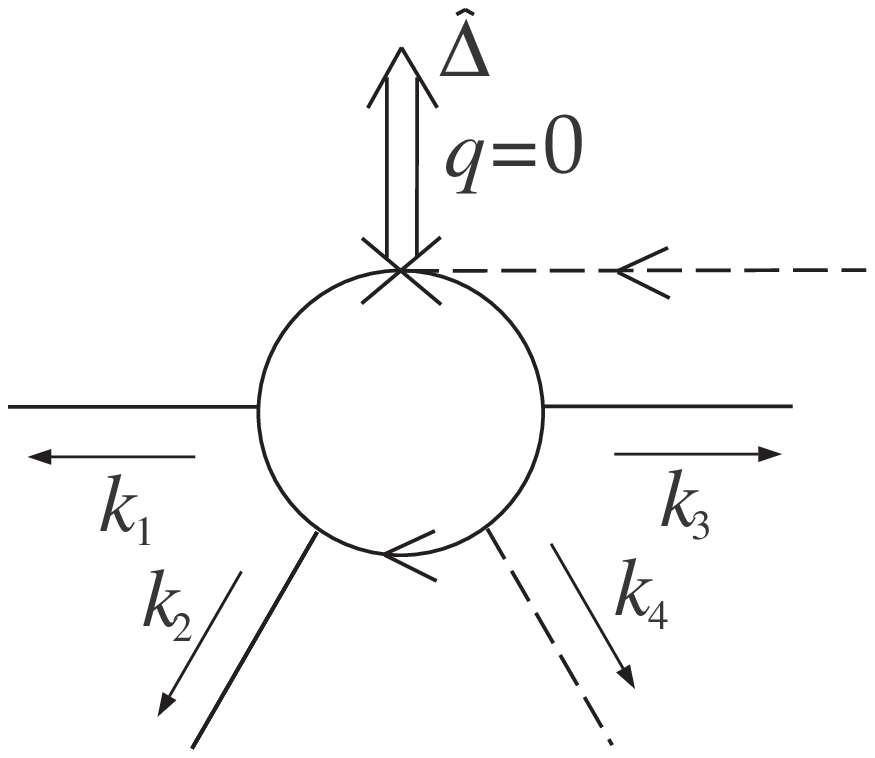}}
         \hfill
	&\hfill\hskip 0.18in
               \epsfxsize=\graphwidth \epsffile{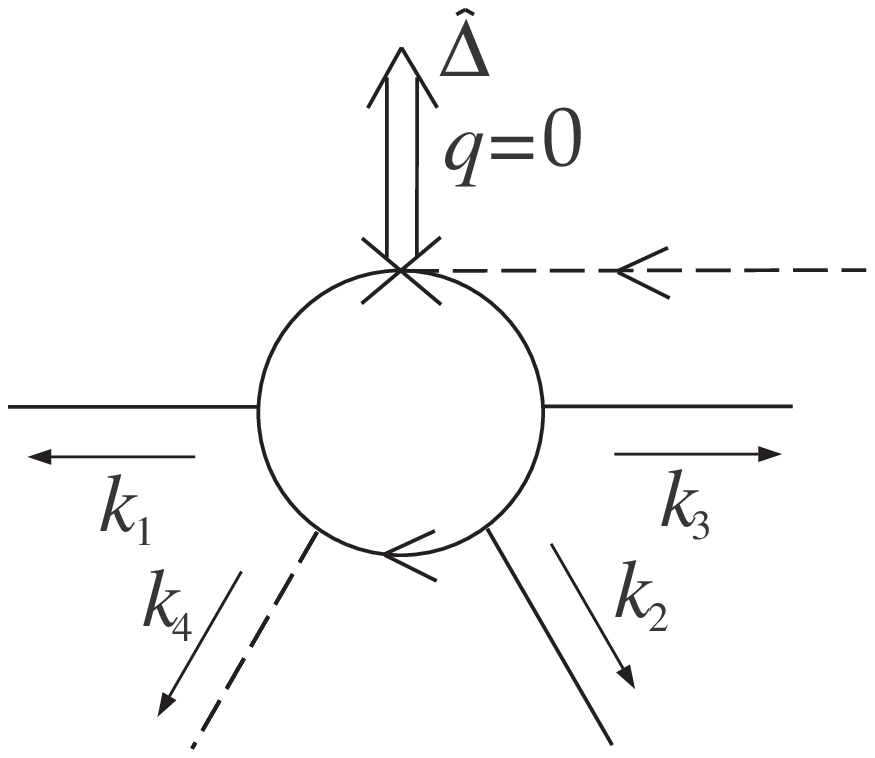}
         \hfill 
	&\hfill\hskip 0.04in
         \vbox{\epsfxsize=\graphwidth \epsffile{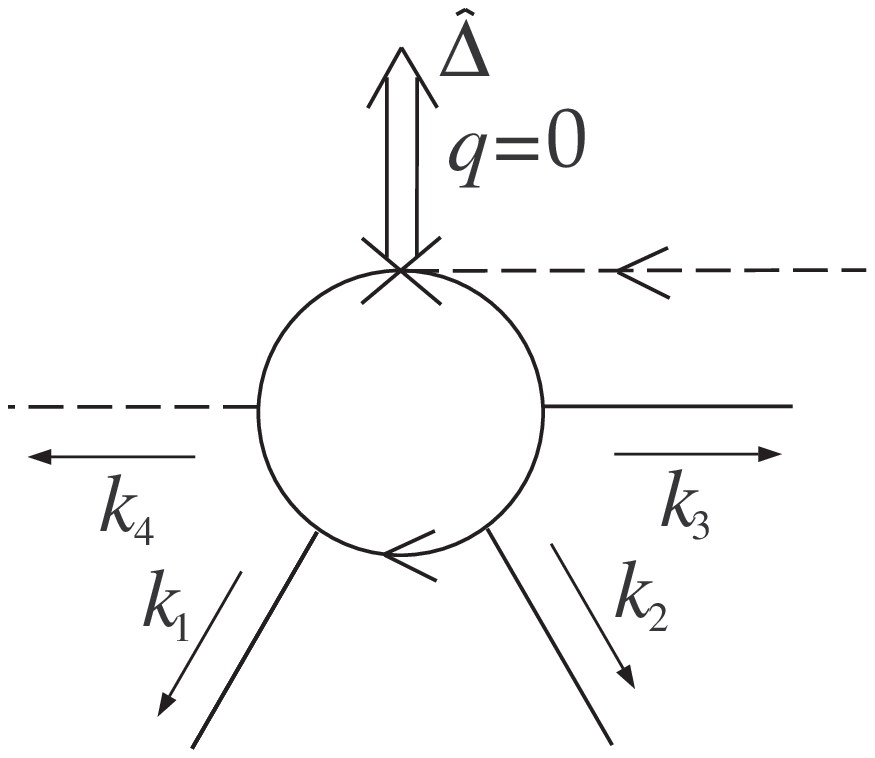}}
         \hskip 0.26in \hfill &\cr
\+\hfill$(i)_{1234}$\hfill
	&\hfill $(ii)_{1234}$\hfill
	&\hfill $(iii)_{1234}$\hfill
	& \hfill $(iv)_{1234}$\hfill &\cr
\+
         \hfill\hskip 0.09in
         \vbox{\epsfxsize=\graphwidth \epsffile{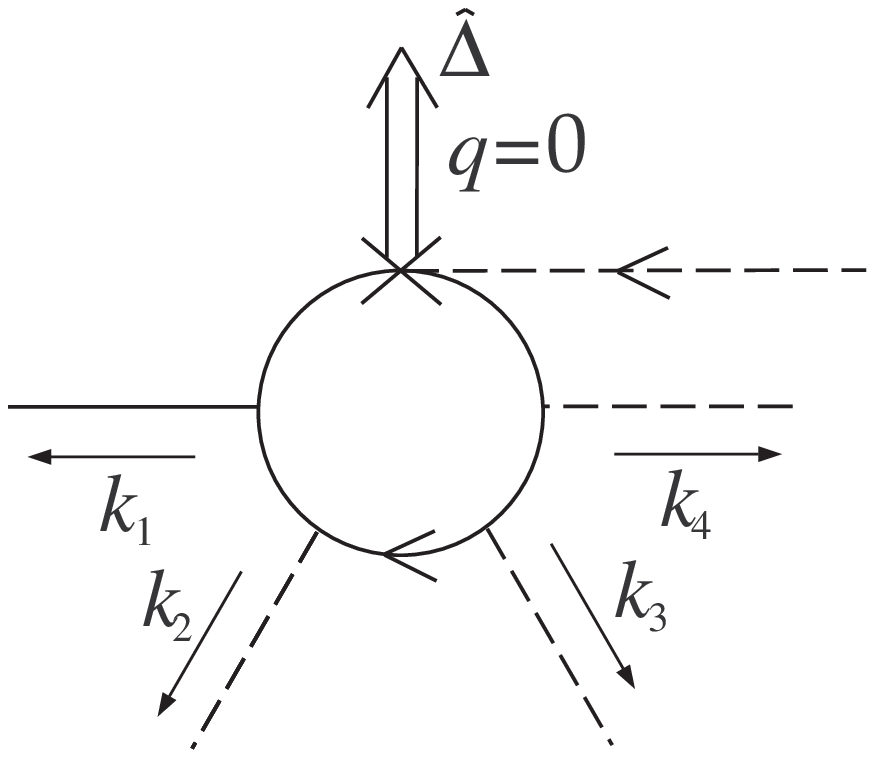}}
         \hfill
	&\hfill\hskip 0.05in
         \vbox{\epsfxsize=\graphwidth \epsffile{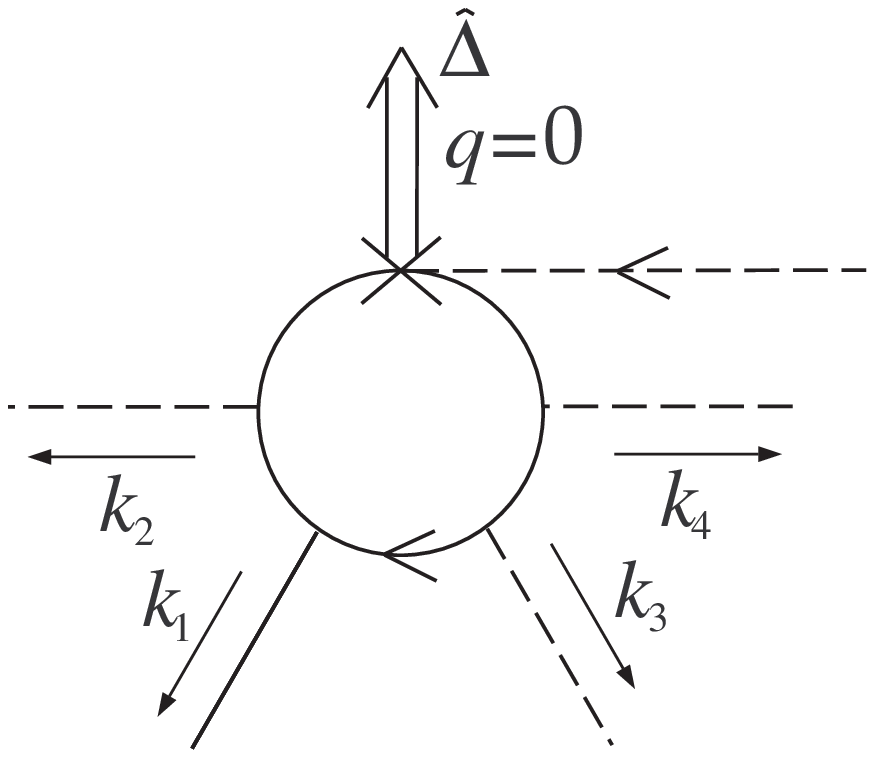}}
         \hfill
	&\hfill\hskip 0.18in
               \epsfxsize=\graphwidth \epsffile{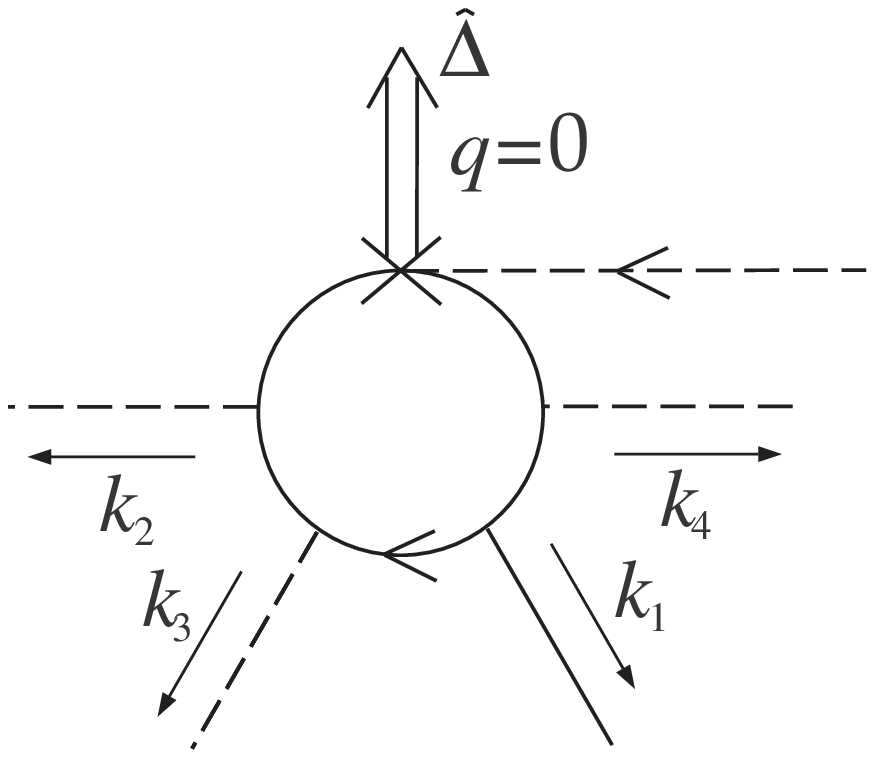}
         \hfill 
	&\hfill\hskip 0.04in
         \vbox{\epsfxsize=\graphwidth \epsffile{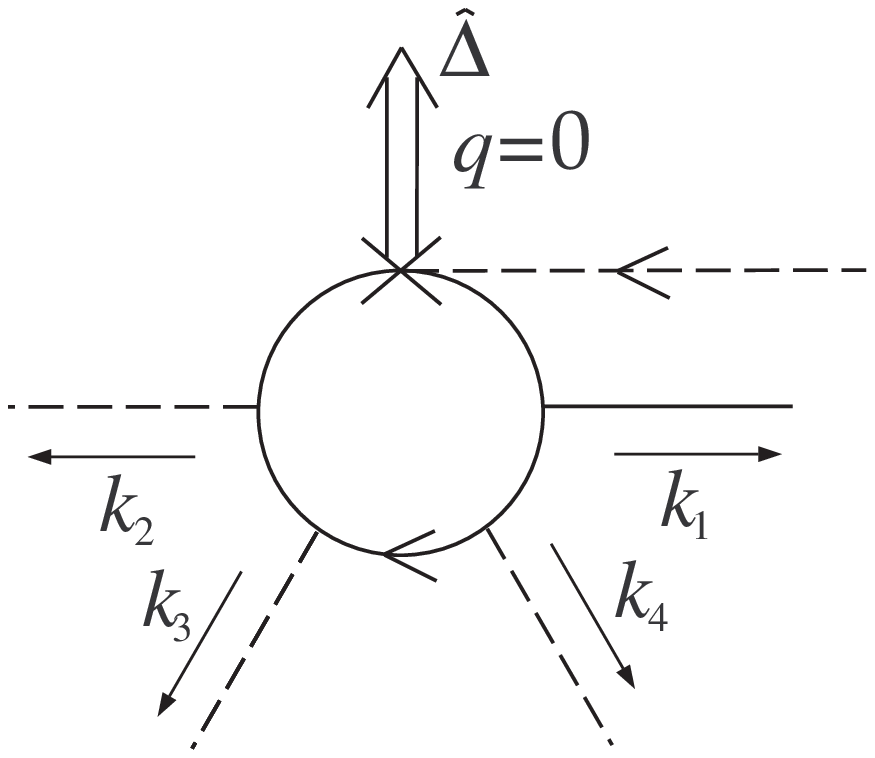}}
         \hskip 0.26in \hfill &\cr
\+\hfill$(v)_{1234}$\hfill
	&\hfill $(vi)_{1234}$\hfill
	&\hfill $(vii)_{1234}$\hfill
	& \hfill $(viii)_{1234}$\hfill &\cr
}

\vskip 12pt
\narrower\noindent {\bf Figure C.4:}
{\eightrm  Feynman diagrams whith four scalar fields
needed to compute the 1PI breaking functions (C.4)}
%
%
\endinsert

{
$$
\eqalignno{\tilde\Gamma_{{\phi_1}{\phi_1}{\phi_1}\phi_2 c;N[\hat\Delta]}^{{\rm R}(1) }
  (k_1,k_2,k_3,k_4)\,
 =\;&(i)_{1234}+(ii)_{1234}+(iii)_{1234} + (iv)_{1234}
    + \hbox{\rm permut. of 123}\,,\cr
\tilde\Gamma_{{\phi_1}{\phi_2}{\phi_2}{\phi_2} c;N[\hat\Delta]}^{{\rm R}(1) }
  (k_1,k_2,k_3,k_4)\, =\;&
  (vi)_{1234} + (vii)_{1234} + (viii)_{1234} + \hbox{\rm permut. of 234}\,,
  &\numeq\cr
}
$$
where, again, all the permutations must be the same.
}

The renormalized result for each diagram of figure C.4 is:

\def\frac#1#2{{\displaystyle #1}\over{\displaystyle #2}}
\def\frac#1#2{{\displaystyle #1}\over{\displaystyle #2}}
$$ 
 \eqalignno{ 
 (4\pi)^2 \; (i)_{1234}=&\;
 -2\,{}\,{f^4} 
  \,,\cr 
 (4\pi)^2 \; (ii)_{1234}=&\;
 -2\,{}\,{f^4} 
  \,,\cr 
 (4\pi)^2 \; (iii)_{1234}=&\;
 -2\,{}\,{f^4} 
  \,,\cr 
 (4\pi)^2 \; (iv)_{1234}=&\;
 -2\,{}\,{f^4} 
  \,,\cr 
 (4\pi)^2 \; (v)_{1234}=&\;
 {\frac{-10\,{}\,{f^4}}{3}} 
  \,,\cr 
 (4\pi)^2 \; (vi)_{1234}=&\;
 2\,{}\,{f^4} 
  \,,\cr 
 (4\pi)^2 \; (vii)_{1234}=&\;
 2\,{}\,{f^4} 
  \,,\cr 
 (4\pi)^2 \; (viii)_{1234}=&\;
 {\frac{-10\,{}\,{f^4}}{3}} 
   \,. \; &\lasteqprime\cr 
 } 
 $$

\subsection{C.5 Diagrams with two or three boson and scalar fields}

\midinsert
{\settabs 4\columns \def\graphwidth{1.8in}   
 \def\graphwidthbis{1.7in}
\eightpoint
\+
         \hfill\hskip 0.13in
         \vbox{\epsfxsize=\graphwidthbis \epsffile{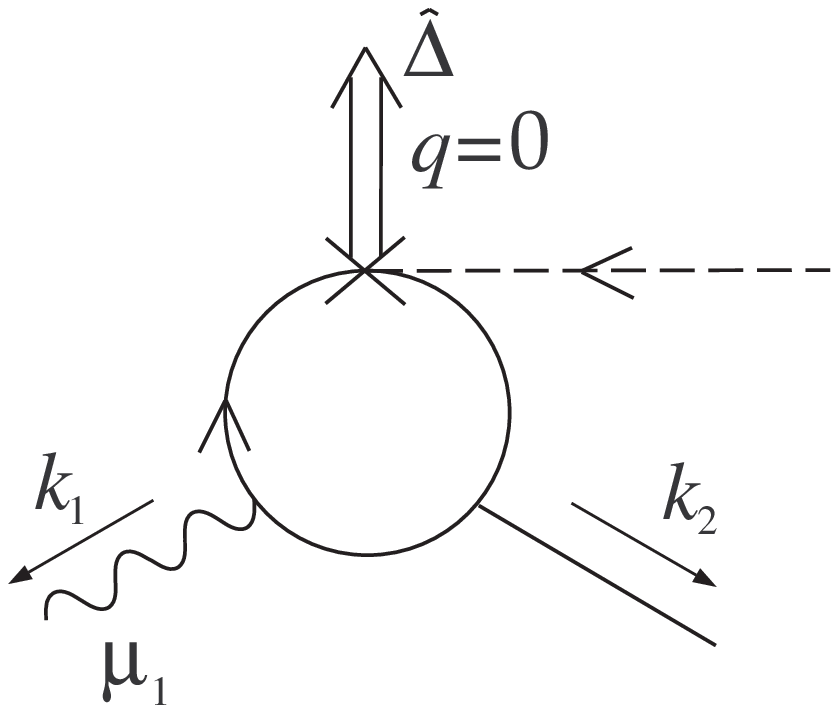}\vskip 0.22in}
         \hfill
	&\hfill\hskip 0.12in
         \vbox{\epsfxsize=\graphwidthbis \epsffile{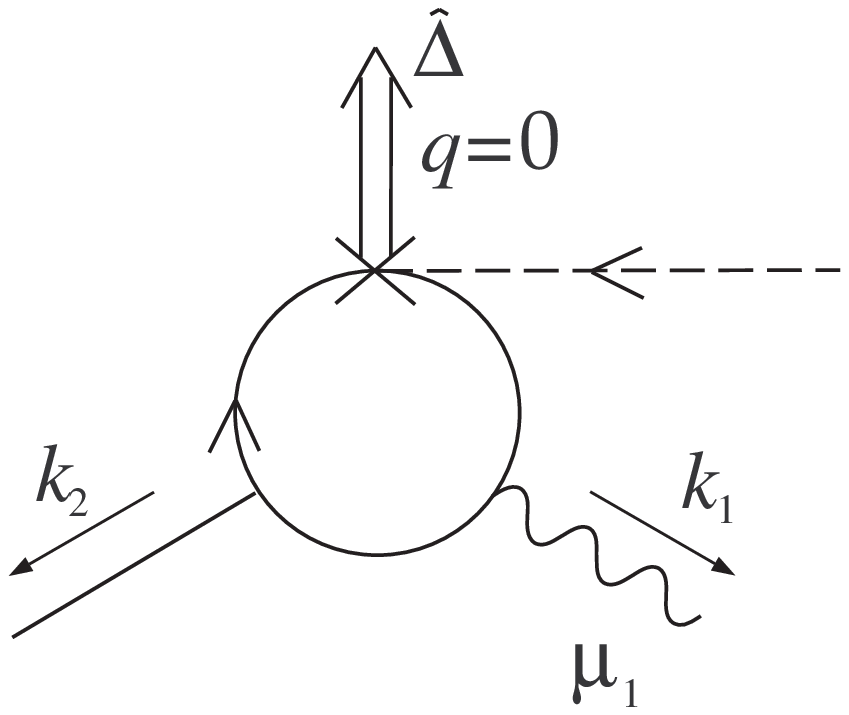}\vskip 0.22in}
         \hfill
	&\hfill
         \vbox{\epsfxsize=\graphwidth \epsffile{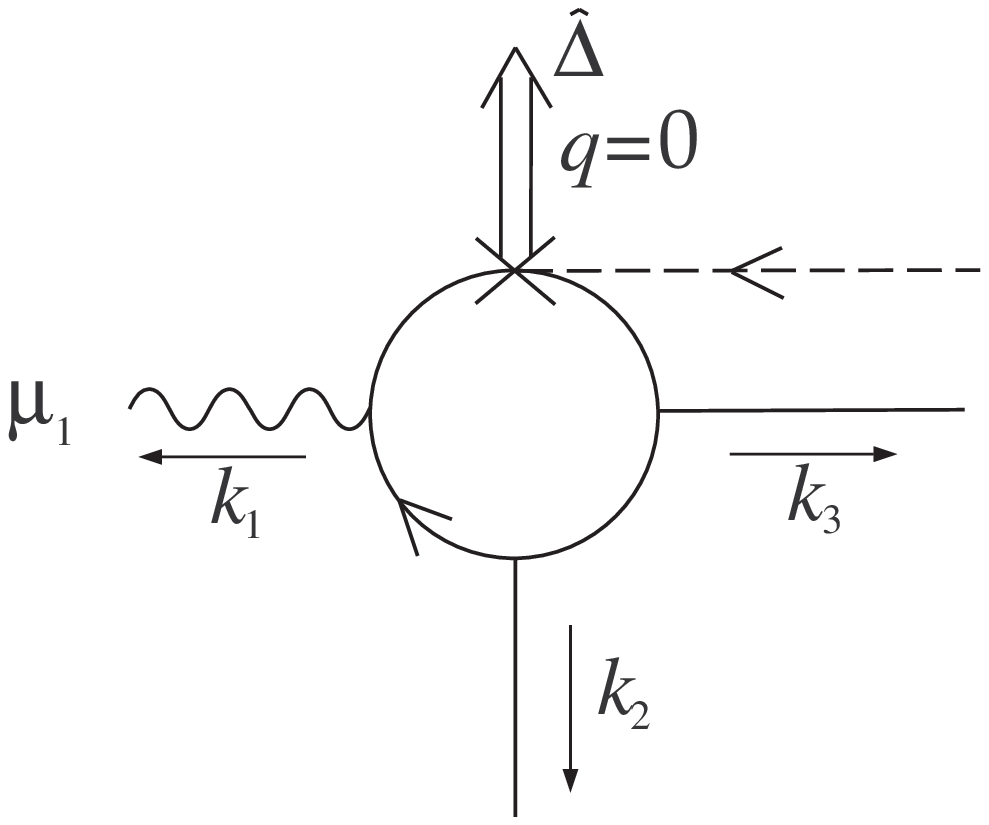}\vskip 0.13in}
         \hskip 0.06in \hfill
	&\hfill
         \vbox{\epsfxsize=\graphwidth \epsffile{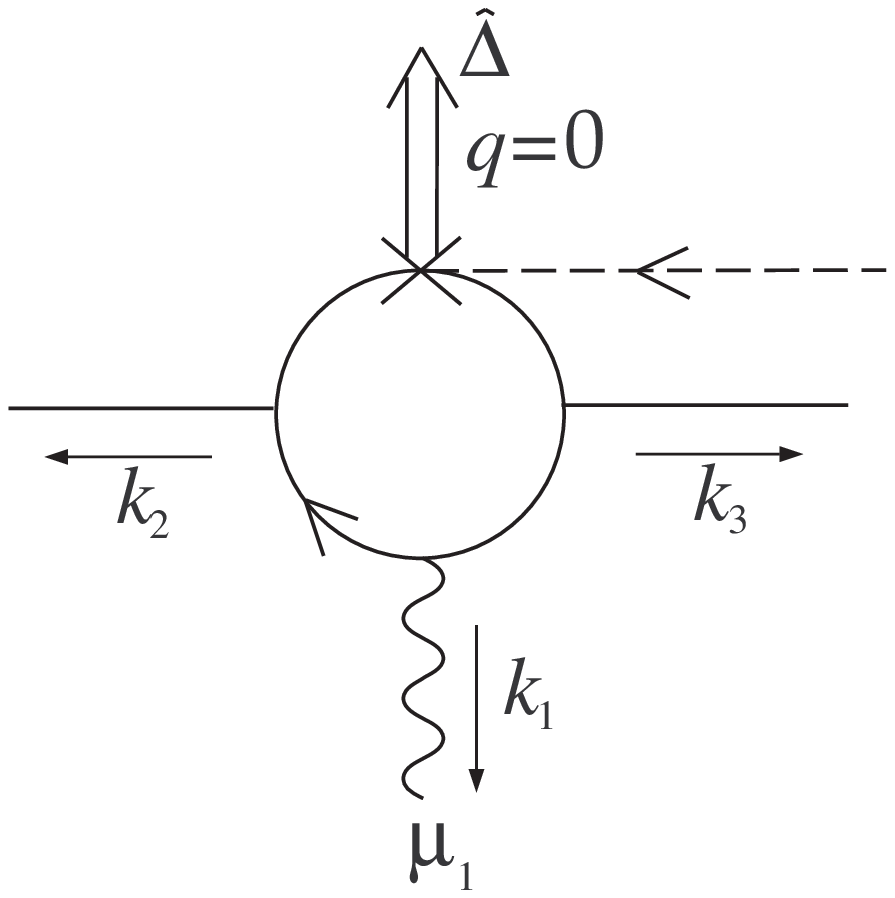}}
         \hskip 0.26in \hfill &\cr
\+\hfill$(i)$\hfill
	&\hfill $(ii)$\hfill
	&\hfill $(iii)_{123}$\hfill
	& \hfill $(iv)_{123}$\hfill &\cr
\+
         \hfill\hskip 0.03in
         \vbox{\epsfxsize=\graphwidth \epsffile{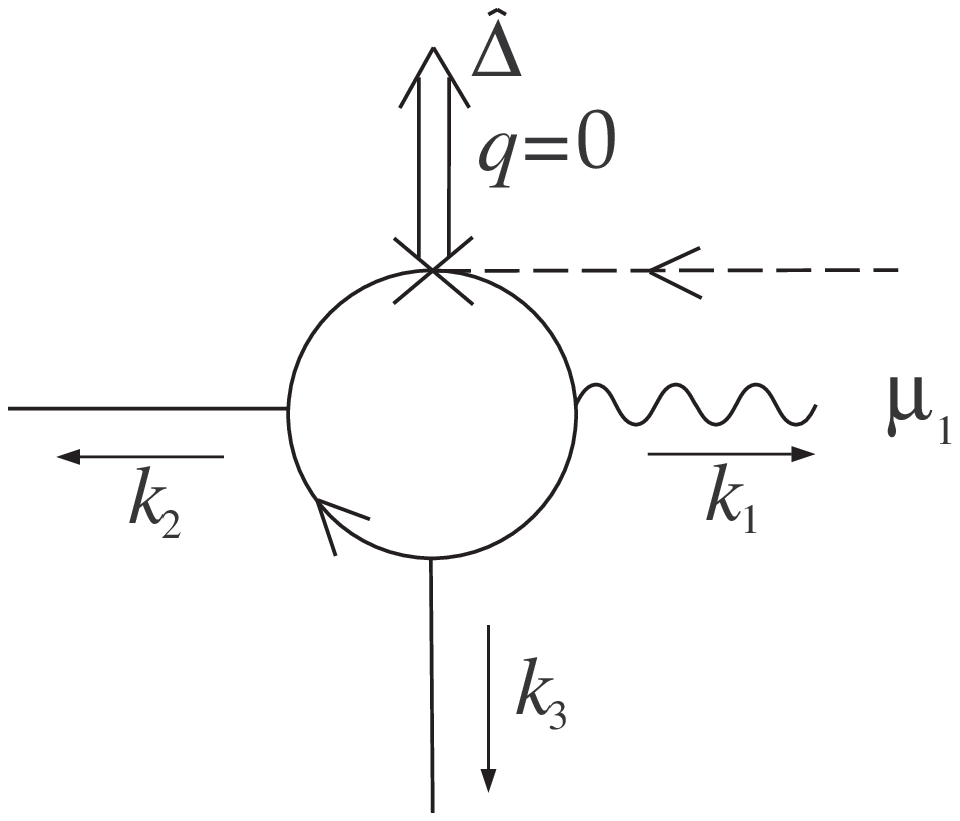}\vskip 0.11in}
         \hfill
	&\hfill
         \vbox{\epsfxsize=\graphwidth \epsffile{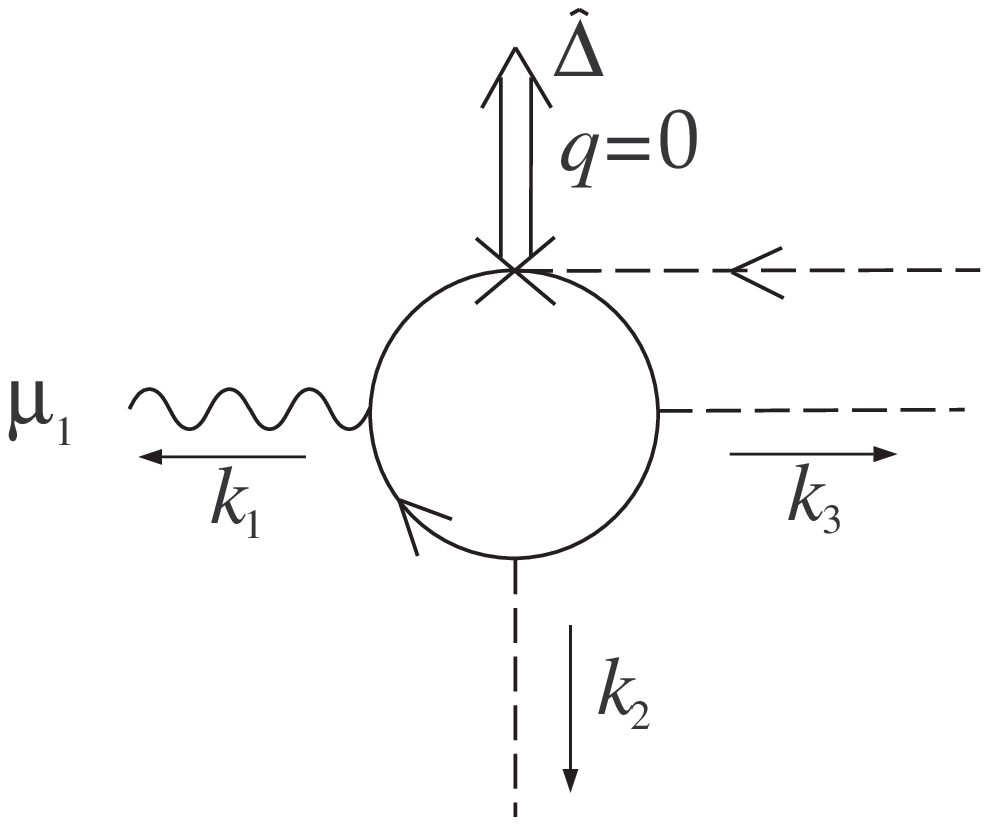}\vskip 0.11in}
         \hfill
	&\hfill\hskip 0.13in
               \epsfxsize=\graphwidth \epsffile{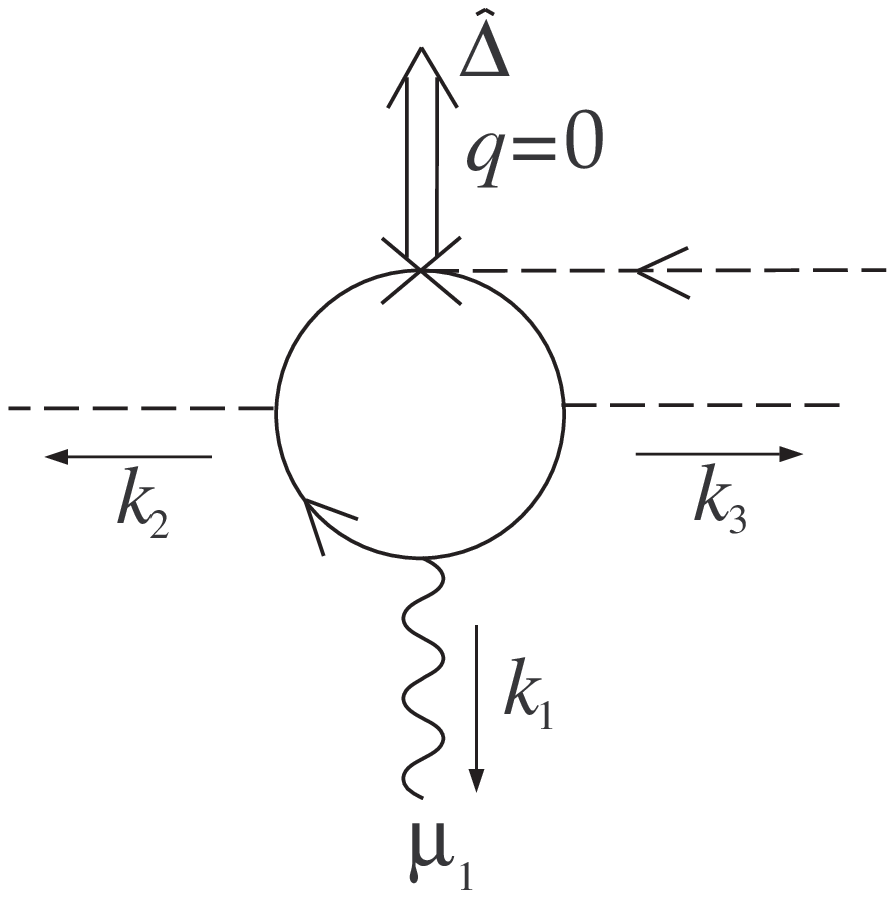}\hfill 
	&\hfill
         \vbox{\epsfxsize=\graphwidth \epsffile{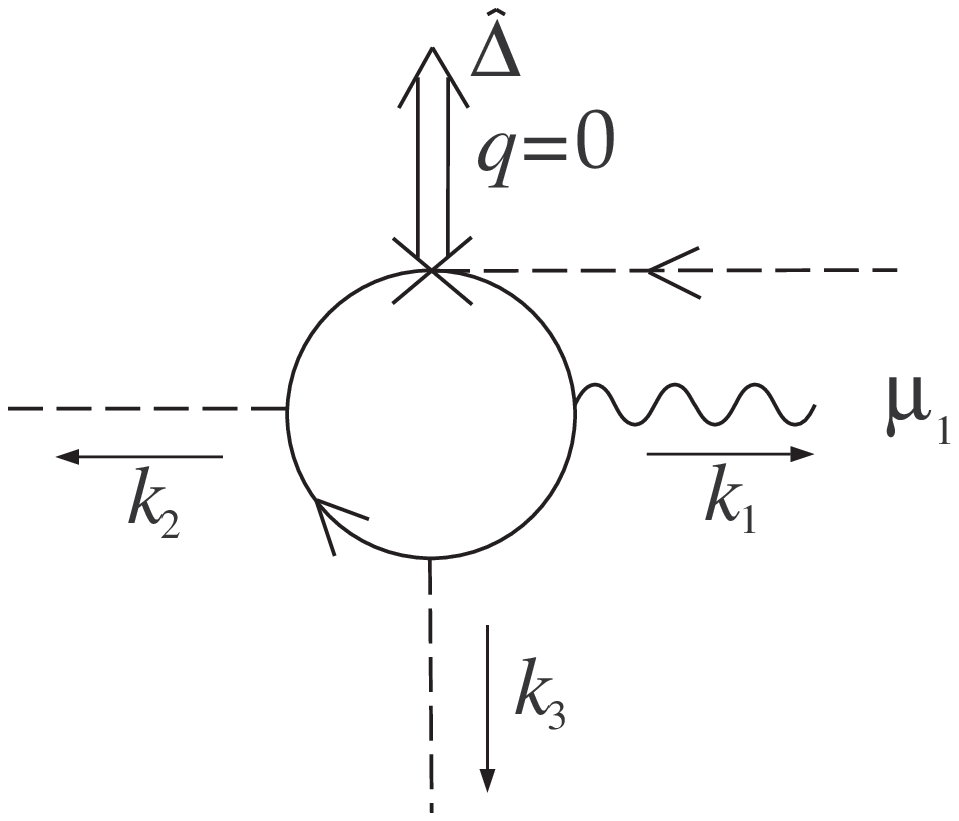}\vskip 0.10in}
         \hskip 0.26in \hfill &\cr
\+\hfill$(v)_{123}$\hfill
	&\hfill $(vi)_{123}$\hfill
	&\hfill $(vii)_{123}$\hfill
	& \hfill $(viii)_{123}$\hfill &\cr
\settabs 3\columns \def\graphwidth{1.8in} 
\+
         \hfill\hskip 0.26in
         \vbox{\epsfxsize=\graphwidth \epsffile{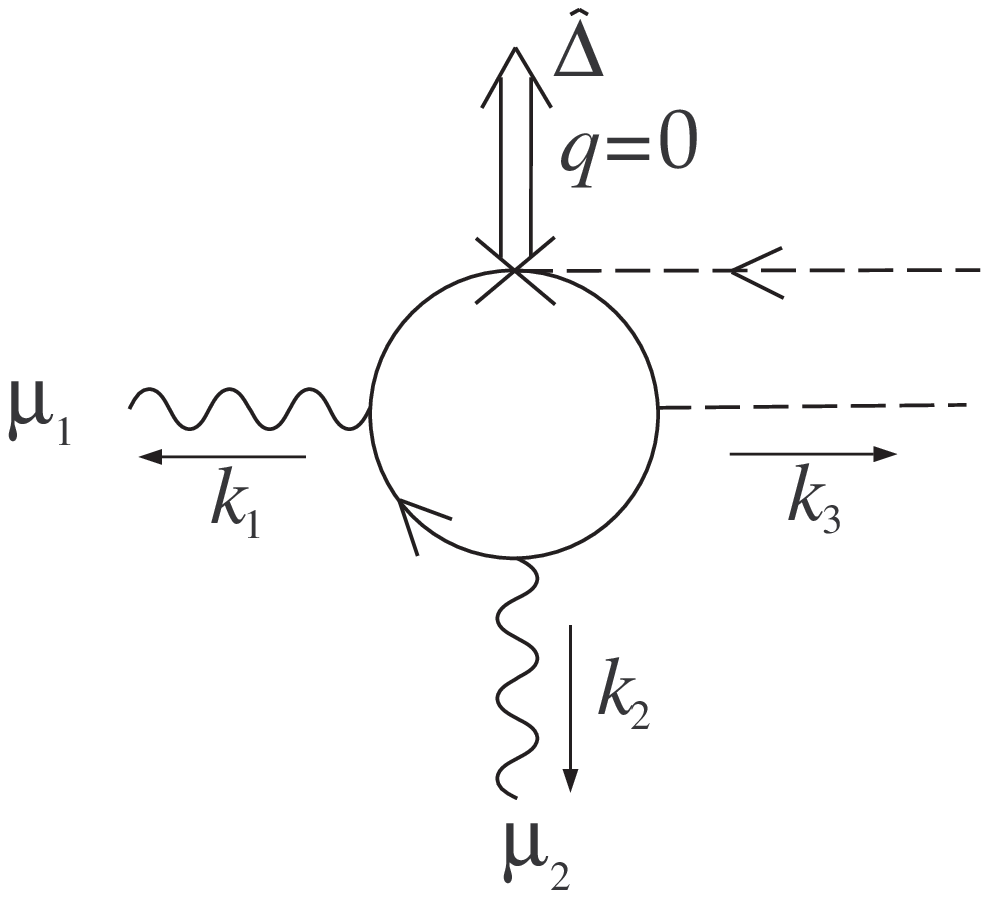}}
         \hfill
	&\hfill\hskip 0.24 in
         \vbox{\epsfxsize=\graphwidth \epsffile{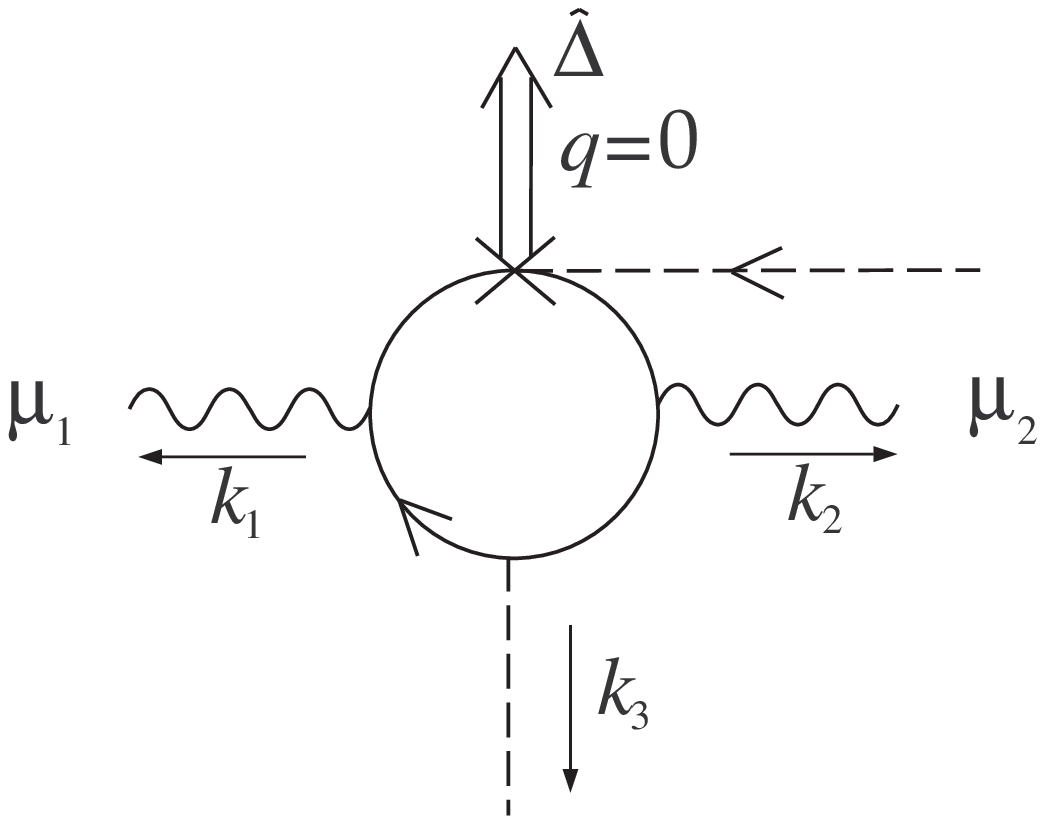}\vskip 0.08in}
         \hfill
	&\hfill
         \vbox{\epsfxsize=\graphwidth \epsffile{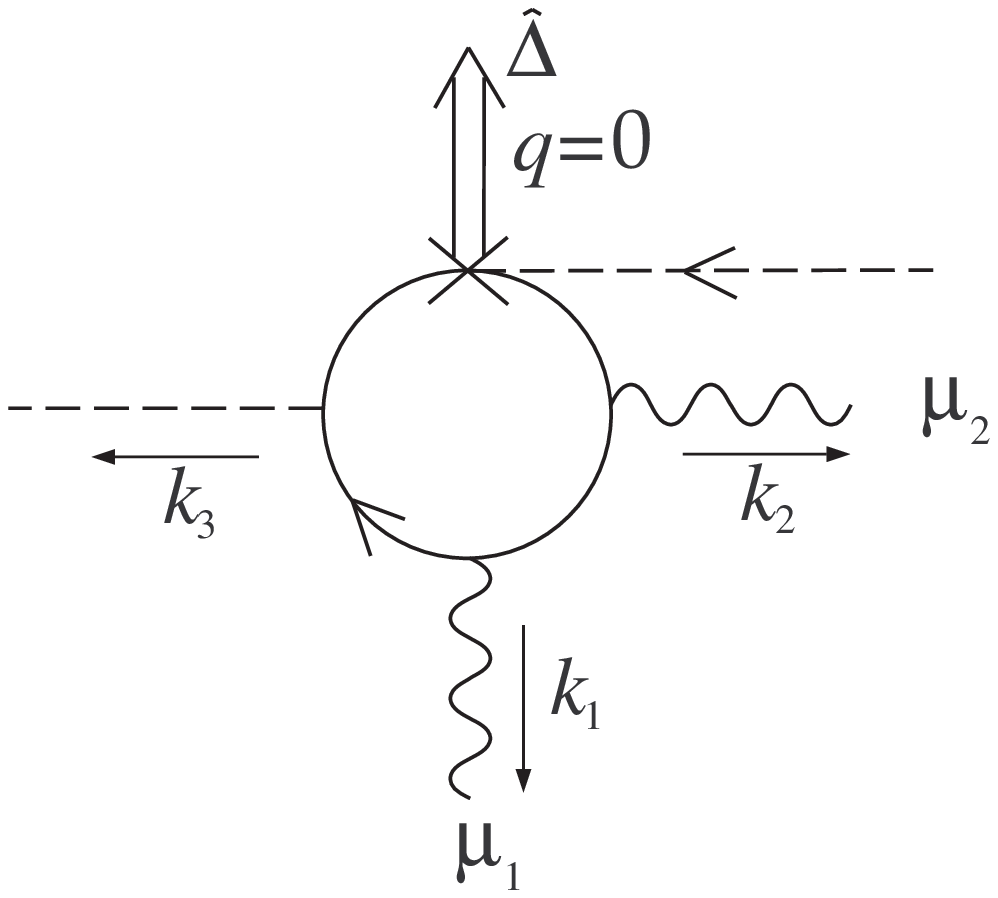}}
         \hskip 0.26in \hfill &\cr
\+\hfill\hskip 0.2in $(ix)_{123}$\hfill
	&\hfill $(x)_{123}$\hfill
	& \hfill $(xi)_{123}$\hskip 0.5in \hfill &\cr
}

\vskip 12pt
\narrower\noindent {\bf Figure C.5:}
{\eightrm  Feynman diagrams whith two or three boson and scalar fields
needed to compute the 1PI breaking functions (C.5)}
%
%
\endinsert

{
$$
\eqalignno{\tilde\Gamma_{A{\phi_1} c;N[\hat\Delta]}^{{\rm R}(1)\;\mu_1 }
  (k_1,k_2)\,
 =\;&(i)+(ii)\,,\cr
\tilde\Gamma_{A{\phi_1}{\phi_1} c;N[\hat\Delta]}^{{\rm R}(1)\;\mu_1 }
  (k_1,k_2,k_3)\, =\;&
  (iii)_{123} + (iv)_{123} + (v)_{123} + \hbox{\rm permut. of 23}\,,\cr
\tilde\Gamma_{A{\phi_2}{\phi_2} c;N[\hat\Delta]}^{{\rm R}(1)\;\mu_1 }
  (k_1,k_2,k_3)\, =\;&
  (vi)_{123} + (vii)_{123} + (viii)_{123} + \hbox{\rm permut. of 23}\,,\cr
\tilde\Gamma_{AA{\phi_2} c;N[\hat\Delta]}^{{\rm R}(1)\;\mu_1\mu_2 }
  (k_1,k_2,k_3)\, =\;&
  (ix)_{123} + (x)_{123} + (xi)_{123} + \hbox{\rm permut. of 12}\,.
  &\numeq\cr
}
$$
}

The renormalized result for each diagram of figure C.5 is:

\def\frac#1#2{{\displaystyle #1}\over{\displaystyle #2}}
\def\frac#1#2{{\displaystyle #1}\over{\displaystyle #2}}
$$ 
 \eqalignno{ 
 (4\pi)^2 \; (i)=&\;
 -2\,i\,{}\,{f^2}\,v\,{k_1}^{{\mu_1}} 
  \,,\cr 
 (4\pi)^2 \; (ii)=&\;
 -2\,i\,{}\,{f^2}\,v\,{k_1}^{{\mu_1}} 
  \,,\cr 
 (4\pi)^2 \; (iii)_{123}=&\;
 -i\,{}\,{f^2}\,\left( {k_1}^{{\mu_1}} +  
     {k_3}^{{\mu_1}} \right)  
  \,,\cr 
 (4\pi)^2 \; (iv)_{123}=&\;
 i\,{}\,{f^2}\,\left( {k_2}^{{\mu_1}} +  
     {k_3}^{{\mu_1}} \right)  
  \,,\cr 
 (4\pi)^2 \; (v)_{123}=&\;
 -i\,{}\,{f^2}\,\left( {k_1}^{{\mu_1}} +  
     {k_2}^{{\mu_1}} \right)  
  \,,\cr 
 (4\pi)^2 \; (vi)_{123}=&\;
 {\frac{-i}{3}}\,{}\,{f^2}\, 
   \left( 3\,{k_1}^{{\mu_1}} +  
     4\,{k_2}^{{\mu_1}} +  
     5\,{k_3}^{{\mu_1}} \right)  
  \,,\cr 
 (4\pi)^2 \; (vii)_{123}=&\;
 -i\,{}\,{f^2}\,\left( {k_2}^{{\mu_1}} +  
     {k_3}^{{\mu_1}} \right)  
  \,,\cr 
 (4\pi)^2 \; (viii)_{123}=&\;
 {\frac{-i}{3}}\,{}\,{f^2}\, 
   \left( 3\,{k_1}^{{\mu_1}} +  
     5\,{k_2}^{{\mu_1}} +  
     4\,{k_3}^{{\mu_1}} \right)  
  \,,\cr 
 (4\pi)^2 \; (ix)_{123}=&\;
 {\frac{4\,{}\,{f^2}\, 
      \left( 1 + {{{\theta}}^2} +  
        {\theta}\,{r} \right) \,v\, 
      g^{{\mu_1} {\mu_2}}}{3}} 
  \,,\cr 
 (4\pi)^2 \; (x)_{123}=&\;
 {\frac{-4\,{}\,{f^2}\, 
      \left( -1 + 2\,{{{\theta}}^2} +  
        2\,{\theta}\,{r} \right) \,v\, 
      g^{{\mu_1} {\mu_2}}}{3}} 
  \,,\cr 
 (4\pi)^2 \; (xi)_{123}=&\;
 {\frac{4\,{}\,{f^2}\, 
      \left( 1 + {{{\theta}}^2} +  
        {\theta}\,{r} \right) \,v\, 
      g^{{\mu_1} {\mu_2}}}{3}} 
   \,. \; &\lasteqprime\cr 
 } 
 $$

\subsection{C.6 Diagrams with four boson and scalar fields}

\midinsert
{\settabs 4\columns \def\graphwidth{1.8in}   
 \def\graphwidthbis{0.9in}
\eightpoint
\+
         \hfill
         \vbox{\epsfxsize=\graphwidth \epsffile{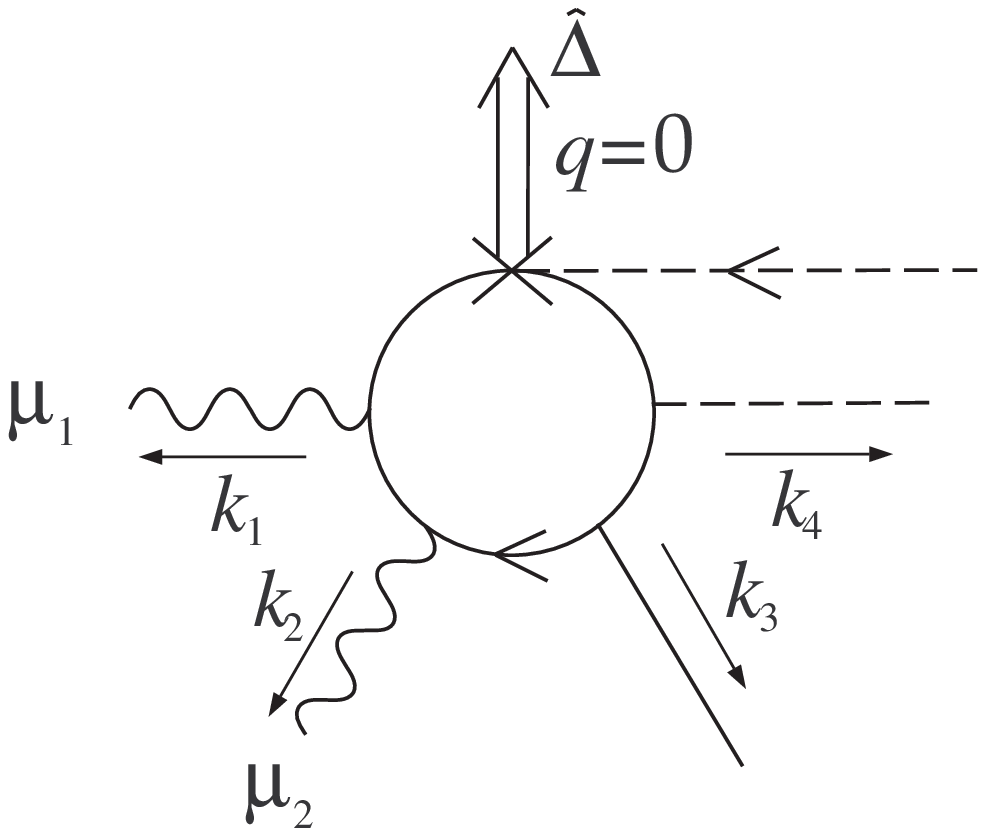}}
         \hskip 0.12in \hfill
	&\hfill
         \vbox{\epsfxsize=\graphwidth \epsffile{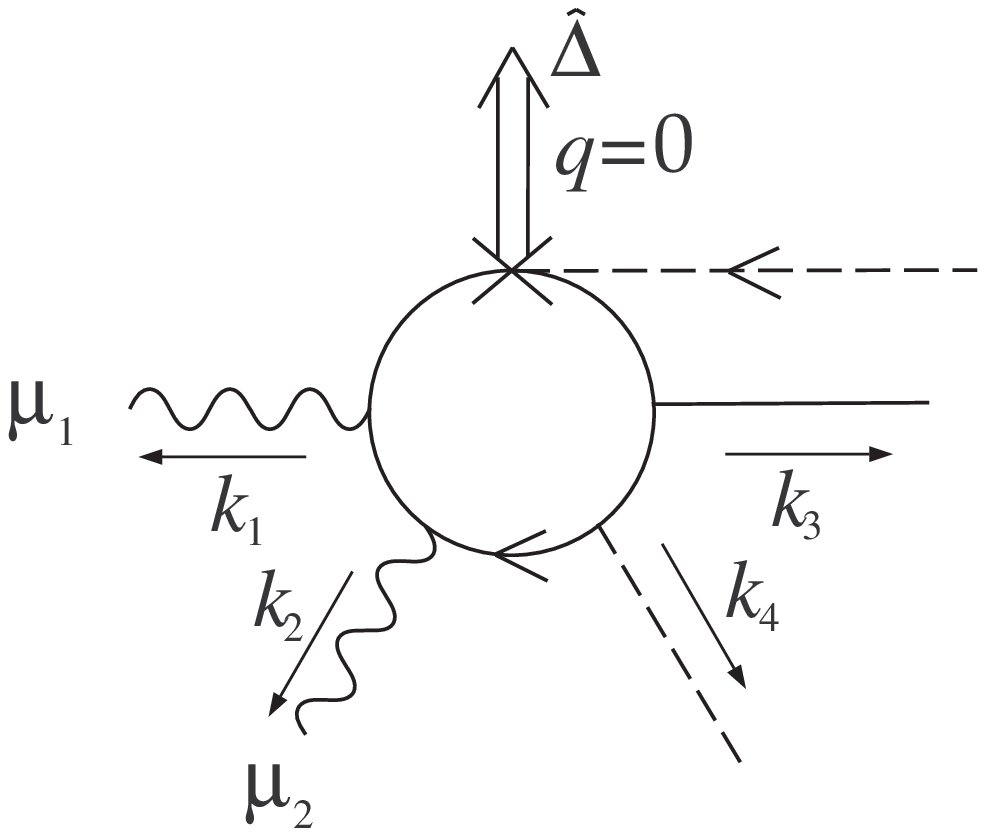}}
         \hskip 0.04in\hfill
	&\hfill
         \vbox{\epsfxsize=\graphwidth \epsffile{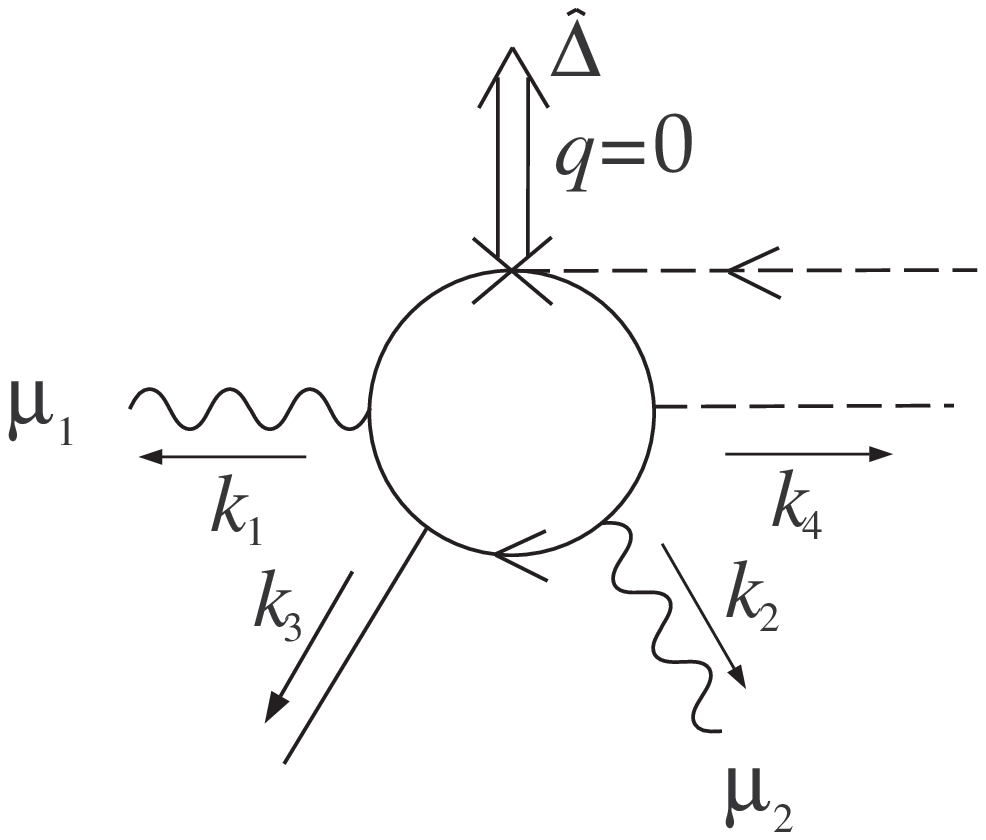}
         \hskip 0.18in} \hfill 
	&\hfill
         \vbox{\epsfxsize=\graphwidth \epsffile{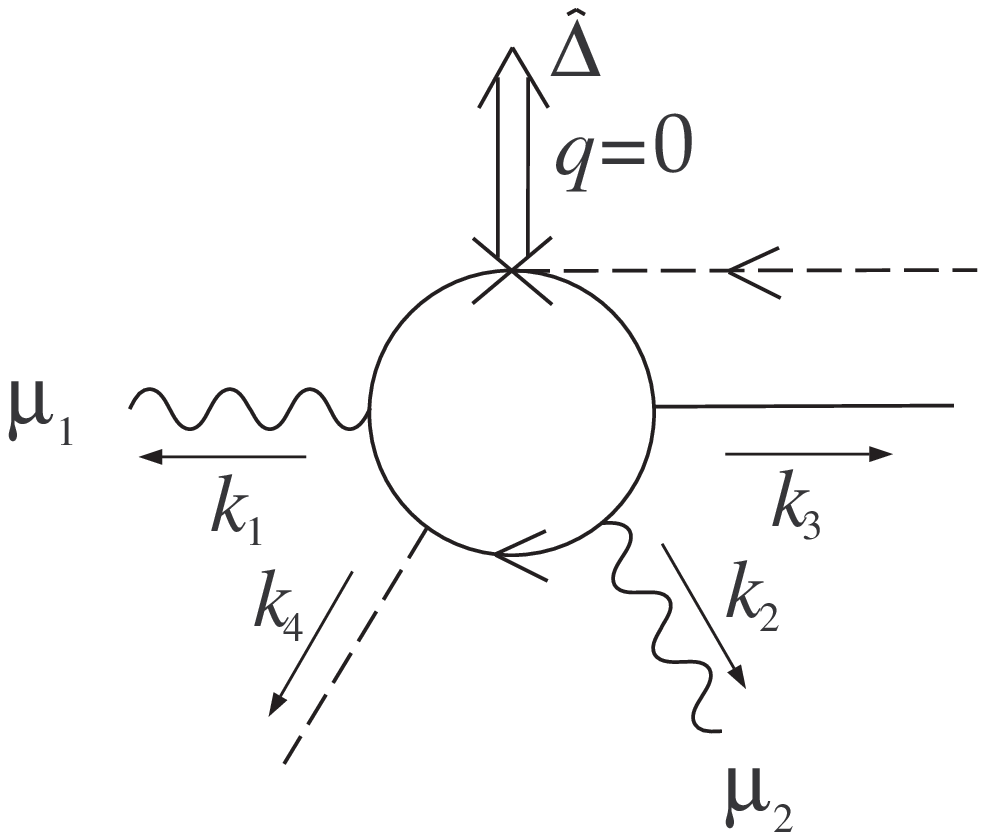}
               \hskip 0.24in}
         \hfill &\cr
\+       \hfill$(i)_{1234}$\hfill
	&\hfill $(ii)_{1234}$\hfill
	&\hfill $(iii)_{1234}$\hfill
	& \hfill $(iv)_{1234}$\hfill &\cr
\+
         \hfill
         \vbox{\epsfxsize=\graphwidth \epsffile{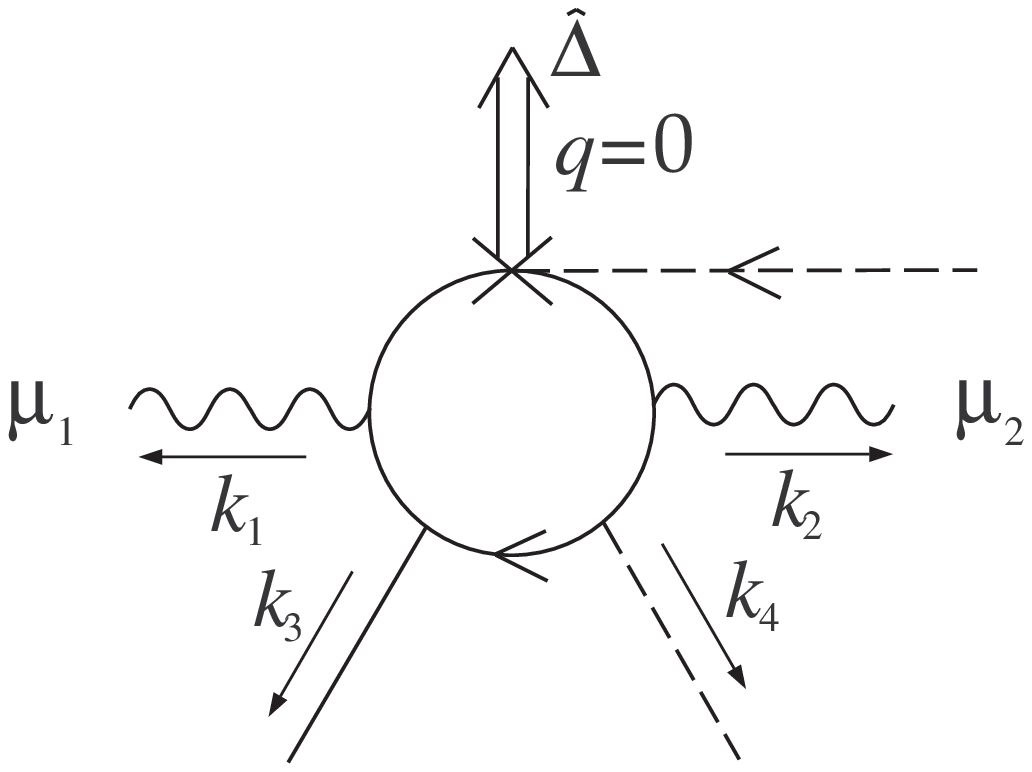}}
         \hskip 0.12in \hfill
	&\hfill
         \vbox{\epsfxsize=\graphwidth \epsffile{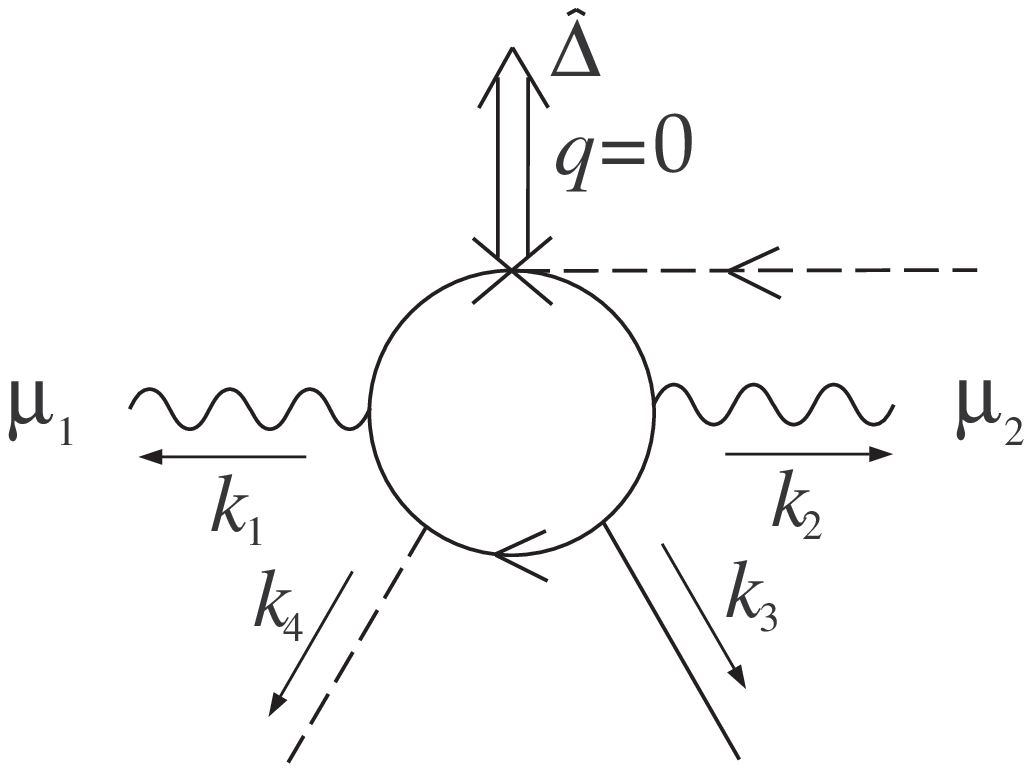}}
         \hskip 0.04in \hfill
	&\hfill\hskip 0.08in
         \vbox{\epsfxsize=\graphwidth \epsffile{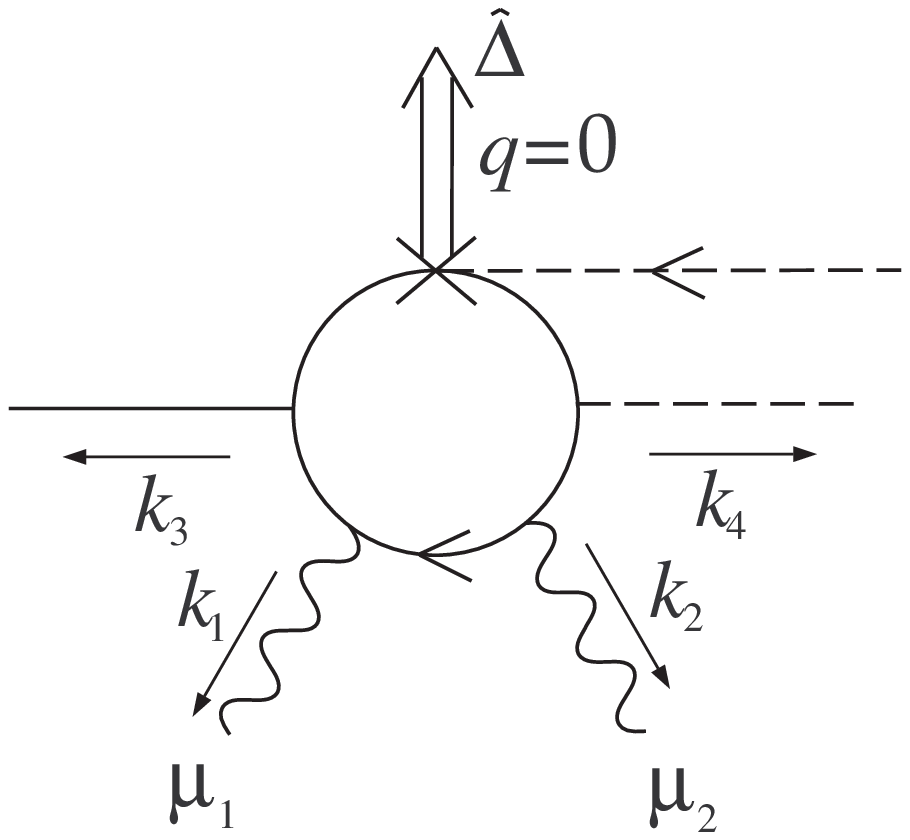}}
         \hfill 
	&\hfill
         \vbox{\epsfxsize=\graphwidth \epsffile{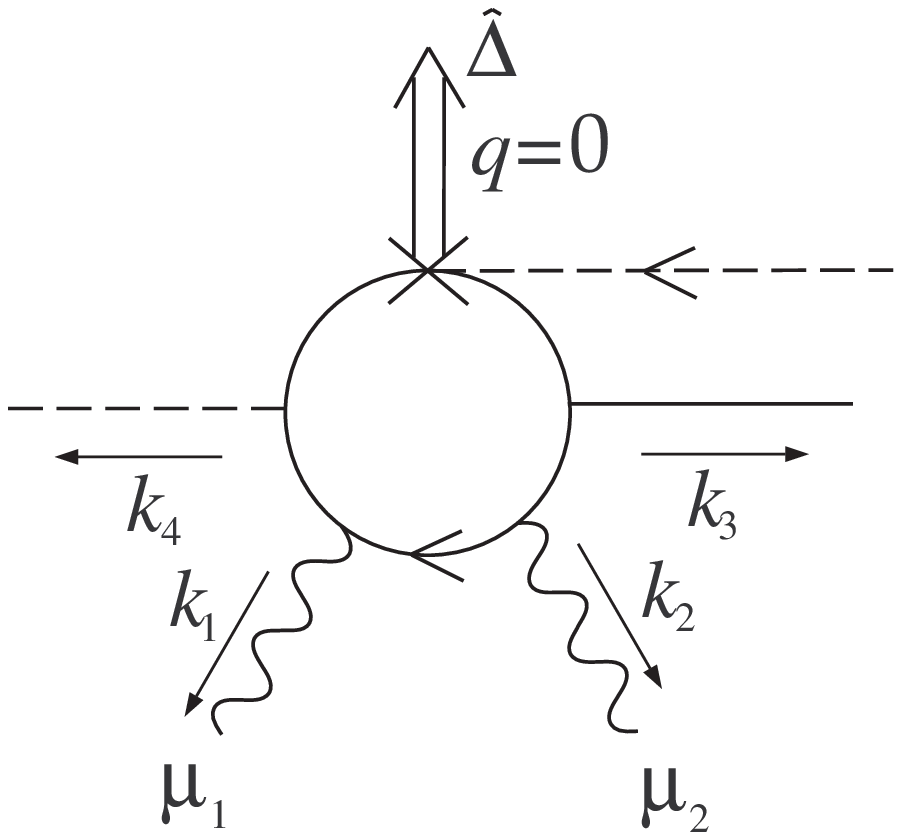}
               \hskip 0.20in }
         \hfill &\cr
\+       \hfill$(v)_{1234}$\hfill
	&\hfill $(vi)_{1234}$\hfill
	&\hfill $(vii)_{1234}$\hfill
	& \hfill $(viii)_{1234}$\hfill &\cr
\+
         \hfill\hskip 0.06in
         \vbox{\epsfxsize=\graphwidth \epsffile{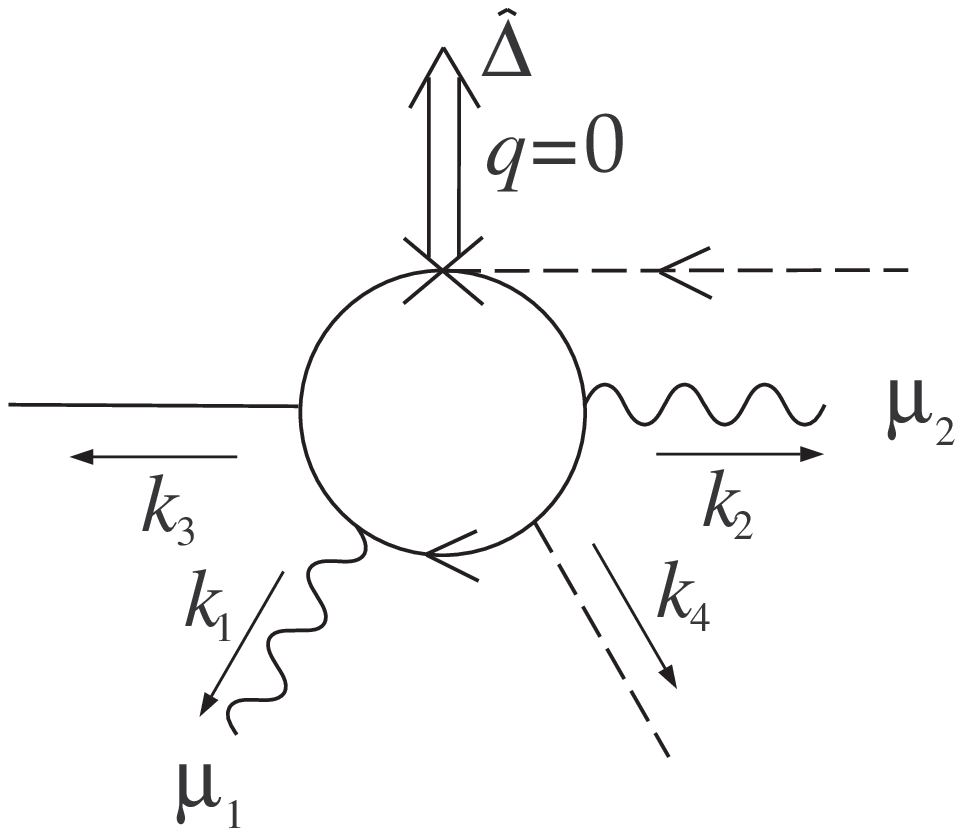}}
         \hfill
	&\hfill\hskip 0.06in
         \vbox{\epsfxsize=\graphwidth \epsffile{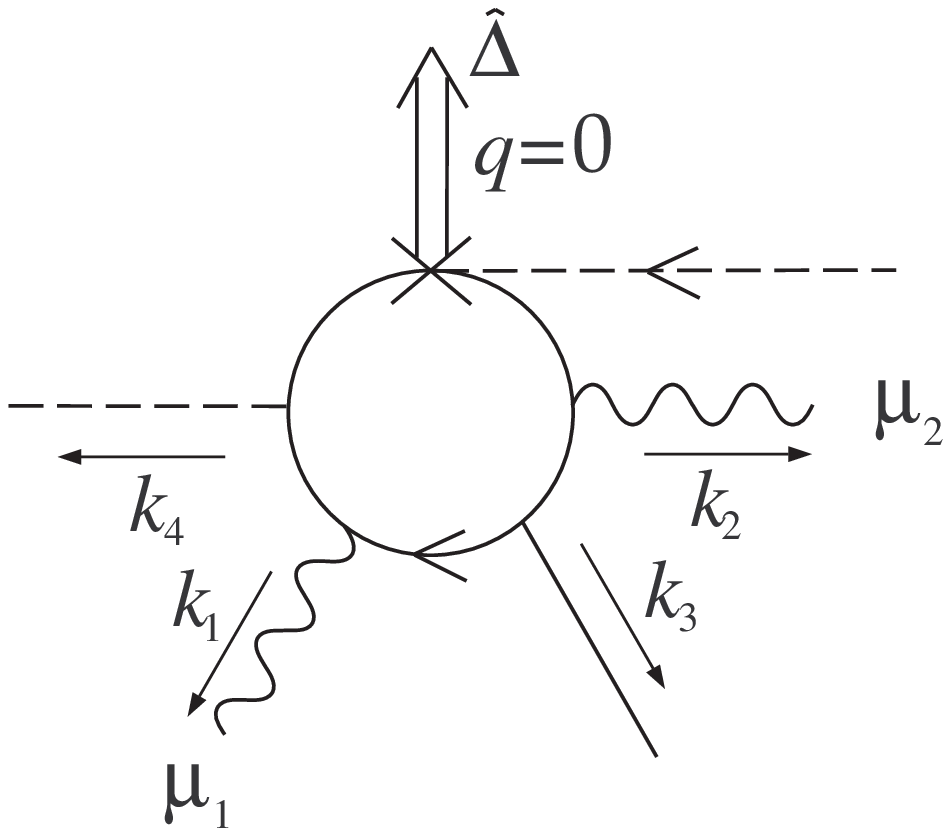}}
         \hfill
	&\hfill
         \vbox{\epsfxsize=\graphwidth \epsffile{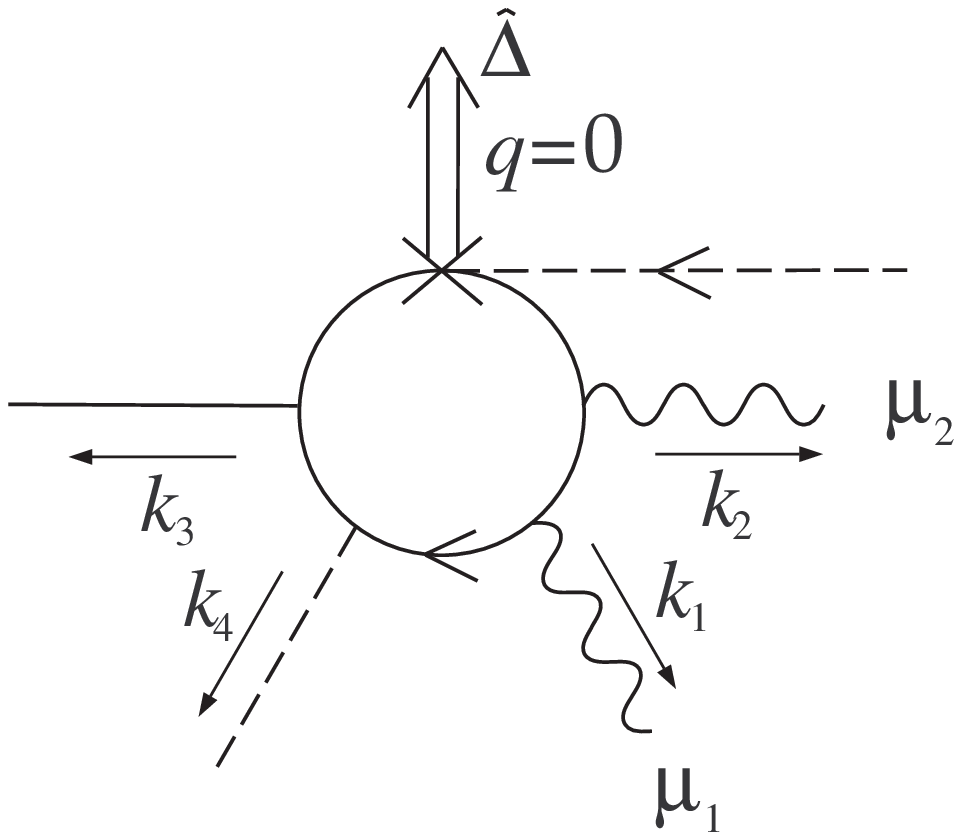}}
         \hfill 
	&\hfill
         \vbox{\epsfxsize=\graphwidth \epsffile{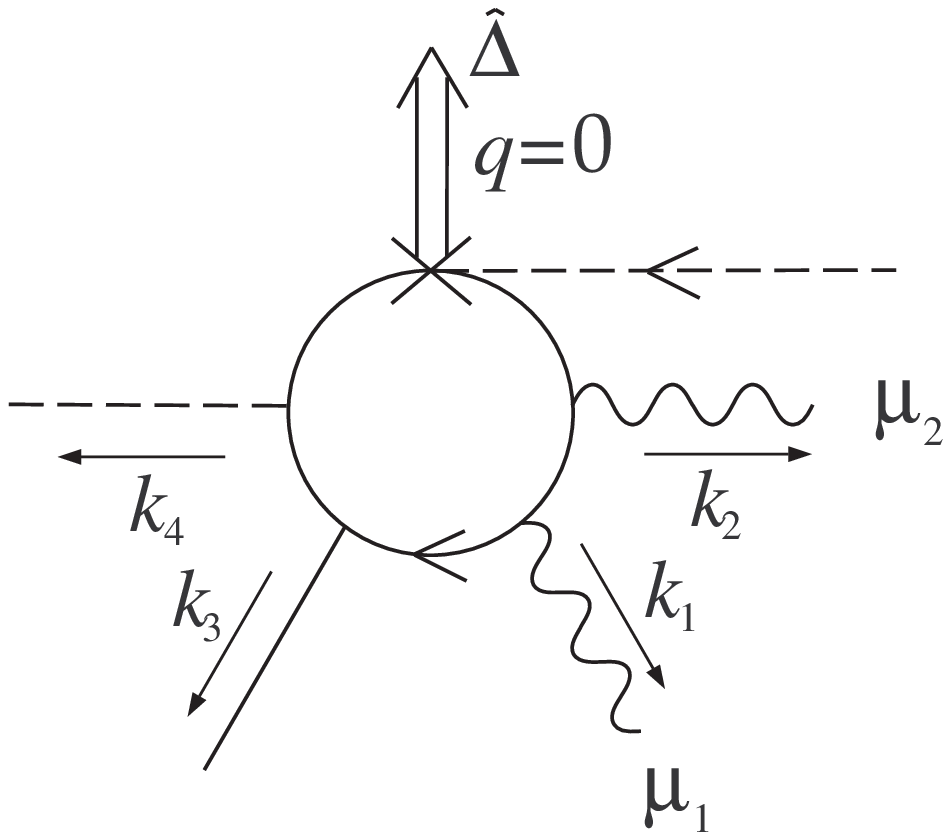}
         \hskip 0.22in } \hfill &\cr
\+       \hfill$(ix)_{1234}$\hfill
	&\hfill $(x)_{1234}$\hfill
	&\hfill $(xi)_{1234}$\hfill
	& \hfill $(xii)_{1234}$\hfill &\cr
}
\vskip 12pt
\narrower\noindent {\bf Figure C.6:}
{\eightrm  Feynman diagrams whith four boson and scalar fields
needed to compute the 1PI breaking functions (C.6)}
%
%
\endinsert

{
$$
\eqalignno{
\tilde\Gamma_{AA\phi_1\phi_2 c;N[\hat\Delta]}^{{\rm R}(1)\;\mu_1\mu_2 }
  (k_1,k_2,k_3,k_4)\, =\;&
  (i)_{1234} + (ii)_{1234} + (iii)_{1234} + (iv)_{1234} +\cr
  &(v)_{1234} + (vi)_{1234} + (vii)_{1234} + (viii)_{1234} + \cr
  &(ix)_{1234} + (xi)_{1234} + (xii)_{1234}
  + \hbox{\rm permut. of 12}\,.
  &\numeq\cr
}
$$
}

The renormalized result for each diagram of figure C.6 is:

\def\frac#1#2{{\displaystyle #1}\over{\displaystyle #2}}
\def\frac#1#2{{\displaystyle #1}\over{\displaystyle #2}}
$$ 
 \eqalignno{ 
 (4\pi)^2 \; (i)_{1234}=&\;
 2\,{}\,{f^2}\,{\theta}\, 
   \left( {\theta} + {r} \right) \, 
   g^{{\mu_1} {\mu_2}} 
  \,,\cr 
 (4\pi)^2 \; (ii)_{1234}=&\;
 {\frac{2\,{}\,{f^2}\, 
      \left( 1 + {{{\theta}}^2} +  
        {\theta}\,{r} \right) \, 
      g^{{\mu_1} {\mu_2}}}{3}} 
  \,,\cr 
 (4\pi)^2 \; (iii)_{1234}=&\;
 -2\,{}\,{f^2}\,{\theta}\, 
   \left( {\theta} + {r} \right) \, 
   g^{{\mu_1} {\mu_2}} 
  \,,\cr 
 (4\pi)^2 \; (iv)_{1234}=&\;
 {\frac{2\,{}\,{f^2}\, 
      \left( 1 + {{{\theta}}^2} +  
        {\theta}\,{r} \right) \, 
      g^{{\mu_1} {\mu_2}}}{3}} 
  \,,\cr 
 (4\pi)^2 \; (v)_{1234}=&\;
 -2\,{}\,{f^2}\,{\theta}\, 
   \left( {\theta} + {r} \right) \, 
   g^{{\mu_1} {\mu_2}} 
  \,,\cr 
 (4\pi)^2 \; (vi)_{1234}=&\;
 -2\,{}\,{f^2}\,{\theta}\, 
   \left( {\theta} + {r} \right) \, 
   g^{{\mu_1} {\mu_2}} 
  \,,\cr 
 (4\pi)^2 \; (vii)_{1234}=&\;
 {\frac{2\,{}\,{f^2}\, 
      \left( 1 + {{{\theta}}^2} +  
        {\theta}\,{r} \right) \, 
      g^{{\mu_1} {\mu_2}}}{3}} 
  \,,\cr 
 (4\pi)^2 \; (viii)_{1234}=&\;
 {\frac{2\,{}\,{f^2}\, 
      \left( 1 + {{{\theta}}^2} +  
        {\theta}\,{r} \right) \, 
      g^{{\mu_1} {\mu_2}}}{3}} 
  \,,\cr 
 (4\pi)^2 \; (ix)_{1234}=&\;
 {\frac{2\,{}\,{f^2}\, 
      \left( 1 + {{{\theta}}^2} +  
        {\theta}\,{r} \right) \, 
      g^{{\mu_1} {\mu_2}}}{3}} 
  \,,\cr 
 (4\pi)^2 \; (x)_{1234}=&\;
 -2\,{}\,{f^2}\,{\theta}\, 
   \left( {\theta} + {r} \right) \, 
   g^{{\mu_1} {\mu_2}} 
  \,,\cr 
 (4\pi)^2 \; (xi)_{1234}=&\;
 {\frac{2\,{}\,{f^2}\, 
      \left( 1 + {{{\theta}}^2} +  
        {\theta}\,{r} \right) \, 
      g^{{\mu_1} {\mu_2}}}{3}} 
  \,,\cr 
 (4\pi)^2 \; (xii)_{1234}=&\;
 2\,{}\,{f^2}\,{\theta}\, 
   \left( {\theta} + {r} \right) \, 
   g^{{\mu_1} {\mu_2}} 
   \,. \; &\lasteqprime\cr 
 } 
 $$

\subsection{C.7 Diagrams with fermion fields}

\midinsert
{\settabs 4\columns \def\graphwidth{1.8in}   
 \def\graphwidthbis{0.9in}
\eightpoint
\+
         \hfill
         \vbox{\epsfxsize=\graphwidth \epsffile{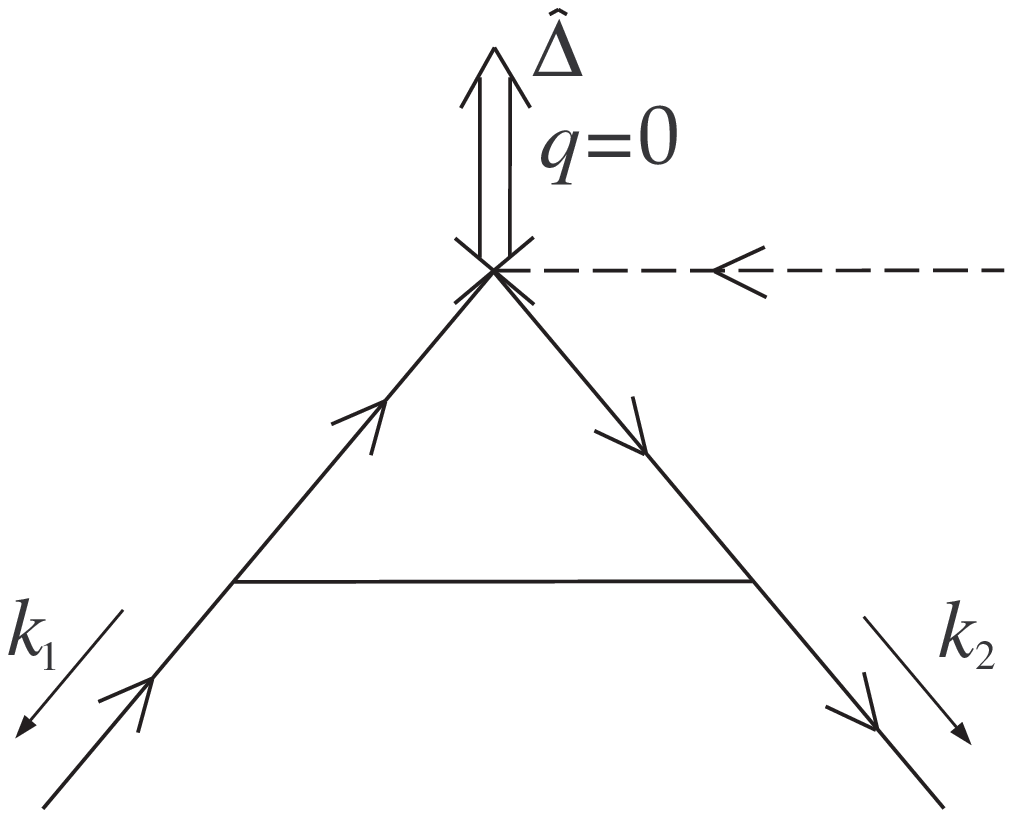}}
         \hfill
	&\hfill
         \vbox{\epsfxsize=\graphwidth \epsffile{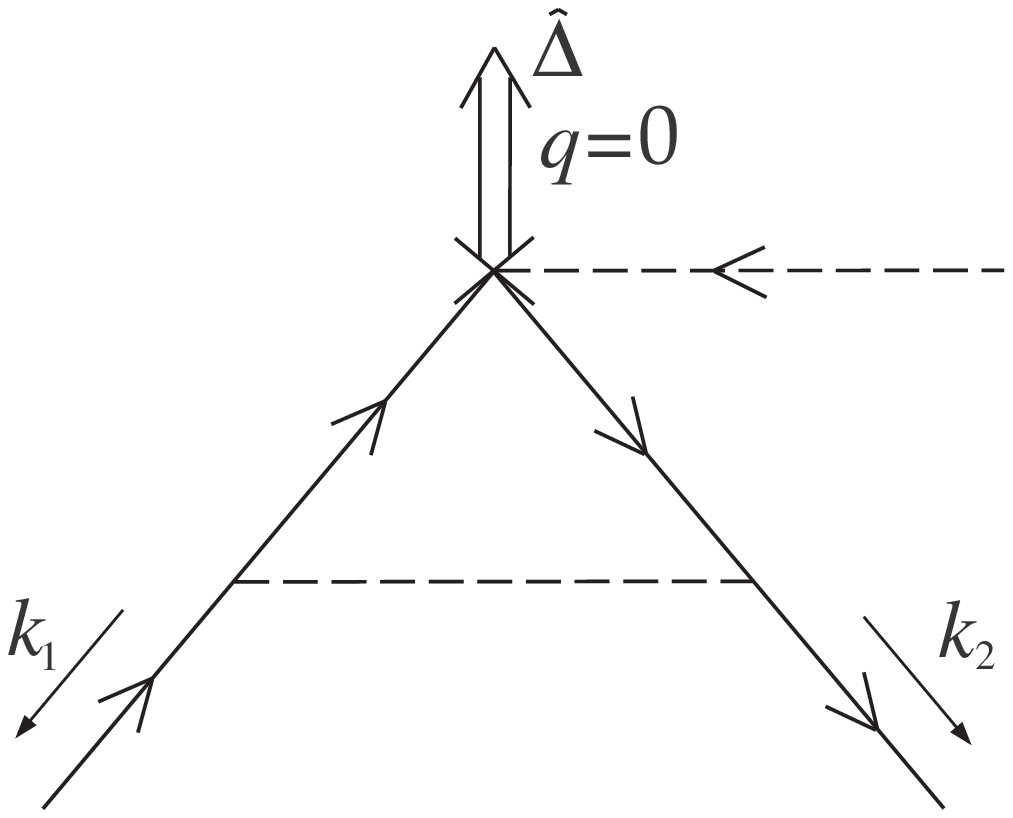}}
         \hfill
	&\hfill
         \vbox{\epsfxsize=\graphwidth \epsffile{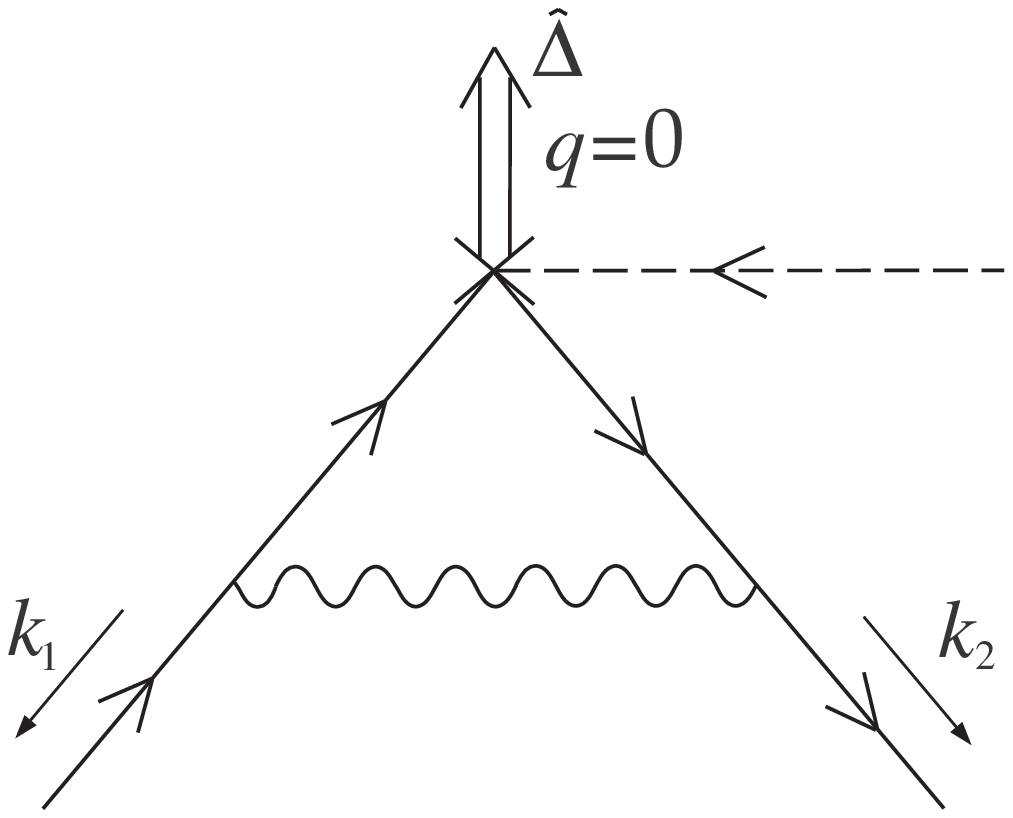}}
         \hfill 
	&\hfill
         \vbox{\epsfxsize=\graphwidth \epsffile{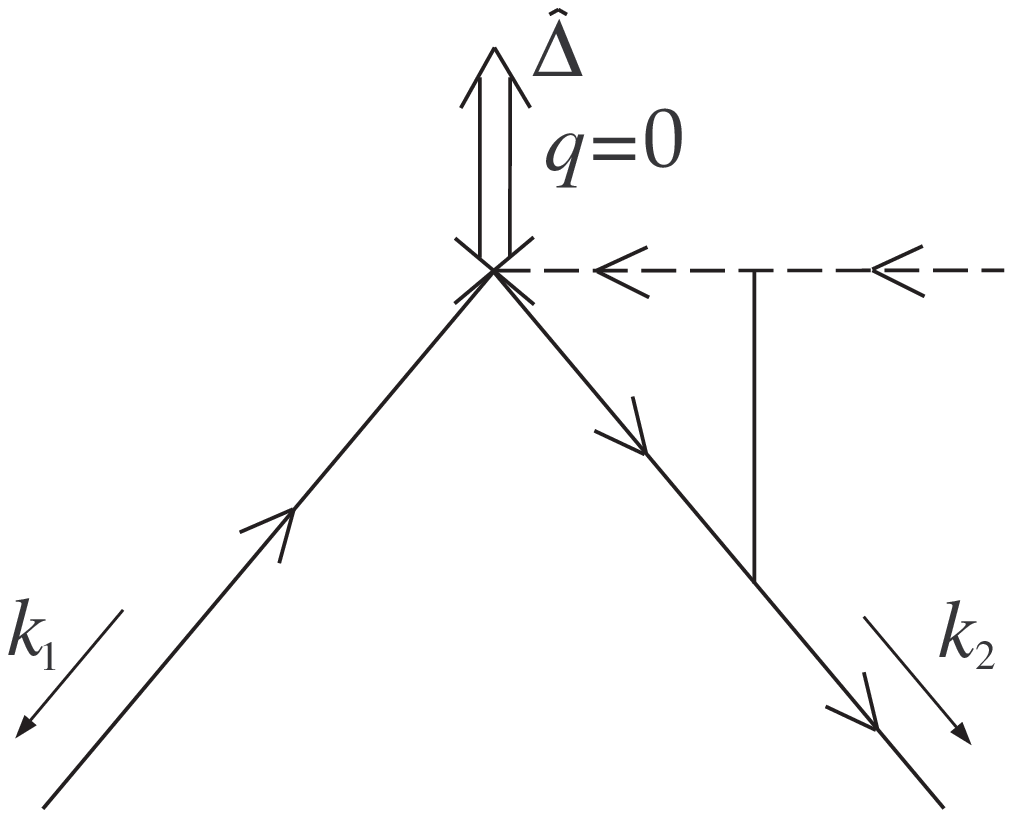}}
         \hskip 0.26in \hfill &\cr
\+       \hfill$(i)$\hfill
	&\hfill $(ii)$\hfill
	&\hfill $(iii)$\hfill
	& \hfill $(iv)$\hfill &\cr
\settabs 3\columns \def\graphwidth{1.8in} 
\+
         \hfill\hskip 0.26in
         \vbox{\epsfxsize=\graphwidth \epsffile{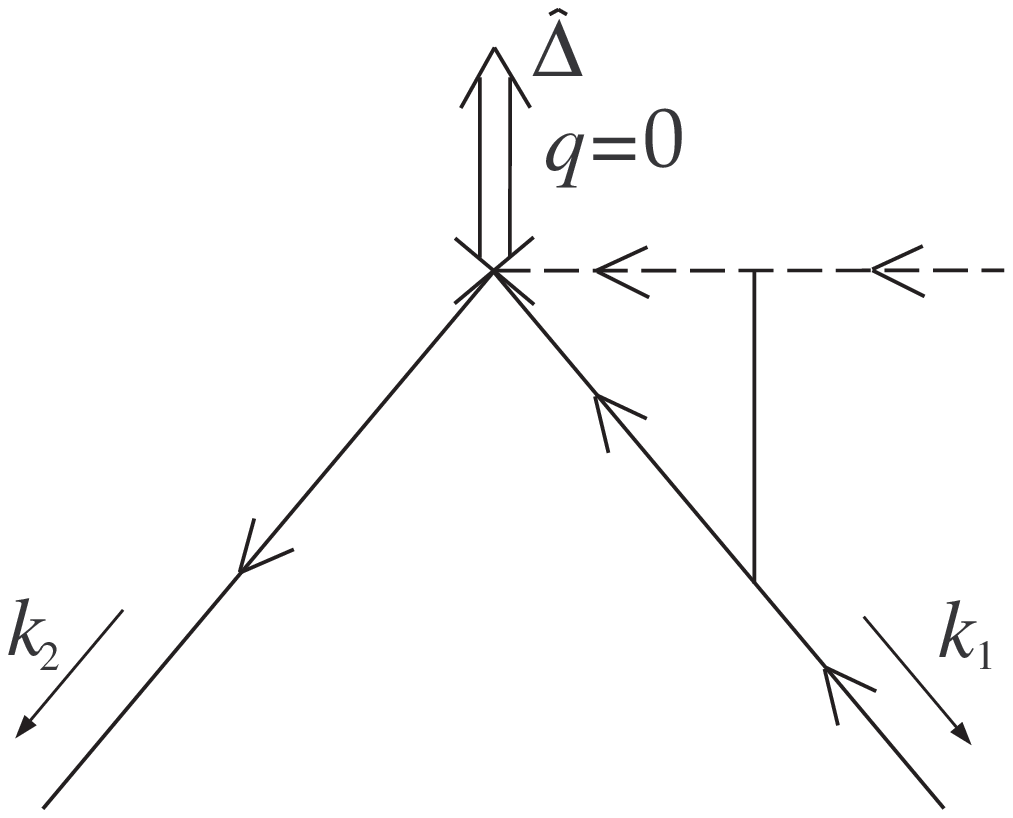}}
         \hfill
	&\hfill\hskip 0.20in
         \vbox{\epsfxsize=\graphwidth \epsffile{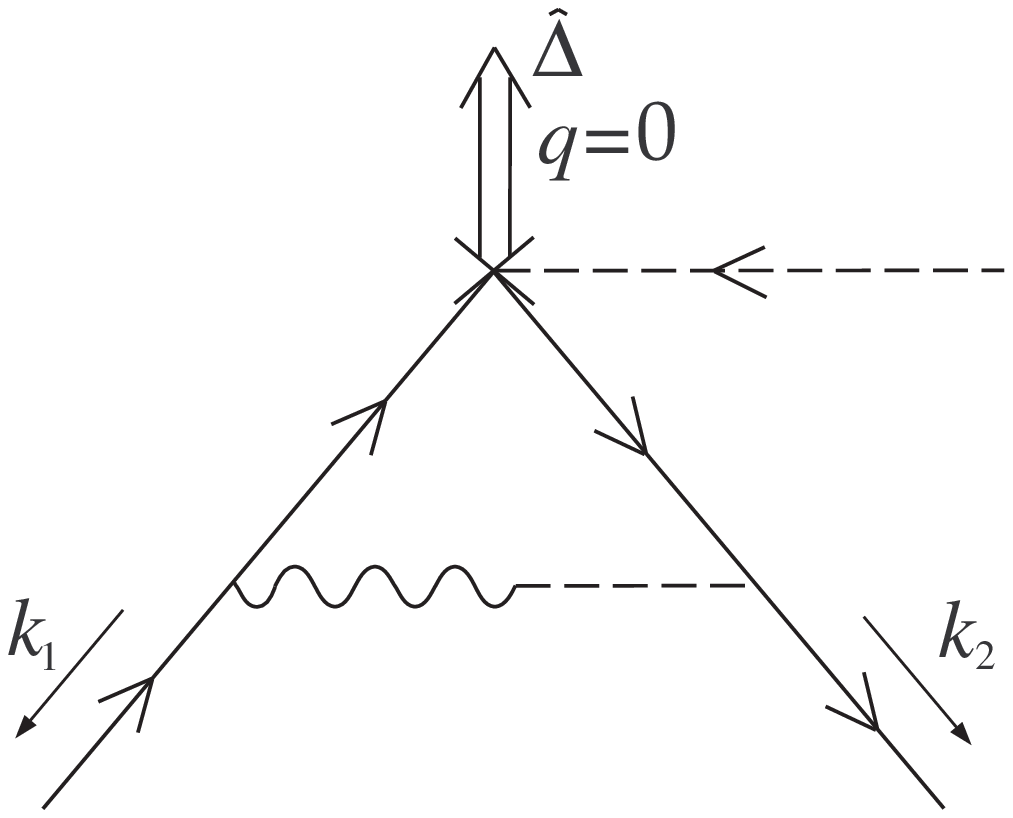}}
         \hfill
	&\hfill\hskip 0.20in
         \vbox{\epsfxsize=\graphwidth \epsffile{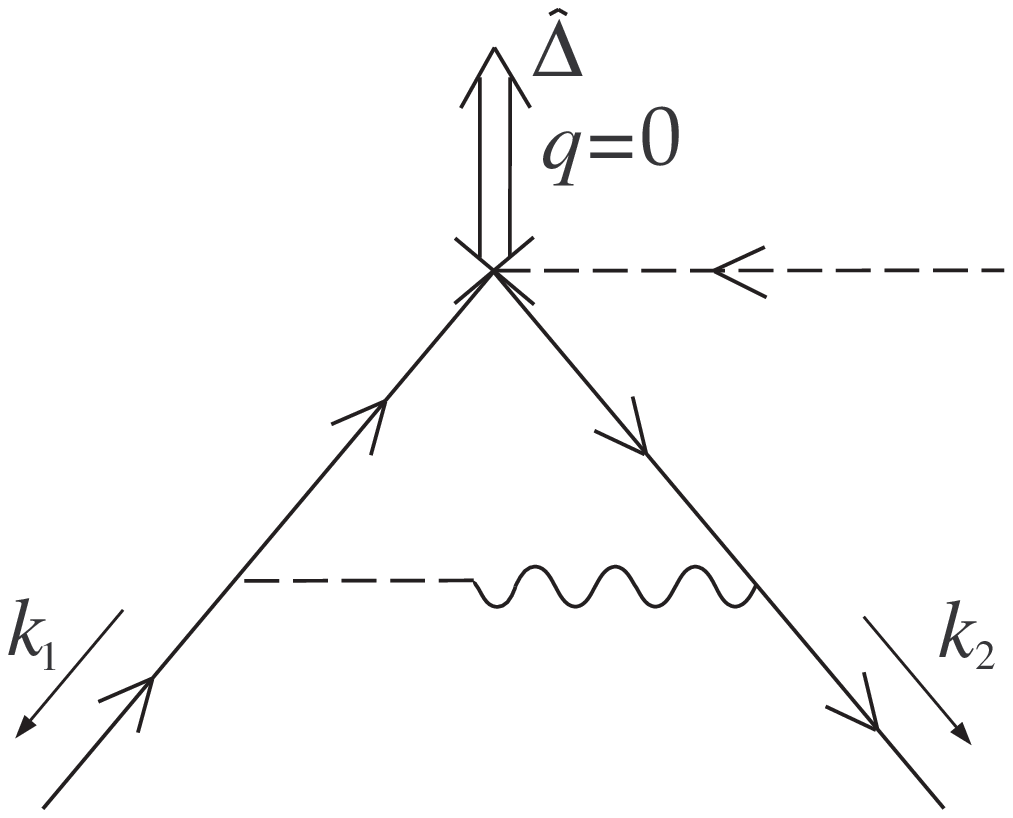}}
         \hfill 
        &\cr
\+       \hfill$(v)$\hfill
	&\hfill $(vi)$\hfill
	&\hfill $(vii)$\hfill
	&\cr
}
\vskip 12pt
\narrower\noindent {\bf Figure C.7:}
{\eightrm  Feynman diagrams whith fermion fields
needed to compute the 1PI breaking functions (C.7)}
%
%
\endinsert

{
$$
\eqalignno{
\tilde\Gamma_{\psi\bar\psi c;N[\hat\Delta]}^{{\rm R}(1)}
  (k_1,k_2)\, =\;&
  (i) + (ii) + (iii) + (iv) + (v) + (vi) + (vii) \,.
  &\numeq\cr
}
$$
}

The renormalized result for each diagram of figure C.7 is:

\def\frac#1#2{{\displaystyle #1}\over{\displaystyle #2}}
\def\frac#1#2{{\displaystyle #1}\over{\displaystyle #2}}
$$ 
 \eqalignno{ 
 (4\pi)^2 \; (i)=&\;
 {\frac{-i}{2}}\,{}\,{f^2}\,{r}\, 
   \left( 2\,f\,v\,\gamma_5  +  \not\! {k_1} \gamma_5  +  
      \not\! {k_2} \gamma_5  \right)  
  \,,\cr 
 (4\pi)^2 \; (ii)=&\;
 {\frac{i}{2}}\,{}\,{f^2}\,{r}\, 
   \left( 2\,f\,v\,\gamma_5  -  \not\! {k_1} \gamma_5  -  
      \not\! {k_2} \gamma_5  \right)  
  \,,\cr 
 (4\pi)^2 \; (iii)=&\;
 {\frac{-2\,i}{3}}\,{}\,f\,{g^2}\, 
    {\theta}\,\left( 1 + {\theta}\,{r} \ 
 \right) \,v\,\left( 5 + {\xi^\prime} \right) \,\gamma_5  \cr & - {\frac{i}{12}}\,{}\,{g^2}\, 
    \left( 2\,{\theta} + {r} \right) \, 
    \left( 5 + {\xi^\prime} \right) \, \not\! {k_1}  \cr & - {\frac{i}{12}}\,{}\,{g^2}\, 
    \left( 2\,{\theta} + {r} \right) \, 
    \left( 5 + {\xi^\prime} \right) \, \not\! {k_2}  \cr & + {\frac{i}{12}}\,{}\,{g^2}\, 
    \left( 2\,{\theta} + {r} +  
      2\,{{{\theta}}^2}\,{r} \right) \, 
    \left( 5 + {\xi^\prime} \right) \, \not\! {k_1} \gamma_5  \ 
 \cr & + {\frac{i}{12}}\,{}\,{g^2}\, 
    \left( 2\,{\theta} + {r} +  
      2\,{{{\theta}}^2}\,{r} \right) \, 
    \left( 5 + {\xi^\prime} \right) \, \not\! {k_2} \gamma_5  
  \,,\cr 
 (4\pi)^2 \; (iv)=&\;
 {\frac{i}{4}}\,{}\,f\,{\rho}\, 
   \left( \left( 2\,{\theta} + {r} \right) \un -  
     {r}\,\gamma_5  \right)  
  \,,\cr 
 (4\pi)^2 \; (v)=&\;
 {\frac{-i}{4}}\,{}\,f\,{\rho}\, 
   \left( \left( 2\,{\theta} + {r} \right) \un +  
     {r}\,\gamma_5  \right)  
  \,,\cr 
 (4\pi)^2 \; (vi)=&\;
 0 
  \,,\cr 
 (4\pi)^2 \; (vii)=&\;
 0 
   \,. \; &\lasteqprime\cr 
 } 
 $$

\subsection{C.8 Diagrams with $\phi_1$ and fermion fields}

\midinsert
{\settabs 4\columns \def\graphwidth{1.8in}   
 \def\graphwidthbis{0.9in}
\eightpoint
\+
         \hfill
         \vbox{\epsfxsize=\graphwidth \epsffile{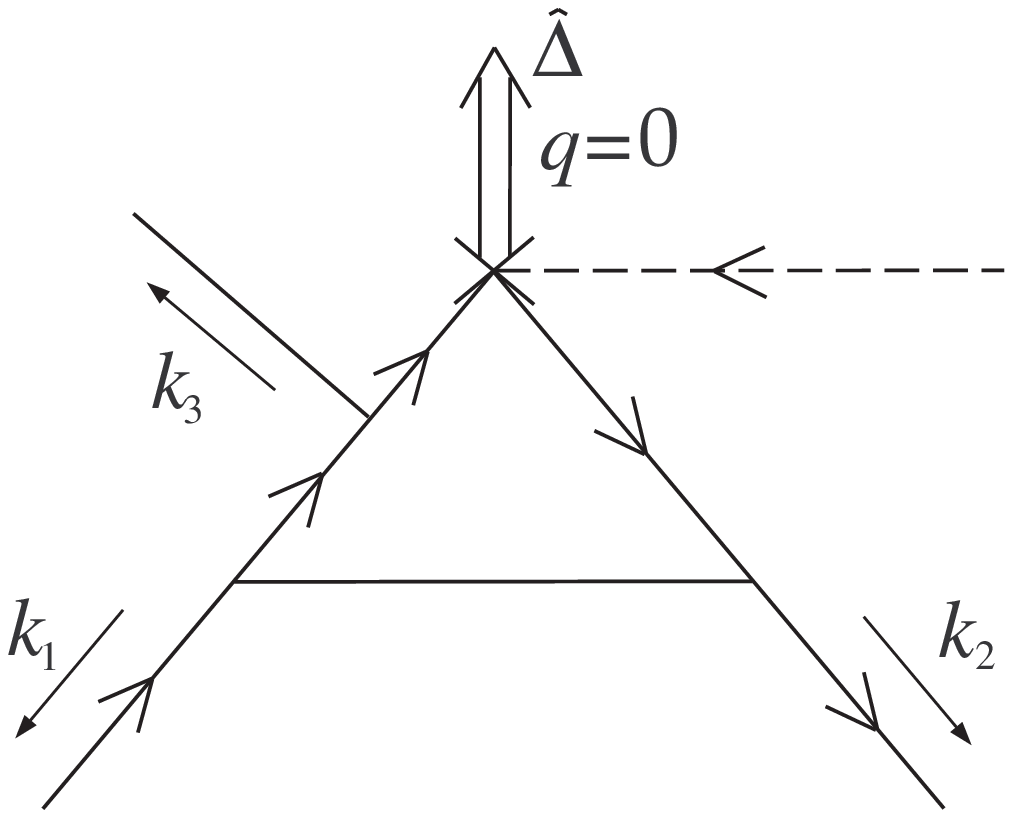}}
         \hfill
	&\hfill
         \vbox{\epsfxsize=\graphwidth \epsffile{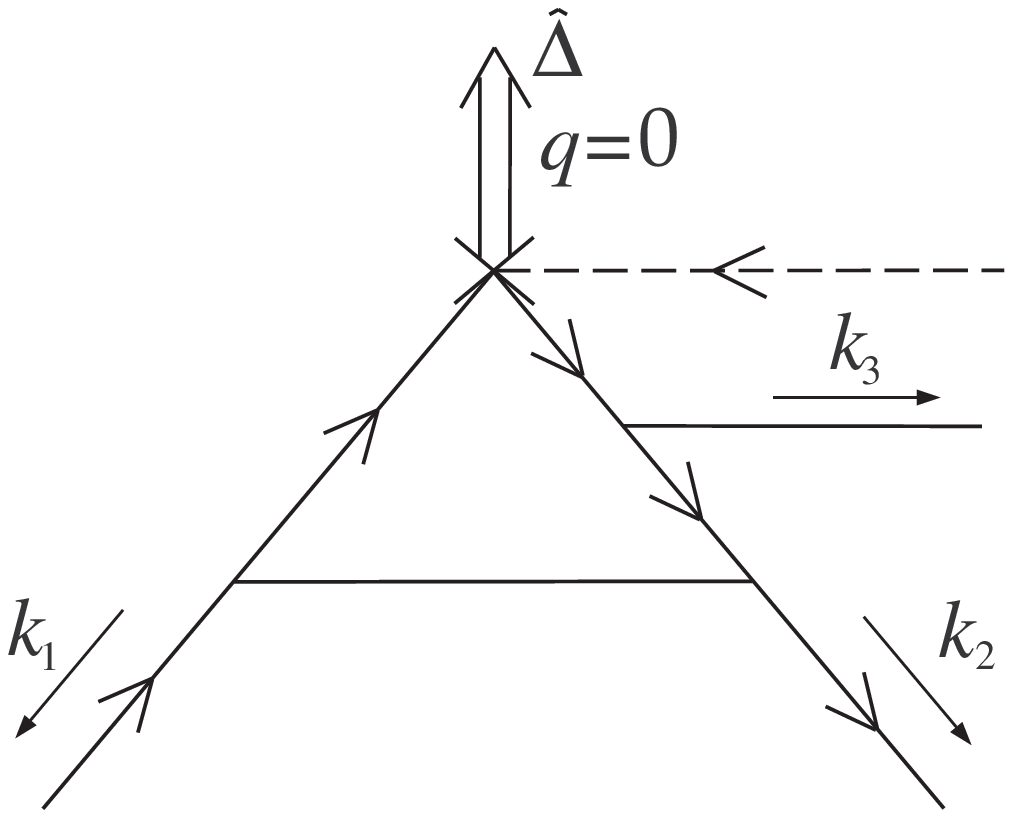}}
         \hfill
	&\hfill
         \vbox{\epsfxsize=\graphwidth \epsffile{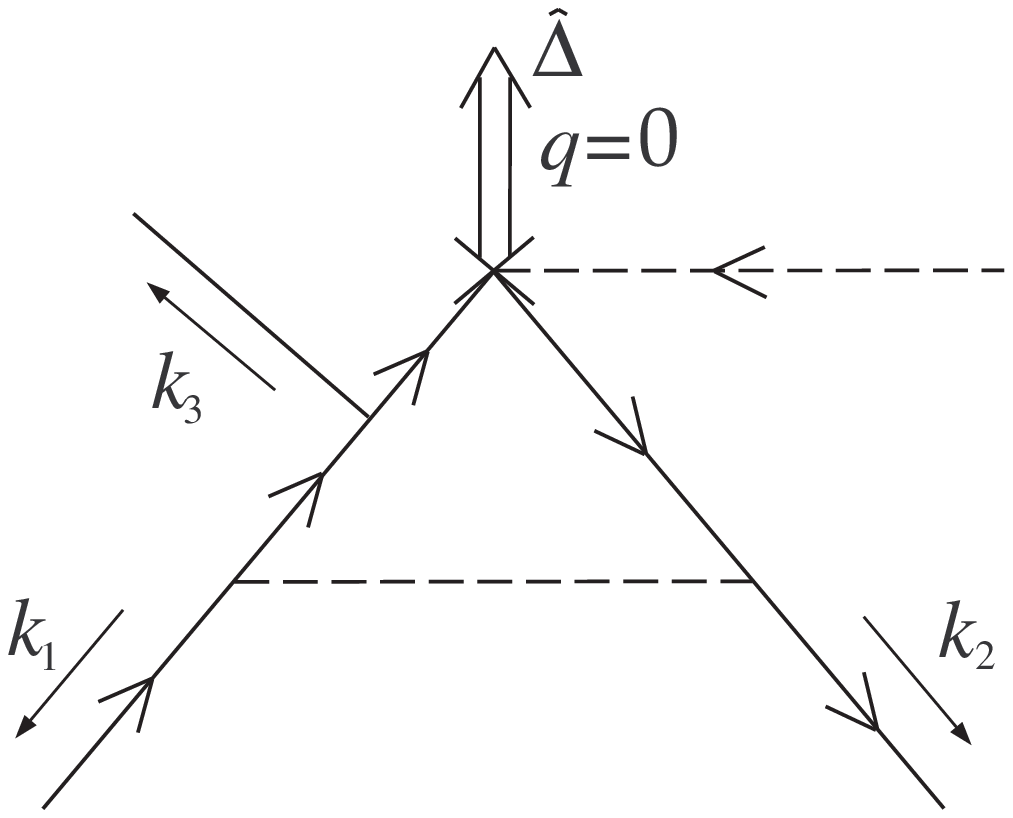}}
         \hfill 
	&\hfill
         \vbox{\epsfxsize=\graphwidth \epsffile{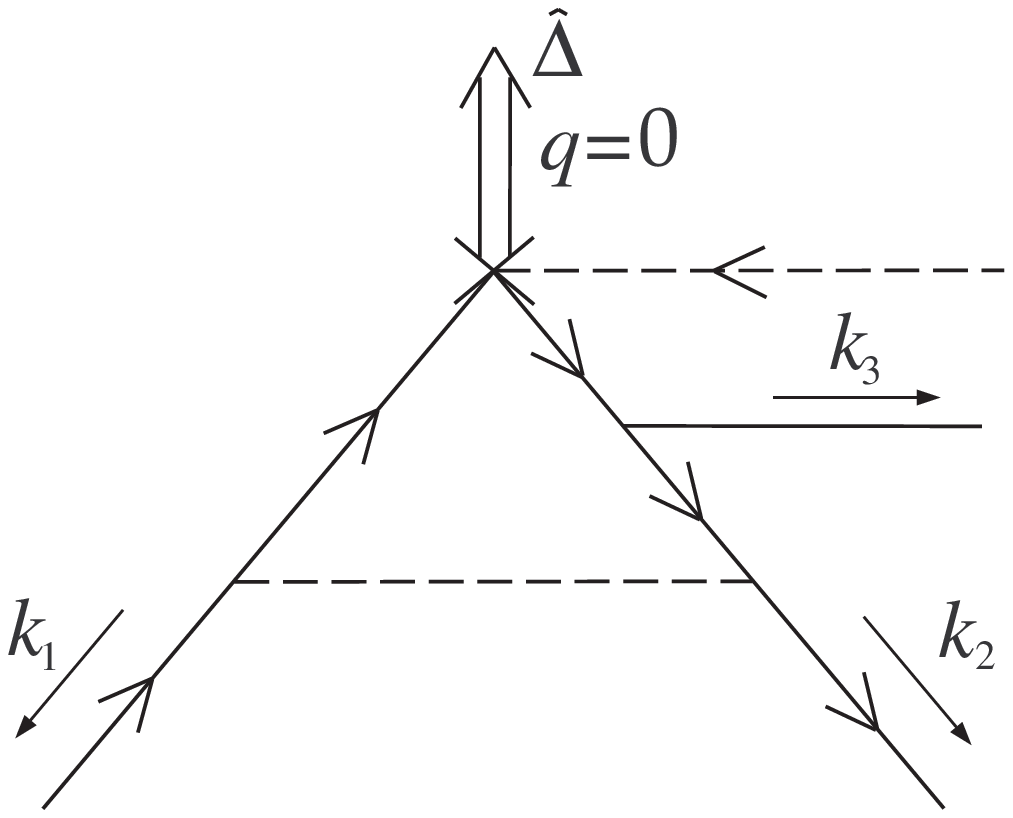}}
         \hskip 0.26in \hfill &\cr
\+       \hfill$(i)$\hfill
	&\hfill $(ii)$\hfill
	&\hfill $(iii)$\hfill
	& \hfill $(iv)$\hfill &\cr
\+
         \hfill
         \vbox{\epsfxsize=\graphwidth \epsffile{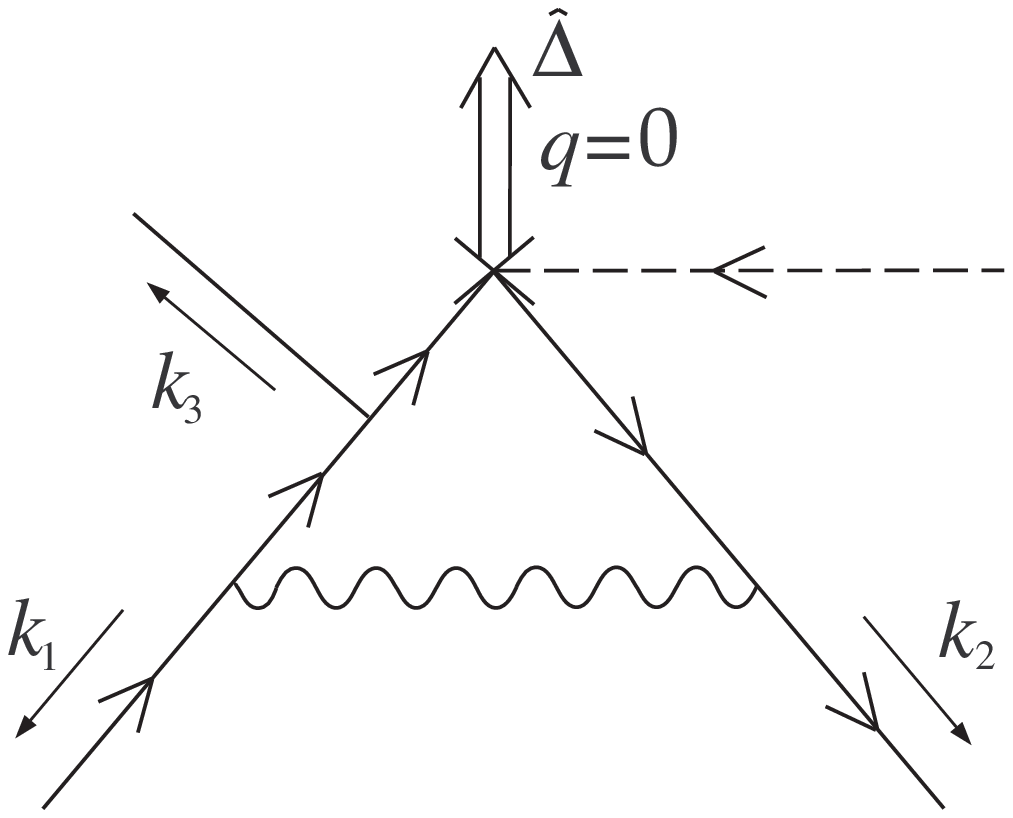}}
         \hfill
	&\hfill
         \vbox{\epsfxsize=\graphwidth \epsffile{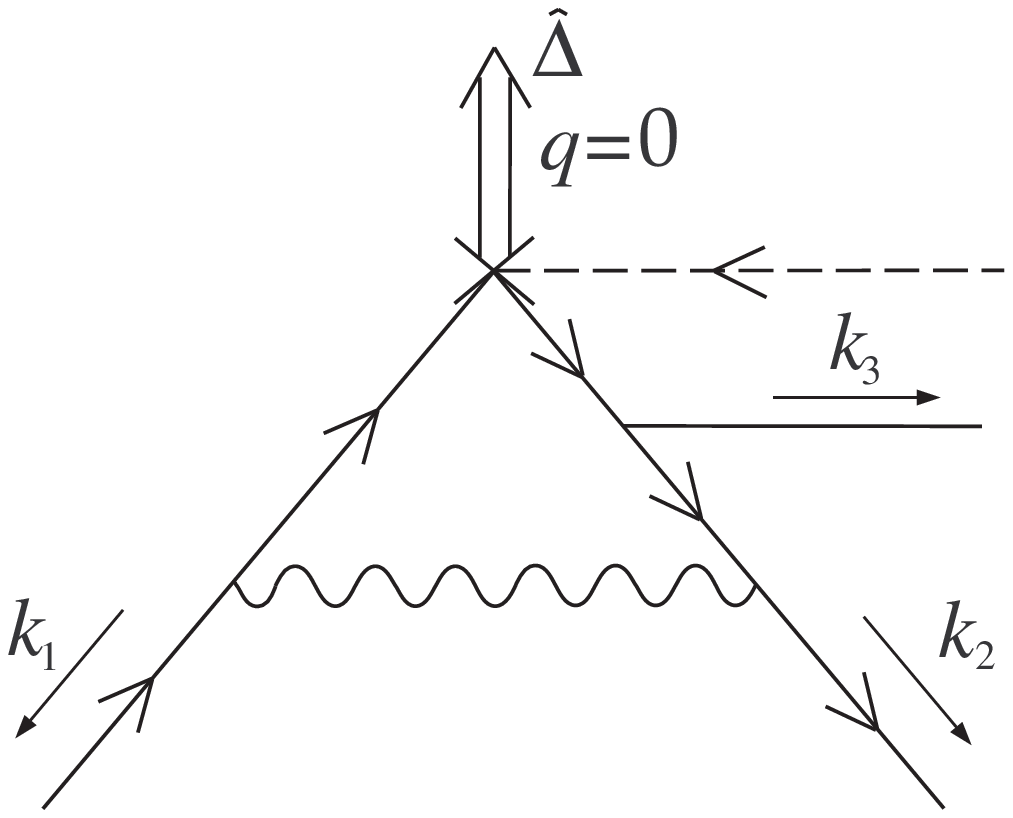}}
         \hfill
	&\hfill
         \vbox{\epsfxsize=\graphwidth \epsffile{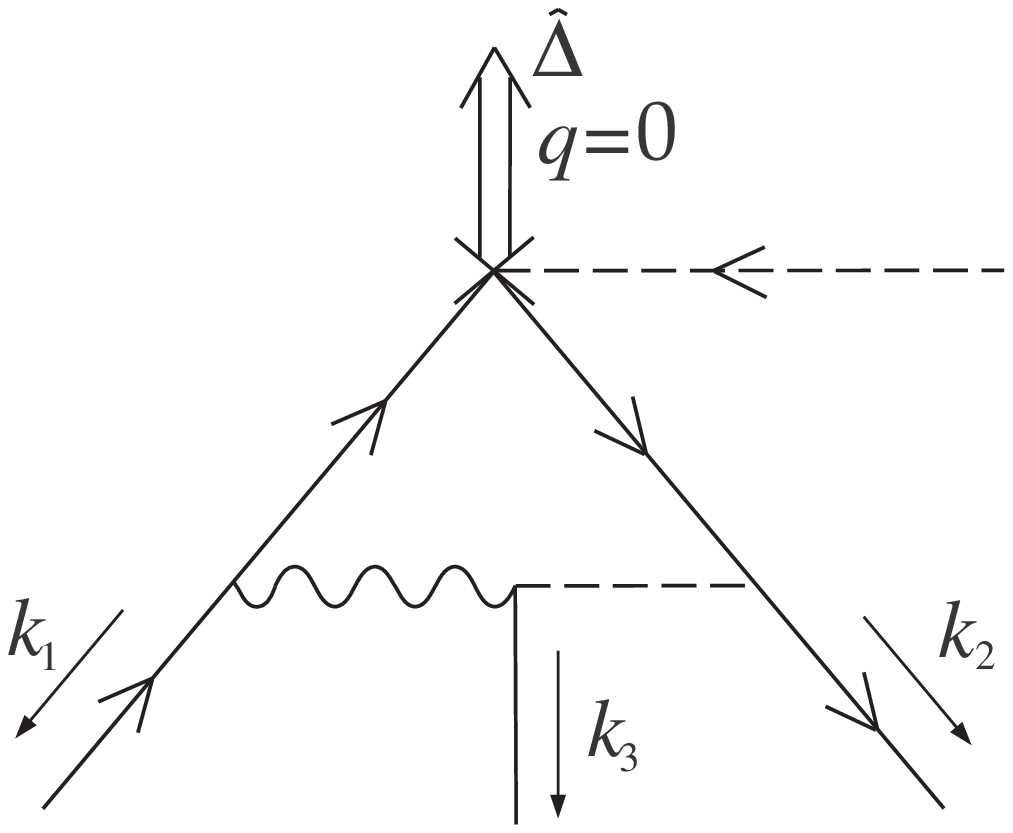}}
         \hfill 
	&\hfill
         \vbox{\epsfxsize=\graphwidth \epsffile{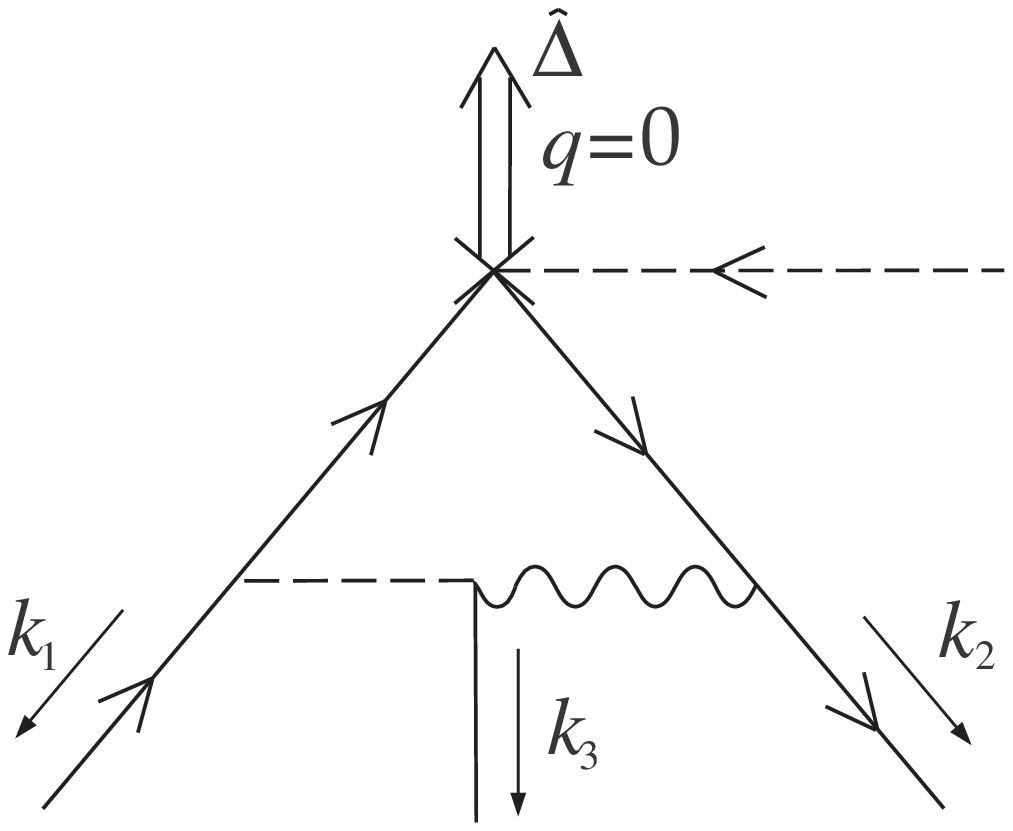}}
         \hskip 0.26in \hfill &\cr
\+       \hfill$(v)$\hfill
	&\hfill $(vi)$\hfill
	&\hfill $(vii)$\hfill
	& \hfill $(viii)$\hfill &\cr
}
\vskip 12pt
\narrower\noindent {\bf Figure C.8:}
{\eightrm  Feynman diagrams whith one scalar and fermion fields
needed to compute the 1PI breaking functions (C.8)}
%
%
\endinsert

{
$$
\eqalignno{
\tilde\Gamma_{\psi\bar\psi\phi_1 c;N[\hat\Delta]}^{{\rm R}(1)}
  (k_1,k_2,k_3)\, =\;&
  (i) + (ii) + (iii) + (iv) + (v) + (vi) + (vii) + (viii)\,.
  &\numeq\cr
}
$$
Note that there are other possible diagrams at one loop with
the same external legs, but due to power counting they are
convergent.
}

The renormalized result for each diagram of figure C.8 is:

\def\frac#1#2{{\displaystyle #1}\over{\displaystyle #2}}
\def\frac#1#2{{\displaystyle #1}\over{\displaystyle #2}}
$$ 
 \eqalignno{ 
 (4\pi)^2 \; (i)=&\;
 {\frac{-i}{2}}\,{}\,{f^3}\,{r}\,\gamma_5  
  \,,\cr 
 (4\pi)^2 \; (ii)=&\;
 {\frac{-i}{2}}\,{}\,{f^3}\,{r}\,\gamma_5  
  \,,\cr 
 (4\pi)^2 \; (iii)=&\;
 {\frac{i}{2}}\,{}\,{f^3}\,{r}\,\gamma_5  
  \,,\cr 
 (4\pi)^2 \; (iv)=&\;
 {\frac{i}{2}}\,{}\,{f^3}\,{r}\,\gamma_5  
  \,,\cr 
 (4\pi)^2 \; (v)=&\;
 {\frac{-i}{3}}\,{}\,f\,{g^2}\,{\theta}\, 
   \left( 1 + {\theta}\,{r} \right) \, 
   \left( 5 + {\xi^\prime} \right) \,\gamma_5  
  \,,\cr 
 (4\pi)^2 \; (vi)=&\;
 {\frac{-i}{3}}\,{}\,f\,{g^2}\,{\theta}\, 
   \left( 1 + {\theta}\,{r} \right) \, 
   \left( 5 + {\xi^\prime} \right) \,\gamma_5  
  \,,\cr 
 (4\pi)^2 \; (vii)=&\;
 0 
  \,,\cr 
 (4\pi)^2 \; (viii)=&\;
 0 
   \,. \; &\lasteqprime\cr 
 } 
 $$

\subsection{C.9 Diagrams with $\phi_2$ and fermion fields}

\midinsert
{\settabs 4\columns \def\graphwidth{1.8in}   
 \def\graphwidthbis{0.9in}
\eightpoint
\+
         \hfill
         \vbox{\epsfxsize=\graphwidth \epsffile{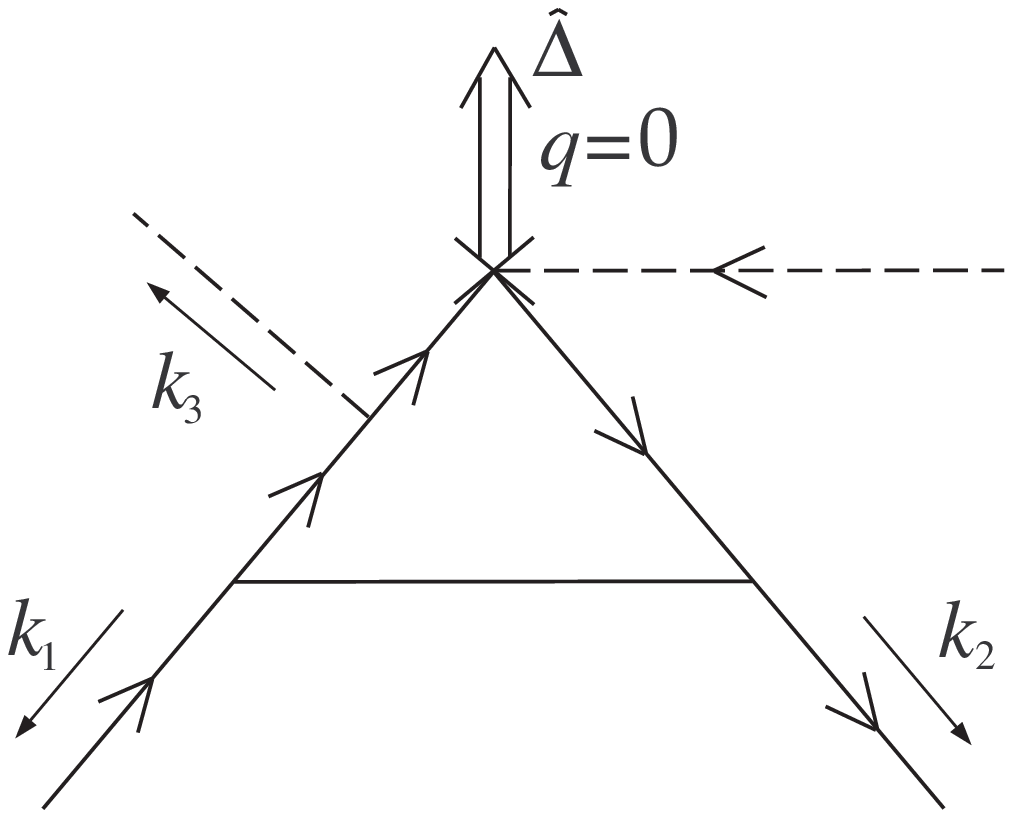}}
         \hfill
	&\hfill
         \vbox{\epsfxsize=\graphwidth \epsffile{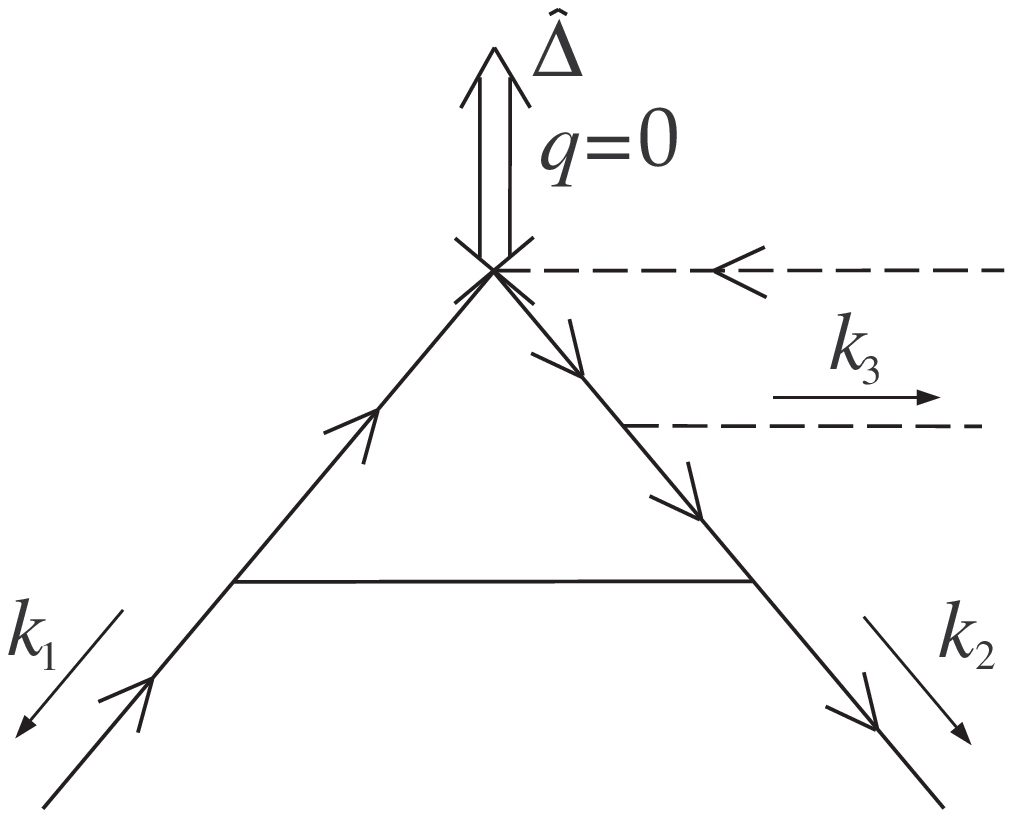}}
         \hfill
	&\hfill
         \vbox{\epsfxsize=\graphwidth \epsffile{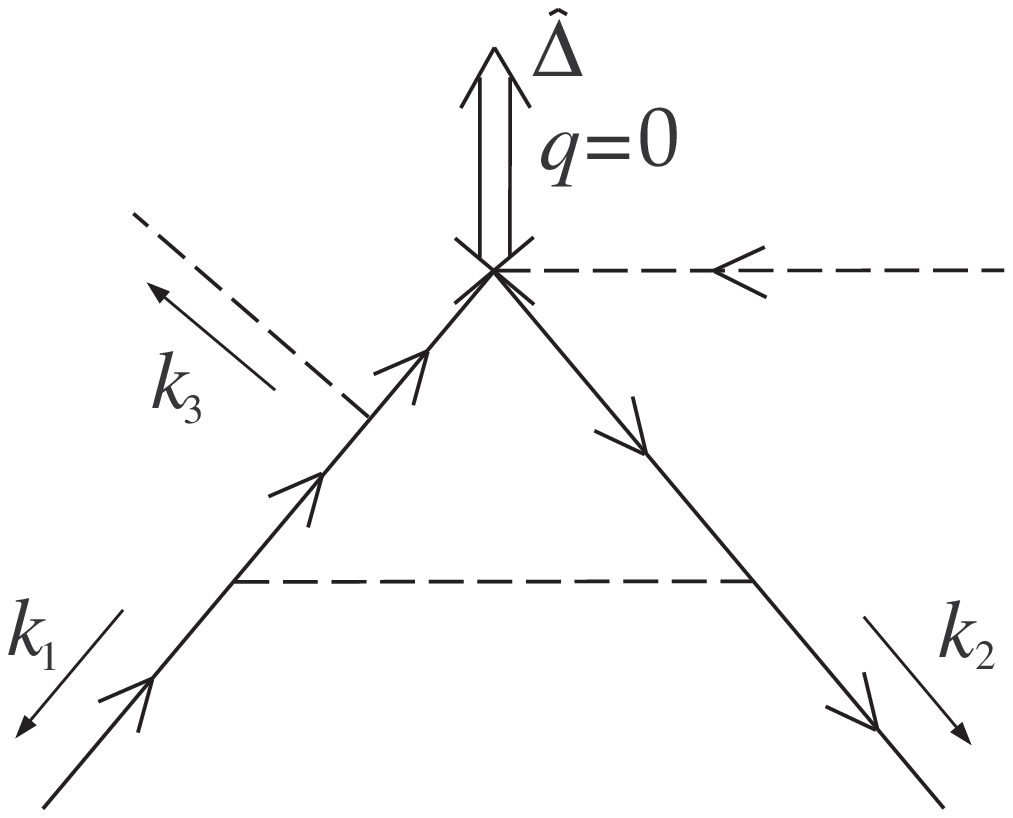}}
         \hfill 
	&\hfill
         \vbox{\epsfxsize=\graphwidth \epsffile{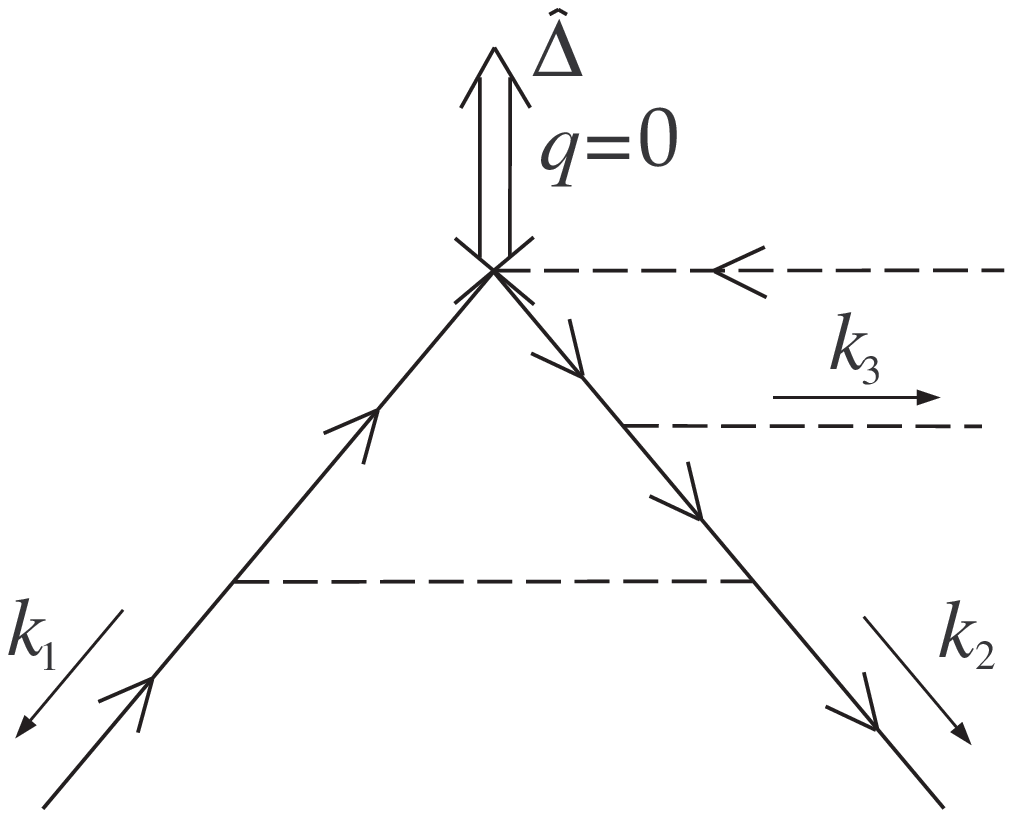}}
         \hskip 0.26in \hfill &\cr
\+       \hfill$(i)$\hfill
	&\hfill $(ii)$\hfill
	&\hfill $(iii)$\hfill
	& \hfill $(iv)$\hfill &\cr
\+
         \hfill
         \vbox{\epsfxsize=\graphwidth \epsffile{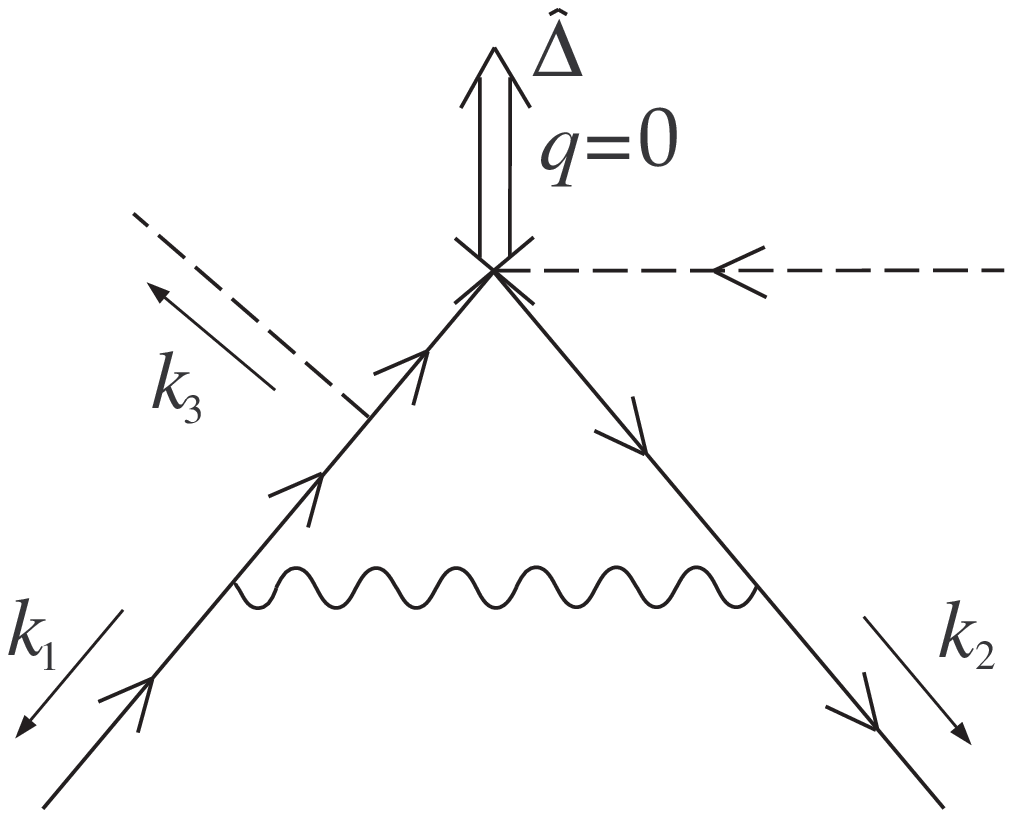}}
         \hfill
	&\hfill
         \vbox{\epsfxsize=\graphwidth \epsffile{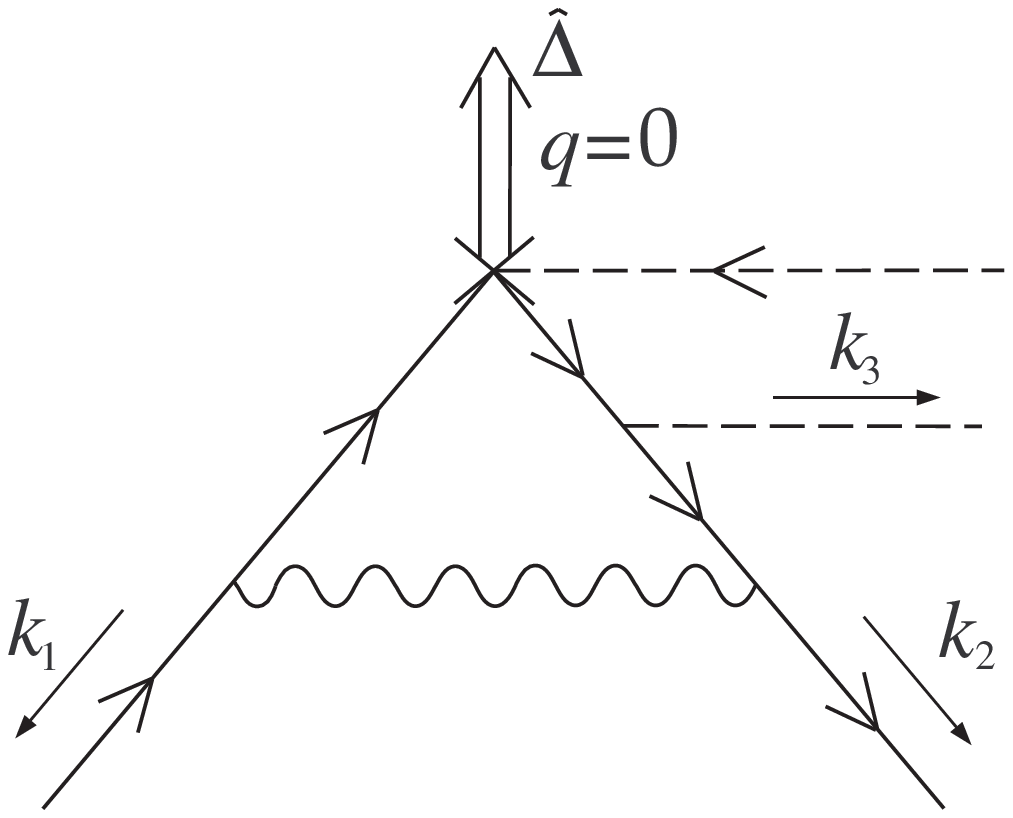}}
         \hfill
	&\hfill
         \vbox{\epsfxsize=\graphwidth \epsffile{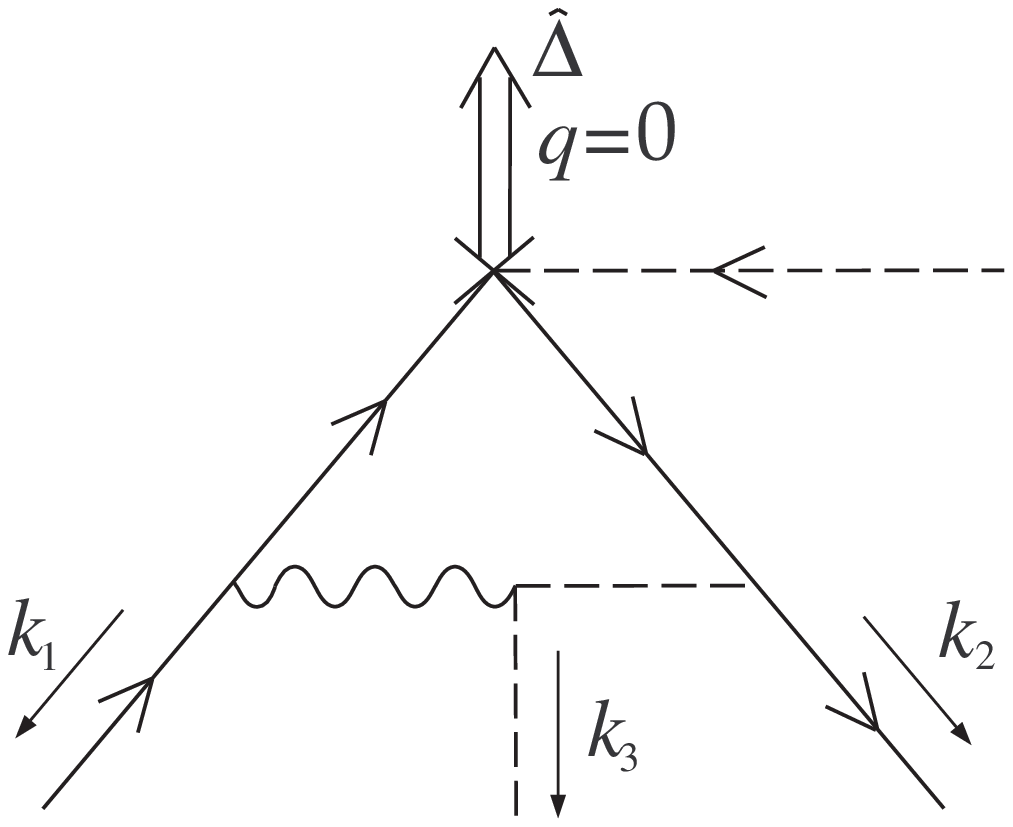}}
         \hfill 
	&\hfill
         \vbox{\epsfxsize=\graphwidth \epsffile{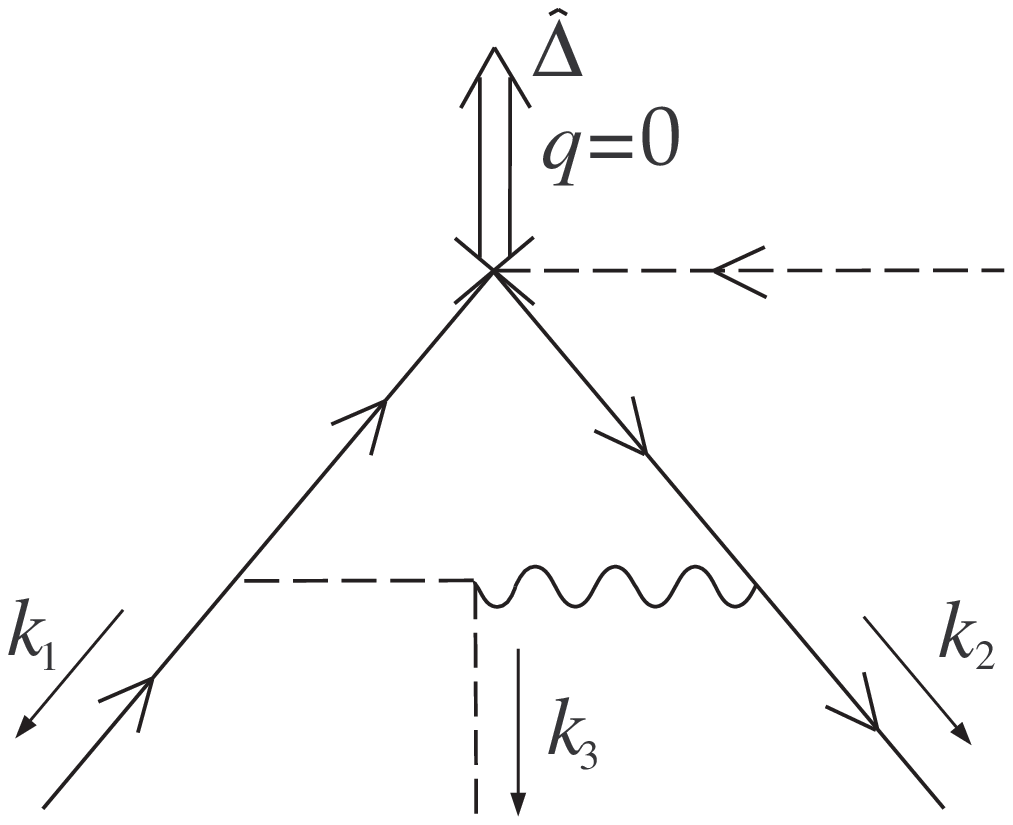}}
         \hskip 0.26in \hfill &\cr
\+       \hfill$(v)$\hfill
	&\hfill $(vi)$\hfill
	&\hfill $(vii)$\hfill
	& \hfill $(viii)$\hfill &\cr
}
\vskip 12pt
\narrower\noindent {\bf Figure C.9:}
{\eightrm  Feynman diagrams whith one scalar and fermion fields
needed to compute the 1PI breaking functions (C.9)}
%
%
\endinsert

{
$$
\eqalignno{
\tilde\Gamma_{\psi\bar\psi\phi_2 c;N[\hat\Delta]}^{{\rm R}(1)}
  (k_1,k_2,k_3)\, =\;&
  (i) + (ii) + (iii) + (iv) + (v) + (vi) + (vii) + (viii)\,.
  &\numeq\cr
}
$$
Again there are other convergent diagrams at one loop with
the same external legs.
}

The renormalized result for each diagram of figure C.9 is:

\def\frac#1#2{{\displaystyle #1}\over{\displaystyle #2}}
\def\frac#1#2{{\displaystyle #1}\over{\displaystyle #2}}
$$ 
 \eqalignno{ 
 (4\pi)^2 \; (i)=&\;
 {\frac{{}\,{f^3}\un}{2}} 
  \,,\cr 
 (4\pi)^2 \; (ii)=&\;
 {\frac{{}\,{f^3}\un}{2}} 
  \,,\cr 
 (4\pi)^2 \; (iii)=&\;
 {\frac{-\left( {}\,{f^3}\un \right) }{2}} 
  \,,\cr 
 (4\pi)^2 \; (iv)=&\;
 {\frac{-\left( {}\,{f^3}\un \right) }{2}} 
  \,,\cr 
 (4\pi)^2 \; (v)=&\;
 {\frac{-\left( {}\,f\,{g^2}\, 
        {\theta}\,\left( {\theta} + {r} \ 
 \right) \,\left( 5 + {\xi^\prime} \right) \un \right) }{3}} 
  \,,\cr 
 (4\pi)^2 \; (vi)=&\;
 {\frac{-\left( {}\,f\,{g^2}\, 
        {\theta}\,\left( {\theta} + {r} \ 
 \right) \,\left( 5 + {\xi^\prime} \right) \un \right) }{3}} 
  \,,\cr 
 (4\pi)^2 \; (vii)=&\;
 0 
  \,,\cr 
 (4\pi)^2 \; (viii)=&\;
 0 
   \,. \; &\lasteqprime\cr 
 } 
 $$

\section{References}

\frenchspacing

{
\refno\Ash.
C.G.Bollini and J.J. Giambiagi,
Phys. Lett. B40 (1972) 566;
J. Ashmore, Lett. Nuovo Cimento 4 (1972) 289;
G.~M.~Cicuta and E.~Montaldi,
Lett.\ Nuovo Cim.\  4 (1972) 329.

\refno\HV.
G. t'Hooft and M. Veltman, Nucl. Phys. B44 (1972) 189.

\refno\BMabc.
P. Breitenlohner and D. Maison, Comm. Math. Phys 52 (1977) 11;
52 (1977) 39;
52 (1977) 55.

\refno\BonneauABC.
G. Bonneau, Nucl. Phys. B167 (1980) 261;
B171 (1980) 477;
B177 (1980) 523.

\refno\Collins.
J. Collins, Renormalization
(Cambridge University Press, Cambridge, 1984).

\refno\Min.
G. `t Hooft, Nucl. Phy. B61 (1973) 455.

\refno\BonneauRemarks.
G. Bonneau, Phys. Lett. B96 (1980) 147.

\refno\naive.
W.A.  Bardeen, R. Gastmans and B. Lautrup, Nucl. Phys. B46 (1972) 319;
M. Chanowitz, M. Furman and I. Hinchliffe, Nucl. Phys. B159 (1979) 225.

\refno\BonneauReview.
G. Bonneau, Int. Journ. Mod. Phys. A5 (1990) 3831.

\refno\facts.
F. Jegerlehner, Eur. Phys. J. C18 (2001) 673,
{\tt arXiv:hep-th/0005255} and references therein.

\refno\Kreimer.
D. Kreimer, Phys. Lett. B237 (1990) 59, {\tt hep-ph/9401354}.

\refno\Delbourgo.
D.~A.~Akyeampong and R.~Delbourgo,
Nuovo Cim.\ A  17 (1973) 578;
Nuovo Cim.\ A  18 (1973) 94;
Nuovo Cim.\ A  19 (1974) 219.

\refno\Tonin.
M. Tonin, Nucl. Phys. B. (Proc. Suppl.) 29B,C (1992) 137.

\refno\PerniciABC.
M. Pernici, M.Raciti and F. Riva, Nucl. Phys. B577  (2000) 
 293, \ \ \ {\tt arXiv:hep-th/9912248};
M. Pernici, Nucl. Phys. B582 (2000) 
 733, {\tt arXiv:hep-th/9912278};
M. Pernici and M.Raciti, Phys. Lett. B513 (2001) 
 421,
{\tt arXiv:hep-th/0003062}.

\refno\Hepp.
K. Hepp, Renormalization theory in statistical mechanics and quantum field
theory, in ``Les Houches XX 1970'' (Gordon and Breach, New York, 1971).

\refno\Colnor.
J.C. Collins, Nucl. Phys. B92 (1975) 477.

\refno\bmym.
C.P. Mart{\'\i}n and D. S\'anchez-Ruiz, Nucl. Phys. B572 (2000) 
387, \ \ \ {\tt arXiv:hep-th/9905076}
  
\refno\nochiral.
C. Schubert, Nucl. Phys. B323 (1989) 478;
A. J. Buras and P. H. Weisz, Nucl. Phys. B333 (1990) 66;
G. Buchalla, Nucl. Phys. B391 (1993) 501;
M. Ciuchini, E. Franco, G. Martinelli, L. Reina and
L. Silvestrini, Nucl. Phys. B415 (1994) 403,
{\tt arXiv:hep-ph/9304257};
M. Ciuchini, E. Franco, L. Reina and
L. Silvestrini, Nucl. Phys. B421 (1994) 41,
{\tt arXiv:hep-ph/9311357}.

\refno\axial.
I.Antoniadis, Phys. Lett. B84 (1979) 223;
T.L.Trueman, Phys. Lett. B88 (1979) 331;
D.R.T. Jones and J.P. Leveille, Nucl. Plys. B206 (1982) 473;
B. A. Ovrut, Nucl. Phys. B213 (1983) 241;
M. A. Rego Monteiro, Lett. Nuov. Cim. 40 (1984) 201;
S.G. Gorishny and S.A. Larin, Phys. Lett. B172 (1986) 109;
S.A. Larin and J.A.M. Vermaseren, Phys. Lett. B259 (1991), 345;
S.A. Larin, Phys. Lett. B303 (1993) 113,
{\tt arXiv:hep-ph/9302240};
T.L. Trueman, Z. Phys. C69 (1996) 525,
{\tt arXiv: hep-ph/9504315}.

\refno\Korner.
J.G. K\"orner, N. Nasrallah and K. Schilcher, Phys. Rev. D41 (1990) 888.

\refno\Ferrari.
R. Ferrari, A. Le Yaouanc, L. Oliver and J.C. Raynal, Phys. Rev. D52 (1995)
3036.

\refno\Weiglein.
A. Freitas, W. Hollik, W. Walter and G. Weiglein, Phys. Lett B495 (2000)
338,
{\tt arXiv:hep-ph/0007091 };
A. Freitas, S. Heinemeyer, W. Hollik, W. Walter 
and G. Weiglein, Nucl. Phys. Proc. Suppl. 89 (2000) 82,
{\tt arXiv:hep-ph/0007129 }

\refno\Susy.
W. Hollik, E. Kraus, D. St\"ockinger, Eur. Phys. J C 20 (2001) 105,
{\tt arXiv:hep-ph/9907393 }

\refno\Epstein.
H. Epstein and V. Glaser, 
Annales Poincare Phys.\ Theor.\ A 19 (1973) 211, and
in ``Statistical Mechanics and Quantum Field Theory'' (C. De Witt
and R. Storea, Edts.). p. 501, Gordon \& Breach, New and Breach, 
New York 1971.

\refno\QAP.
Y.M.P. Lam, Phys. Rev. D 6 (1972), 2145;
D 7 (1973), 2943;
J. Lowenstein, Comm. Math. Phys. 24 (1971), 1:
J. Lowenstein and B. Schroer, Phys. Rev. D 7 (1975), 1929;
W. Zimmermann, Comm. Math. Phys. 39 (1974). 81

\refno\BRSah.
C. Becchi, A. Rouet and R. Stora, Commun. Math. Phys. 42 (1975), 127.

\refno\BRSann.
C. Becchi, A. Rouet and R. Stora, Ann. Phys. 98 (1976), 287;
C. Becchi, Lectures on Renormalization of Gauge Theories, in
``Relativity, group and topology II'', Les Houches, 1983.

\refno\Piguet.
O. Piguet and A. Rouet, Phys. Rep. 76 (1981) 1;
O. Piguet and S. Sorella, Algebraic renormalization
(Springer-Verlag, Berlin, 1995) 
and references therein.

\refno\ARSM.
E. Krauss, K. Sibold, Nucl. Phys. Proc. Suppl. 51C (1996) 81-87,\ \ \ 
{\tt arXiv:hep-th/9608143}; 
E. Krauss, 
Ann. Phys. (NY) 262 (1998) 155,
{\tt arXiv:hep-th/9709154};
Acta. Phys. Pol. B29 (1998), 2647,
{\tt arXiv:hep-th/9807102};
``Renormalization of the
Electroweak Standard Model''
in \ Lectures of Saalburg Summer
School, 1997, \ \
{\tt arXiv: hep-th/9809069};
P. A. Grassi,
Nucl. Phys. B537 (1999) 527,
{\tt arXiv:hep-th/9804013};
Nucl. Phys. B560 (1999) 499,
{\tt arXiv:hep-th/9908188};
P. Gambino, P. A. Grassi,
Phys. Rev. D62 (2000) 076002,
{\tt arXiv:hep-th/9907254}.

\refno\GrassiAH.
R. Ferrari and P.A. Grassi, Phys. Rev. D60 (1999) 065010,
{\tt arXiv:hep-th/9807191}.

\refno\GrassiPAR.
P.A. Grassi, T. Hurth and M. Steinhauser, Ann. Phys. 288 (2001) 
197, 
{\tt arXiv: hep-ph/9907426};
JHEP 0011 (2000), 037,
{\tt arXiv: hep-ph/0011067};
Nucl. Phys. B 610 (2001) 215,
{\tt arXiv: hep-ph/0102005 1067};
P.A. Grassi and T. Hurth,
Talk at 5th International Symposium on Radiative Corrections
(RADCOR-2000),
{\tt arXiv: hep-ph/0101183};

\refno\abhk.
P. Higgs, Phys. Rev. 145 (1966) 1156;
T.W. Kibble, Phys. Rev. 155 (1967) 1554;
B.W. Lee, Phys. Rev. D 5 (1972), 823;
T. Applequist and H. Quinn, hys.\ Lett.\ B 39 (1972) 229;
Y. Yao, Phys. Rev. D7 (1973) 1647;
J. Lowenstein, M. Weinstein and W. Zimmermann, Phys. Rev. D 10 (1974)
1854; D 10 (1974) 2500;
O. Piguet, Comm. Math. Phys. 37 (1974), 19;
E. Kraus, Z. Phys. C. 68 (1995) 331,
{\tt arXiv: hep-th/9503140};
C75 (1997) 739,
{\tt arXiv: hep-th/9608160}.

\refno\GrossJackiw.
D.J. Gross and R. Jackiw, Phys. Rev. D 6 (2972) 477.

\refno\Mathematica.
S. Wolfram, Mathematica, A System for Doing Mathematics by Computer,
Addison Wesley (1991).

\refno\Tracer.
M. Jamin and M.E. Lautenbacher, Comput. Phys. Commun. 74 (1993) 265.

}

\bye